\documentclass[journal]{IEEEtran}
\usepackage{amssymb,stmaryrd,amsmath,amsfonts,rotating}%amsfonts,amsthm 
\usepackage[noadjust]{cite}
\usepackage{color}%temporary packages
\usepackage[vflt]{floatflt}
\usepackage{epic}

\definecolor{TODO}{rgb}{0.6,0.6,0.6} % TO DO!!!!

\definecolor{TOCHECK}{rgb}{0.8,0.8,0.8} % TO CKECK!!!!

\newtheorem{theorem}{Theorem}
\newcommand{\btheo}{\begin{theorem}}
\newcommand{\etheo}{\end{theorem}}
\newcommand{\bproof}{\begin{proof}}
\newcommand{\eproof}{\end{proof}}
\newtheorem{definition}{Definition}
\newcommand{\bdefi}{\begin{definition}}
\newcommand{\edefi}{\end{definition}}
\newtheorem{fact}{Fact}
\newcommand{\bprop}{\begin{fact}}
\newcommand{\eprop}{\end{fact}}
\newtheorem{corollary}{Corollary}
\newcommand{\bcor}{\begin{corollary}}
\newcommand{\ecor}{\end{corollary}}
\newtheorem{example}{Example}
\newcommand{\bex}{\begin{example}}
\newcommand{\eex}{\end{example}}
\newtheorem{lemma}{Lemma}
\newcommand{\blemma}{\begin{lemma}}
\newcommand{\elemma}{\end{lemma}}
\newtheorem{remark}{Remark}
\newcommand{\bremark}{\begin{remark}}
\newcommand{\eremark}{\end{remark}}
\newtheorem{conj}{Conjecture}
\newcommand{\reals}{\ensuremath{\mathbb{R}}}
\newcommand{\naturals}{\ensuremath{\mathbb{N}}}
\newcommand{\expectation}{\ensuremath{\mathbb{E}}}

\newcommand{\defas}{\ensuremath{\stackrel{{\vartriangle}}{=}}} 
\def\0{{\tt 0}} % Ex.: BPSK modulation => 0 is encoded into +1
\def\1{{\tt 1}} % Ex.: BPSK modulation => 1 is encoded into -1
\def\?{{\tt *}} % erasure symbol
\newcommand{\graph}{{\ensuremath{\tt G}}}
\newcommand{\ldpc}{{{\ensuremath{\text{LDPC}}}}}
\newcommand{\ddp}{dd~pair~} %degree distribution pair %d.d.
\newcommand{\n}{\ensuremath{n}} % code length
\newcommand{\drate}{r} % design rate
\newcommand{\ledge}{\ensuremath{\lambda}} % ledge (edge perspective) 
\newcommand{\redge}{\ensuremath{\rho}} % redge (edge perspective
\newcommand{\lnode}{\ensuremath{\Lambda}} % lnode (node perspective) 
\newcommand{\ldegree}{{\ensuremath{\ensuremath{\tt l}}}} % left degree (regular codes) 
\newcommand{\rdegree}{{\ensuremath{\ensuremath{\tt r}}}} % right degree (regular codes) 
\newcommand{\ih}{\ensuremath{{\tt{h}}}} % intrinsic (or input or channel) entropy
\newcommand{\cp}{\epsilon} % channel parameter 
\newcommand{\xh}{\ensuremath{h}} % extrinsic (or EXIT) entropy
\newcommand{\MAP}{\ensuremath{\text{MAP}}} % Maximum A Posteriori
\newcommand{\BP}{\ensuremath{\text{BP}}} % Belief Propagation 
\newcommand{\EBP}{\ensuremath{\text{EBP}}} % Extended Belief Propagation
\newcommand{\Sh}{\ensuremath{\text{Sh}}} % Shannon
\newcommand{\MAPsmall}{\ensuremath{\text{\tiny MAP}}} % Maximum A Posteriori
\newcommand{\BPsmall}{\ensuremath{\text{\tiny BP}}} % Belief Propagation
\newcommand{\EBPsmall}{\ensuremath{\text{\tiny EBP}}} % Extended Belief Propagation
\newcommand{\Shsmall}{\ensuremath{\text{\tiny Sh}}} % Shannon
\newcommand{\itersmall}{{\tiny \text{$\ell$}}} % 

\newcommand{\xl}{\ensuremath{{\tt{x}}}}%\xi}} %decoding parameter, FIND A LETTER FOR IT!!!
\renewcommand{\mid}{\,|\,}
\newcommand{\tablespace}{\vspace{10pt}}

\newcommand{\Tc}{T}
\newcommand{\Td}{R}
\newcommand{\de}{{\text d}}

\newcommand{\exit}{\ensuremath{\text{EXIT}}}
\newcommand{\gexit}{\ensuremath{\text{GEXIT}}}

\newcommand{\dens}[1]{\mathsf{#1}}
\newcommand{\Ldens}[1]{\dens{#1}}
\newcommand{\absdens}[1]{\absd{\dens{#1}}}
\newcommand{\absLdens}[1]{\absdens{#1}}
\newcommand{\Ddens}[1]{\mathfrak{{#1}}}
\newcommand{\BEC}{\ensuremath{\text{BEC}}}
\newcommand{\EC}{\ensuremath{\text{EC}}}
\newcommand{\BSC}{\ensuremath{\text{BSC}}}
\newcommand{\BAWGNC}{\ensuremath{\text{BAWGNC}}}
\newcommand{\BMS}{\ensuremath{\text{BMS}}}
\newcommand{\BM}{\ensuremath{\text{BM}}}
\newcommand{\BSCsmall}{\ensuremath{\text{\tiny BSC}}}
\newcommand{\BMSsmall}{\ensuremath{\text{\tiny BMS}}}
\newcommand{\BECsmall}{\ensuremath{\text{\tiny BEC}}}
\newcommand{\BAWGNCsmall}{\ensuremath{\text{\tiny BAWGNC}}}

\newcommand{\absd}[1]{|#1|}
\newcommand{\absDdens}[1]{\absd{\Ddens{#1}}}

\newcommand{\exitkl}[1]{l(#1)}  % EXIT kernel in L domain
\newcommand{\gexitkl}[2]{l^{#1}(#2)}  % GEXIT kernel in L domain
\newcommand{\gexitkabsd}[2]{|d|^{#1}(#2)}  % GEXIT kernel in |D| domain
\newcommand{\exitf}[1]{h^{#1}}  % EXIT function 
\newcommand{\exitfc}[2]{h^{#1}_{#2}}  % EXIT function
\newcommand{\exitfi}[1]{h^{#1}_i}  % EXIT function
\newcommand{\gexitf}[1]{g^{#1}}  % GEXIT function 
\newcommand{\gexitfc}[2]{g^{#1}_{#2}}  % GEXIT function
\newcommand{\gexitfi}[1]{g_i^{#1}}  % GEXIT function
\newcommand{\gexitfroot}[1]{g_{\text{r}}^{#1}}  % GEXIT function
\newcommand{\gexitfleaf}[1]{g_{\text{l}}^{#1}}  % GEXIT function

\newcommand{\newcaption}[1]{\caption{\footnotesize{#1}}}
\newcommand{\eldpc}[3]{{\rm LDPC}(#1, #2, #3)}

\def\cX{{\cal X}}
\def\cY{{\cal Y}}

\def\0t{{\tt 0}}
\def\1t{{\tt 1}}

\def\snr{{\sf snr}}
\def\R{{\mathbb R}}
\def\E{{\mathbb E}}

\newcommand{\iter}{\ell}
\newcommand{\drmax}{{\tt r}_{\text{max}}}

\newcommand{\entropy}{H} % entropy operator 
\newcommand{\gentropy}{G} % entropy operator
\newcommand{\conv}{\star}
\DeclareMathOperator{\perr}{\mathfrak{E}}       % probability of error operator
\DeclareMathOperator{\batta}{\mathfrak{B}}      % Battacharya operator
\newcommand{\ent}{\ensuremath{{\tt{h}}}}
\newcommand{\logtwo}{\log_2}
\newcommand{\lognat}{\log}

\begin{document}
\title{The Generalized Area Theorem and Some of its Consequences}
\author{
	Cyril~M\'easson,$^\dagger$ %
	\thanks{$\dagger$ EPFL, School for Computer and Communication Sciences, CH-1015 Lausanne, Switzerland. E-mail: cyril.measson@epfl.ch} %
	Andrea~Montanari,$^*$%
	\thanks{$*$ ENS, Laboratoire de Physique Th\'eorique, F-75231 Paris, France. E-mail: montanar@lpt.ens.fr} %
	Tom Richardson,$^+$%
	\thanks{$+$ Flarion Technologies, Bedminster, USA E-mail: tjr@flarion.com} %
	and R\"udiger Urbanke%
	$^\ddagger$\thanks{$\ddagger$ EPFL, School for Computer and Communication Sciences, CH-1015 Lausanne, Switzerland. E-mail: ruediger.urbanke@epfl.ch} %
	\thanks{Parts of the material were presented in \cite{MMRU04, MMRU05}.}
} %

\maketitle

\begin{abstract}
There is a fundamental relationship between belief propagation and 
maximum a posteriori decoding. The case of transmission over the 
binary erasure channel was investigated in detail in a companion paper. 
This paper investigates the extension to general 
 memoryless channels (paying special attention to the binary case). 
An area theorem for transmission over general memoryless channels is  
introduced 
and some of its many consequences are discussed. We show that 
this area theorem gives rise to an upper-bound
on the maximum a posteriori threshold for sparse graph codes.
In situations where this bound is tight, the extrinsic
soft bit estimates delivered by the belief propagation decoder coincide with
the correct a posteriori probabilities above the maximum a 
posteriori threshold.
More generally, it is conjectured that the fundamental relationship between the
maximum a posteriori  and the belief propagation decoder which
was observed for transmission over the binary erasure channel
carries over to the general case. We  finally demonstrate
that in order for the design rate of an ensemble 
to approach the capacity under belief propagation  decoding the 
component codes have to be perfectly matched, a statement which
is well known for the special case of transmission over the binary
erasure channel.
\end{abstract}

\begin{keywords}
belief propagation, maximum a posteriori, maximum likelihood, Maxwell construction, 
threshold, phase transition, Area Theorem, $\exit$ curve, entropy
\end{keywords}

%\IEEEpeerreviewmaketitle

\section{Introduction}
\begin{figure}[hbt]
\vspace{25bp}
\centering
\setlength{\unitlength}{0.5bp}
\begin{picture}(105,153)
\put(-45,0){\includegraphics[scale=0.62]{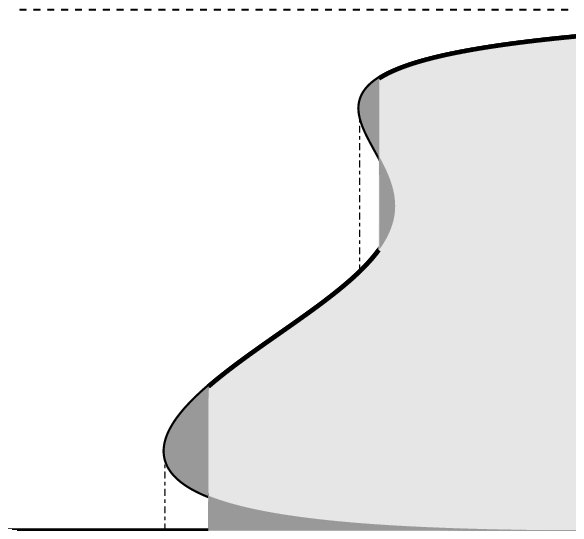}}
\put(36,-7){\makebox(0,0){\small{$\ent^{\MAPsmall}$}}}
\put(7,-7){\makebox(0,0){\small{$\ent^{\BPsmall}$}}}
\put(116,88){\makebox(0,0){\small{$\exitf {} (\ent)$}}}
\put(46,105){\makebox(0,0){\small{$\exitf{\BPsmall}(\ent)$}}}
\end{picture}
\caption{\label{fig:multijump} 
The $\EBP$ $\exit$ curve (double ``S''-shaped curve), the corresponding
$\BP$ $\exit$ curve (dashed and solid line; 
the ``envelope'' of the $\EBP$ $\exit$ curve) 
and the $\MAP$ $\exit$ curve (thick solid line; 
constructed by ``cutting'' $\EBP$ $\exit$ at the
two ``S''-shaped spots in such a way that there is a local balance of the 
areas shown in gray) 
for the ensemble LDPC($\frac{3x+3x^2+4x^{13}}{10},x^6$).  }
\vspace{5bp}
\end{figure}
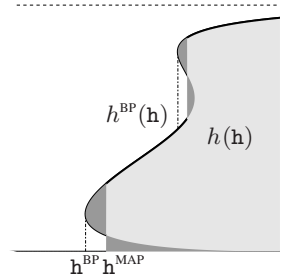

\PARstart{I}{t} was shown in \cite{MeU03,MMU04,MMU05} that, when transmission takes place over the
binary erasure channel ($\BEC$) using sparse graph codes, there exists a surprising and fundamental relationship
between the belief propagation ($\BP$) and the maximum a posteriori ($\MAP$) decoder. 
This relationship emerges in the limit
of large blocklengths. Operationally, this relationship is furnished
for the $\BEC$ by the so-called Maxwell decoder. This decoder bridges the gap between
$\BP$ and $\MAP$ 
decoding by augmenting the $\BP$ decoder with an additional ``guessing''
device. Analytically, the relationship between $\BP$ and $\MAP$ decoding
is given in terms of the so-called extended $\BP$ $\exit$ ($\EBP$ $\exit$) function. 
Fig.~\ref{fig:multijump} shows this curve (double ``S''-shaped curve)
for transmission over the $\BEC$ and
the ensemble LDPC($\frac{3x+3x^2+4x^{13}}{10},x^6$) (the degree distributions are
from an edge perspective).
The $\BP$ $\exit$ curve 
is the ``envelope'' of the $\EBP$ $\exit$ curve (let a ball run slowly down the slope). 
The $\MAP$ $\exit$ curve on the other hand
is conjecture to be derived in general from the 
$\EBP$ $\exit$ curve by the so-called Maxwell
construction. This Maxwell construction consists of converting the 
$\EBP$ $\exit$ curve
into a single-valued function by ``cutting'' the $\EBP$ $\exit$ curve at the
two ``S''-shaped spots in such a way that there is a local balance of the cut areas. 
A detailed discussion of this relationship in the case of transmission
over the $\BEC$ can be found in \cite{MMU05}.
Let us summarize. For transmission over the $\BEC$ using sparse
graph codes from long ensembles, 
$\BP$ decoding is asymptotically characterized
by its $\BP$ $\exit$ curve and $\MAP$ decoding is characterized by its $\MAP$ $\exit$ curve.
These two curves are linked via the  $\EBP$ $\exit$ curve.

\subsection{Overview of Results}

The pleasing picture shown in Fig.~\ref{fig:multijump}  
seems to have a fairly complete analog in the general setting.
Unfortunately we are not able to prove this claim in any generality.
But we show how several of the key ingredients can be suitably extended
to the general case and we will be able to prove some of their fundamental
properties.

Namely, we introduce a general area theorem (GAT). This
area theorem, when applied to the $\BEC$, leads back to the notion of $\exit$ functions
as shown in the companion paper \cite{MMU05}.  
For the general case however, it is necessary to use a distinct 
function (but similar in many respects to $\exit$).
We call it 
the generalized $\exit$ ($\gexit$) function. We then show that $\gexit$ functions share
some of the key properties with $\exit$ functions. In particular, we are able
to extend the upper-bound on the $\MAP$ threshold presented in \cite{MeU03} (or, more generally, the lower
bound on the conditional entropy) to general channels.

In \cite{GSV04c,GSV05} Guo, Shamai and Verd{\'u} showed that for Gaussian channels
the derivative
(with respect of the signal-to-noise ratio) of the mutual information,
is equal to the mean square error (MSE),
and in \cite{GSV04c} they showed that a similar
relationship holds for Poisson channels.
One can think of $\gexit$ functions as providing such
a relationship in a more general
setting (where the generalization
is with respect to the admissible channel families).
For some channel families, 
$\gexit$ functions have particularly nice interpretations.
E.g., for Gaussian channels,
we not only have the interpretation of the derivative in terms of the MSE
detector, but this interpretation can be simplified 
even further in the binary case: the derivative of the mutual information
can be seen as the ``magnetization'' of the system
as was shown by
Macris in \cite{Mac05}.
The results in \cite{GSV05i}, which have appeared since the introduction 
of GEXIT functions in \cite{MMRU04}, can be reformulated
to give an interpretation of GEXIT functions
for the class of additive channels (see also \cite{Zak05}). 
It is likely that interpretations for other classes
of channels will be found in the future.

\subsection{Paper Outline}
In Section \ref{sec:review} we review the necessary background material
and in particular recall the GAT first stated in \cite{MMRU04}. 
Starting from this GAT, we introduce in Section \ref{sec:gexit}
$\gexit$ functions. We will see that for transmission over the $\BEC$,
$\gexit$ functions coincide with  standard $\exit$ functions, but that
this is no longer true for general channels.  
In Section \ref{sec:asymptotic} we then concentrate on 
LDPC ensembles. In particular we define the quantities which
appear in the asymptotic setting.
In Section \ref{sec:basicproperties} we
then prove one of the fundamental properties of $\gexit$ functions, namely
that $\gexit$ kernels preserve the ordering implied by physical degradation.
This fact is then exploited in Section \ref{sec:upperbound}, where we show
how to compute an upper bound on the threshold under $\MAP$ decoding
(or, more generally, a lower bound on the conditional entropy) by considering
the \BP\ $\gexit$ function, which results 
from the regular $\gexit$ function if we substitute
the MAP density by its equivalent BP density.
In Section \ref{sec:egexit} we define extended BP $\gexit$ (EBP $\gexit$) functions which
include the unstable branches, present several examples of these function
and discuss how they provide a bridge between belief propagation and 
maximum a posteriori decoding.
Several properties of EPB $\gexit$ functions are discussed
in Section \ref{sec:HowToEBP} together with a numerical procedure for 
constructing them. We show that they satisfy an area theorem as well.
Section \ref{sec:Regularity} presents some partial results on the smoothness
and uniqueness of EBP $\gexit$ functions.
In Section \ref{sec:mapversusbp} we show the surprising fact that, in case the
previously computed upper bound on the $\MAP$ threshold is tight, then the
a posteriori probabilities on the bits are equal to the corresponding 
$\BP$ estimates. Section \ref{sec:Matching} contains a proof
that iterative coding systems cannot achieve reliable communication
above capacity, using only density evolution and the area theorem 
(and not the standard Fano inequality).
A matching condition for component codes of capacity achieving
sequences follows. 
In the appendices we collect some technical derivations and
a discussion of several equivalent forms of the $\gexit$ 
functions for Gaussian channels. 
We finally conclude
with some remarks in Section \ref{sec:conclusion}.
%
%********************************************************************
%
\section{Review and Notations}
\label{sec:review}
Let ${\cal{X}}$ denote the channel input alphabet 
(which we always assume finite)
and ${\cal{Y}}$ the channel output alphabet (typically, ${\cal{Y}}={\mathbb{R}}$).
All channels considered in this paper are {\em memoryless} (M).
Rather than looking at a single memoryless channel, we usually consider
{\em families} of memoryless channels parameterized by a real-valued
parameter $\cp$, which we denote by $\{\text{M}(\cp)\}_{\cp}$. 
Each channel from such
a family is characterized by its transition probability density
$p_{Y \mid X}(y \mid x)$ (where $x\in\cX$ and $y\in\cY$).
We adopt here the convention of formally denoting channels by 
their transition density even when such a density does not exist,
and write $\int f(y) p_{Y \mid X}(y \mid x)\text{d}y$ as a proxy for
the corresponding expectation.

Transmission over {\em binary}-input
{\em memoryless} output-{\em symmetric}\footnote{A binary 
memoryless channels is said to be symmetric 
(or, more precisely, output-symmetric) when 
the transition probability verifies 
$p_{Y|X}(y|+1)=p_{Y|X}(y|-1)$.} ($\BMS$) channel plays
a particularly important role. 
In this case, it will be convenient to 
assume that the input bit $X_i$ takes values 
$x_i\in {\cal{X}}\defas\{+1,-1\}$. 
The channel indexed by parameter $\cp$ is
generically denoted by $\BMS(\cp)$. 

In the sequel
we will often assume that the {\em channel} family $\{\BMS(\cp)\}_{\cp}$
is {\em ordered by physical degradation} (see
\cite{RiU05} for a discussion of this concept). 
It is well known that the standard families
$\{\BEC(\cp)\}_{\cp=0}^{1}$ (binary erasure
channels with erasure parameter $\epsilon$),
$\{ \text{BSC}(\cp)\}_{\cp=0}^{\frac{1}{2}}$
(binary symmetric channels with cross-over probability $\epsilon$), and
$\{ \text{BAWGNC}(\sigma)\}_{\sigma=0}^{\infty}$
(binary-input additive white Gaussian noise channels $Y=X+N$
where $X$ takes values in ${\cal{X}}$ and the noise $N$
has standard deviation $\sigma$ and zero-mean) all have
this property. For notational simplicity we will use a shorthand
and say that a channel family is {\em degraded}.

In the binary case, an important role is played by the distribution of
the log-likelihood ratio
$L \defas \log\frac{p_{Y|X}(Y|+1)}{p_{Y|X}(Y|-1)}$, assuming $X=1$.
We denote the corresponding density by $\Ldens{c}(l)$ and call it
an $L$-density.
In fact, without loss of generality we can assume that
the log-likelihood ratio ($L$) mapping, 
$y\mapsto \log\frac{p_{Y|X}(y|+1)}{p_{Y|X}(y|-1)}$,
is already included in the channel description.
This is justified since the 
random variable $L$ constitutes a 
sufficient statistic. 
This inclusion of the $L$-processing is equivalent to assuming
that $p_{Y|X}(y|+1)= \Ldens{c}(l)$.
Further facts regarding $\BMS$ channels can be found in \cite{RiU05}. 
As far as LDPC and iterative coding systems are concerned, 
we will keep the formalism introduced in the
companion paper \cite{MMU05} and which is found, e.g., 
in \cite{LMSS01,LMSS01b,RiU01,RSU01}.

In the case of  a {\em non-binary} input alphabet ${\cal X}$,
the log-likelihood mapping will be replaced by the 
`canonical' representation of the channel output 
$y\mapsto \nu(y) \defas\{p_{Y|X}(y|x)/z(y)\, :\, x\in{\cal X} \}$,
where $z(y) \defas \sum_{x\in {\cal X}} p_{Y|X}(y|x)$.
Notice that $\nu(y)$ belongs to the $(|{\cal X}|-1)$-dimensional
simplex $S_{|{\cal X}|-1}$. 
In the binary case, the log-likelihood ratio is just
a particular parametrization of the one-dimensional simplex.

In what follows we will often be concerned with how certain
quantities (e.g., the conditional entropy $H(X \mid Y)$)
behave as we change the channel parameter. In order
to ensure that the involved objects exits we need to impose 
some regularity conditions on the channel family with respect
to the channel parameter. This can be done in various ways, but
to be concrete we will impose the following restriction. 
\bdefi[Channel Smoothness]
Consider a family of memoryless channels with input and output alphabets 
${\cal X}$ and ${\cal Y}$, respectively, 
and characterized by their transition
probability $p_{Y|X}(y | x)$ (with $y$ taking the canonical form described 
above). 
Assume that the family is parameterized by $\cp$,
where $\cp$ takes values in some interval $I \subseteq {\mathbb R}$.
The channel family is said to be {\em smooth}
with respect to the parameter $\cp$ if for all $x\in {\cal X}$ and
all bounded continuously differentiable functions
$f(y)$ on $S_{|{\cal X}|-1}$, the integral 
$\int f(y) p_{Y | X}(y | x) \text{d} y$ exists
and is a continuously differentiable function with respect to $\cp$, $\cp \in I$.
\edefi
In the sequel we often say as a shorthand  
that a {\em channel} $\BMS(\cp)$ is smooth
to mean that we are transmitting over the channel
$\BMS(\cp)$ and that the {\em channel family} $\{\BMS(\cp)\}_{\cp}$ is
smooth at the point $\cp$. If $\BMS(\cp)$ is smooth, the derivative
$\frac{\text{d}\phantom{\cp}}{\text{d}\cp}\int f(y) p_{Y | X}(y | x) 
\text{d} y$ exists and 
is a linear functional of $f$. It is therefore consistent to formally
{\em define} the derivative of $ p_{Y | X}(y | x)$ with respect 
to $\cp$ by setting
\begin{align}
\frac{\text{d}\phantom{\cp}}{\text{d}\cp}\int\! f(y) p_{Y | X}(y | x) 
\, \text{d} y \defas
\int\! f(y) \frac{\text{d}p_{Y | X}(y | x)}{\text{d}\cp} \,
\text{d} y\, .\label{eq:FormalDerivative}
\end{align}
For a large class of channel families it is straightforward
to check that they are smooth.
This is e.g. the case if $\{\cal Y\}$ is finite and the
transition probabilities are differentiable functions of $\cp$,
or if it admits a density with respect to the Lebesgue measure,
and the density is differentiable for each $y$.
In these cases, the formal derivative (\ref{eq:FormalDerivative})
coincides with the ordinary derivative.

\bex[Smooth Channels] \label{ex:channeldefinition}
It is straightforward to check that the families
$\{\BEC(\cp)\}_{\cp=0}^{1}$,
$\{ \text{BSC}(\cp)\}_{\cp=0}^{\frac{1}{2}}$, and
$\{ \text{BAWGNC}(\sigma)\}_{\sigma=0}^{\infty}$ 
are all smooth.
\eex

In the case of transmission over a $\BMS$ channel
it is useful to parameterize
the channels in such a way that the parameter reflects
the channel entropy. More precisely, we denote 
by $\ent$ the conditional entropy $H(X|Y)$ when
the channel input $X$ is chosen uniformly at random from $\{+1,-1\}$, and 
the corresponding output is $Y$. Consider a family of
$\BMS$ channels characterized by their $L$-densities.
We then write this family of $L$-densities as
$\{\Ldens{c}_{\ent}\}_{\ent}$ if
$\entropy(\Ldens{c}_{\ent})=\ent$, where the {\em entropy
operator} is defined as (see, e.g., \cite{RiU05})
\begin{align}
\label{equ:entropy}
\entropy(\Ldens{c}) & \defas \int_{-\infty}^{\infty} 
\Ldens{c}(y) \logtwo(1+e^{-y}) \text{d}y =
\int_{-\infty}^{\infty}
\Ldens{c}(y) \exitkl y \text{d}y.
\end{align}
This integral always exists as can be seen by writing it
in the equivalent form as Rieman-Stieltjes integral
$ \int_{0}^{\infty}
h_2\left(\frac{e^{-y}}{1+e^{-y}} \right)\text{d}\absLdens{C}(y)$.
In the above definition we have introduced the {\em kernel}
$\exitkl y \defas \logtwo(1+e^{-y})$. For reasons that
will become clearer in Lemma \ref{lem:exitlinear}, 
we call $\exitkl y$ the $\exit$ kernel.

The channel family is said to be {\em complete} if 
$\ent$ ranges from $0$ to $1$. For the binary erasure channel the natural parameter $\cp$ 
(the erasure probability) already
represents an entropy. Nevertheless, to be consistent we will write
in the future $\BEC(\ent)$. By some abuse of notation,
we write BSC$(\ent)$ to denote the BSC with cross-over probability
equal to $\cp(\ent)=h_2^{-1}(\ent)$, 
where $h_2(x) \defas -x \log_2(x)-(1-x) \log_2(1-x)$, the 
binary entropy function. In the same manner,
BAWGNC$(\ent)$ denotes the BAWGNC with a standard deviation of the
noise such that the channel entropy is equal to $\ent$.

We will encounter cases where it is useful to allow each
bit of a codeword to be transmitted through a different (family of) 
$\BMS$ channel(s). By some abuse of notation, we will denote the
$i^{\text{th}}$ channel family by $\{\BMS(\ent_i)\}_{\ent_i}$. 
%and its corresponding family of $L$-density
%by $\Ldens{c}_i$. 
A situation in which this
more general view appears naturally is when we consider 
punctured ensembles. We can describe this case by assuming that
some bits are passed through an erasure channel with erasure
probability equal to one, whereas the remaining bits are passed
through some other $\BMS$ channel. In such cases it is convenient
to assume that all individual families $\{\BMS(\ent_i)\}_{\ent_i}$ 
are parameterized 
in a smooth (differentiable) way by a single
real parameter, call it $\cp$, i.e., $\ent_i=\ent_i(\cp)$.
In this way, by changing $\cp$ all channels change according to $\ent_i(\cp)$
and they describe a path through ``channel space''.

The general area theorem (GAT), first introduced in \cite{MMRU04}, 
plays center stage in the remainder of this paper. 
\btheo[General Area Theorem]
\label{theo:generalareatheorem}
Let $X$ be chosen with probability $p_X(x)$ from ${\cal X}^n$. 
Let the channel from $X$ to $Y$ be memoryless, where
$Y_i$ is the result of passing $X_i$ through the smooth 
family $\{\text{M}(\cp_i)\}_{\cp_i}$, $\cp_i \in I_i$.
Let $\Omega$ be a further observation of $X$ so that
$p_{\Omega \mid X, Y}(\omega \mid x, y) = p_{\Omega \mid X}(\omega \mid x)$.
Then
\begin{align}
\label{eq:generalareatheorem}
\text{d} H(X \mid Y,\Omega) & =
\sum_{i=1}^\n \frac{\partial H(X_i \mid Y,\Omega ) }{\partial \cp_i} \text{d}\cp_i.
\end{align}
\etheo
\bproof
For $i \in [\n]$, the entropy rule gives $H(X \mid Y,\Omega)=H(X_i \mid Y,\Omega)+H(X_{\sim i} \mid X_i,Y,\Omega)$.
We claim that
\begin{align}
\label{equ:factorization}
p(X_{\sim i} \mid X_i,Y,\Omega) = p(X_{\sim i} \mid X_i,Y_{\sim i},\Omega),
\end{align}
which is true since the channel is memoryless and $p_{\Omega \mid X, Y}(\omega \mid x, y) = p_{\Omega \mid X}(\omega \mid x)$.
Furthermore $H(X_i \mid Y,\Omega)$ is differentiable with
respect to $\cp_i$ as a consequence of the channel smoothness
(it is straightforward to write the conditional entropy as expectation
of a differentiable kernel, cf.~Lemma \ref{lemma:gexitBM} and remarks below).
Therefore, $H(X_{\sim i} \mid X_i,Y,\Omega)=H(X_{\sim i} \mid X_i,Y_{\sim i},\Omega)$ and
$\frac{\partial H(X \mid Y,\Omega ) }{\partial \cp_i} = \frac{\partial H(X_i \mid Y,\Omega) }{\partial \cp_i}$.
From this the total derivate as stated in 
(\ref{eq:generalareatheorem}) follows
immediately.
\eproof

\section{$\gexit$ Functions}
\label{sec:gexit}
Let $X$ be chosen with probability $p_X(x)$ from ${\cal X}^n$.
Assume that the $i^{\text{th}}$ component of $X$ is transmitted
over a memoryless erasure channel (not necessarily binary) with
erasure probability $\cp_i$, denote it by $\EC(\cp_i)$.
Then 
$H(X_i \mid Y) = \bar{\cp}_i H(X_i \mid Y_i=X_i, Y_{\sim i}) + \cp_i 
H(X_{i} \mid Y_i=?, Y_{\sim i}) = \cp_i
H(X_{i} \mid Y_{\sim i})$.
Apply equation (\ref{eq:generalareatheorem}) in 
Theorem \ref{theo:generalareatheorem}
assuming that $\cp_i=\cp$, $i \in [n]$.
To remind ourselves that $Y$ is a function of the
parameter $\cp$ we write $Y(\cp)$.
Then
\begin{align*}
\frac{1}{n} \frac{\text{d}\phantom{\cp}}{\text{d} \cp} H(X \mid Y(\cp)) 
& = \frac{1}{n} \sum_{i=1}^{n} H(X_{i} \mid Y_{\sim i}(\cp)).
\end{align*}
The function $h^{}_i(\cp) \defas H(X_{i} \mid Y_{\sim i}(\cp))$
is known in the literature as the $\exit$ 
function associated to the  $i^{\text{th}}$  bit
of the given code and 
$h^{}(\cp) \defas \frac{1}{n} \sum_{i=1}^{n} H(X_{i} \mid Y_{\sim i}(\cp))$
is the ({\em average}) $\exit$ function.\footnote{\label{foo:exit} More precisely,
$\exit$ functions are usually defined as 
$I(X_i \mid Y_{\sim i}(\epsilon))=H(X_i)-H(X_i \mid Y_{\sim i}(\epsilon))$, which differs
from our definition only in a trivial way.}
We conclude that {\em for transmission over $\EC(\cp)$},
$h^{}(\cp) = \frac{1}{n} \frac{\text{d} \phantom{\cp}}{\text{d} \cp} H(X \mid Y(\cp))$. 
If we integrate this relationship with respect to
$\cp$ from $0$ to $1$ and note that
$H(X \mid Y(0))=0$ and 
$H(X \mid Y(1))=H(X)$, then we get the basic form of the area theorem
for the $\EC(\cp)$:
$\int_{0}^{1} h^{}(\cp) \text{d} \cp = H(X)/n$.
This statement was first proved, in the binary case, by Ashikhmin, Kramer, and
ten Brink in \cite{AKtB04} using a different framework.
\bex[Area Theorem for Repetition Code and $\BEC$]
Consider the {\em binary} repetition code with parameters $[\n, 1, \n]$,
where the first component describes the blocklength, the second
component denotes the dimension of the code, and the final component
gives the minimum (Hamming) distance.
By symmetry $h^{}_i(\ent)=h^{}(\ent)=
\ent^{n-1}$ for all $i \in [n]$.
We have $\int_{0}^{1} h^{}(\ent) \text{d} \ent=\frac{1}{n}=H(X)/n$,
as predicted.
\eex 
The above scenario can easily be generalized by allowing the various components
of the code to be transmitted over different erasure channels. Consider, e.g.,
a binary repetition code of length 
$n$ in which the first component is transmitted through $\BEC(\delta)$,
where $\delta$ is constant, but the remaining components are passed through
$\BEC(\ent)$. In this case we have 
$\int_{0}^{1} h^{}(\ent) \text{d} \ent=
(H(X \mid Y(\delta, 1, \cdots, 1))-H(X \mid Y(\delta, 0, \cdots, 0)))/n=\delta/n$
(assuming that $X$ is chosen uniformly at random from the set of codewords).
We will get back to this point shortly when we introduce $\gexit$ functions
in Definition \ref{def:gexit}.

The concept of $\exit$ functions 
extends to general channels in the natural way.
To simplify notation somewhat let us focus on the binary case.
\bdefi[$\exitf {}$ for $\BMS$ Channels]
Let $X$ be a binary vector of length $n$ chosen with probability $p_X(x)$.
Assume that transmission takes place over the family $\{\BMS(\ent)\}_{\ent}$. 
Then
\begin{align*}
\exitfi {}(\ent) & \defas H(X_i \mid Y_{\sim i}(\ent)), \\
\exitf {}(\ent) & \defas \frac{1}{n} \sum_{i=1}^{n} H(X_i \mid Y_{\sim i}(\ent))
= \frac{1}{n} \sum_{i=1}^{n} \exitfi {}(\ent).
\end{align*}
\edefi
This is the definition of the $\exit$ function introduced by ten Brink 
\cite{teB99a,teB99b,tBr00,teB00,teB01} (see footnote \ref{foo:exit}).

We get a more explicit representation if we consider transmission
using binary {\em linear} codes. In this context recall that a binary
linear code is {\em proper} if it possess a generator matrix with no zero columns.
As a consequence, in a proper binary linear code half the codewords
take on the value 
$+1$ and half the value $-1$ in each given position.
\blemma[$\exitf {}$ for Linear Codes and $\BMS$ Channels]
\label{lem:exitlinear}
Let $X$ be chosen uniformly at random from a proper binary linear code
and assume that transmission takes place over the family 
$\{\BMS(\ent)\}_{\ent}$.
Define
\begin{align}
\label{equ:phidef}
\phi_i(y_{\sim i}) & \defas
\lognat
\Bigl(
\frac{p_{X_i \mid Y_{\sim i}}(+1 \mid y_{\sim i})}{p_{X_i \mid Y_{\sim i}}(-1 \mid y_{\sim i})}
\Bigr),
\end{align}
and $\Phi_i \defas \phi_i(Y_{\sim i})$.
Let $\Ldens{a}_i$ denote the density of $\Phi_i$,
assuming that the all-one codeword was transmitted, and let
$\Ldens{a} \defas \frac{1}{n} \sum_{i=1}^{n} \Ldens{a}_i$.
Then
\begin{xalignat*}{3}
\exitfi {}(\ent)  & =
\entropy(\Ldens{a}_i), & 
\exitf {}(\ent)  & =
\entropy(\Ldens{a}),
\end{xalignat*}
where $\entropy(\cdot)$ is the entropy operator 
introduced in (\ref{equ:entropy}).
\elemma
\begin{proof}
Note that $X_i \rightarrow \Phi_i \rightarrow Y_{\sim i}$ forms
a Markov chain.\footnote{
For $z \in \reals$, let $y_{\sim i}$ be an element of
$\bigl(\phi_i\bigr)^{-1}(z)$ so that
$z = \phi_i(y_{\sim i})$. Then
$p_{X_i \mid Y_{\sim i}, \Phi_i}(x_i \mid y_{\sim i}, z)
= \frac{(1+x_i) +(1-x_i) e^{z}}{2(1+e^{z})}=
p_{X_i \mid \Phi_i}(x_i \mid z)$.
From this we conclude that 
$p_{Y_{\sim i} \mid X_i, \Phi_i}(y_i \mid x_{\sim i}, z)=
p_{Y_{\sim i} \mid \Phi_i}(y_{\sim i} \mid z)$.}
Equivalently,  we claim that $\Phi_i$ is a sufficient
statistic for $X_i$. From this we conclude that (see \cite[Section 2.8]{CoT91})
\begin{align*}
H(X_i \mid Y_{\sim i}) & =  H(X_i \mid \Phi_i).
\end{align*}
Now note that since we assume that $X$ was chosen
uniformly at random from a proper binary linear codes, it follows
that the prior for each $X_i$ is the uniform one.
Therefore, $\Phi_i$ is in fact a log-likelihood ratio.
It is shown in \cite[Lemma 3.37]{RiU05} that, assuming that $X$ is chosen
uniformly at random from a proper binary linear code, the binary
``channel'' $p(\phi_i \mid x_i)$ is symmetric.
%Further, note 
%the $L$-density of $\Phi_i$ conditioned on $X_i=1$
%is equal to the density of $\Phi_i$ conditioned on $X_i=1$.
Further, note that the density of $\Phi_i$ conditioned on $X_i=1$
is equal to the density of $\Phi_i$ 
conditioned that the all-one codeword was transmitted.\footnote{To see this,
note that, using the symmetry of the channel and the equal prior on the codewords,
we can write 
$p_{X_i\mid Y}(x_i \mid y) = c(y) 
\sum_{\tilde{x} \in {\cal C}: \tilde{x}_i=x_i} 
p_{Y \mid X}(y \tilde{x} \mid \underline{1})$, 
where $c(y)$ is a constant independent of $x_i$, ${\cal C}$ denotes
the code, and $\underline{1}$ denotes
the all-one codeword.
In the same manner, if $x' \in {\cal C}$, then 
$p_{X_i\mid Y}(x_i \mid y x') = c'(y) 
\sum_{\tilde{x} \in {\cal C}: \tilde{x}_i=x_i x_i'} 
p_{Y \mid X}(y x' \tilde{x} \mid x')$.
Compare the density of the log-likelihood ratio
assuming that the all-one codeword was transmitted to the one assuming that
the codeword $x'$ was transmitted. The claim follows by noting that
for any $y \in {\cal Y}$, $p_{Y \mid X}(y \mid \underline{1})=
p_{Y \mid X}(y x' \mid x')$, and that in this case also
$p_{Y \mid X}(y \tilde{x} \mid \underline{1})=
p_{Y \mid X}(y x' \tilde{x} \mid x')$.
}
By assumption this $L$-density is equal to  $\Ldens{a}_i$.
We conclude that $H(X_i \mid \Phi_i)=\entropy(\Ldens{a}_i)$.
\end{proof}

\begin{figure}[htp]
\centering
\setlength{\unitlength}{0.75bp}%
\begin{picture}(130,130)
\put(0,0){\includegraphics[scale=0.75]{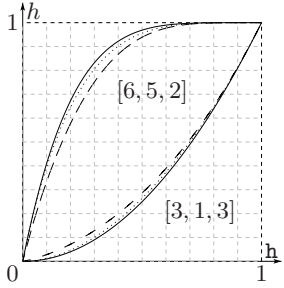}}
\put(-2, -2){\makebox(0,0)[rt]{\small $0$}}
\put(120,-2){\makebox(0,0)[t]{\small $1$}}
\put(-2,120){\makebox(0,0)[r]{\small $1$}}
\put(2,122){\makebox(0,0)[bl]{\small{$\exitf {}$}}}
\put(122, 2){\makebox(0,0)[bl]{\small{$\ent$}}}
\put(47,85){\makebox(0,0)[l]{\small $[6, 5, 2]$}}
\put(71,25){\makebox(0,0)[l]{\small $[3, 1, 3]$}}
\end{picture}
\caption{\label{fig:exitchartbecexamples} The $\exit$  function of
the $[3, 1, 3]$ repetition code and the $[6, 5, 2]$ parity-check code
for the $\BEC(\ent)$ (solid curve),
 BSC$(\ent)$ (dashed curve) and BAWGNC$(\ent)$ (dotted
curve).}
\end{figure}

As the next example shows, the $\exit$ function does {\em not}
fulfill the area theorem in the general case.
\bex[$\exit$ Function for General $\BMS$ Channels]
Fig.~\ref{fig:exitchartbecexamples} shows the $\exit$ function for
the $[3, 1, 3]$ repetition code as well as for the 
$[6, 5, 2]$ single parity-check code for  $\BEC(\ent)$,  
BSC$(\ent)$, and BAWGNC$(\ent)$. E.g., the $\exit$ function 
for the $[\n,\n-1,2]$ single parity-check code over  BSC$(\ent)$ 
is given by
\begin{align*}
\exitfi {}(\ent) = \exitf {}(\ent) = h_2\Bigl(\frac{1-(1-2\cp(\ent))^{\n-1}}{2}  \Bigr),
\end{align*}
where
$\cp(\ent)=h_2^{-1}(\ent)$. Note that these $\exit$ functions 
are ``ordered.'' More precisely, for a repetition code we get the highest
extrinsic entropy at the output for the channel 
family $\{ \text{BSC}(\ent)\}_{\ent}$ and we get the lowest such entropy if
we use instead the family $\{ \text{BEC}(\ent)\}_{\ent}$. Indeed, one can show
that these two families are the {\em least} and {\em most} ``informative'' 
family
of channels over the whole class of $\BMS$ channels for a repetition code, 
\cite{LHHH03,HuH03,SSZ03}.
The roles are exactly exchanged at a check node.
Since we know
that the $\exit$ function for the $\BEC$ fulfills the area theorem, it follows
from this extremality properties that the $\exit$ functions
for the $\BSC$ and
the $\BAWGNC$ do {\em not} fulfill the area theorem.
Indeed, for a single parity-check code with $n=3$ and the $\BSC(\ent)$ the area under
the $\exit$  function is given by 
\begin{align*}
\int_{0}^{1} h_2\left(\frac{1-(1-2\cp(\ent))^{2}}{2}  \right) \;\text{d}\ent \approx 0.643704 < 2/3.
\end{align*}
\eex
Although the above fact might be disappointing it is not surprising.
As it should be clear from the discussion at the beginning of 
this section, the $\exit$ function is related to the GAT
only in the case of the erasure channel.
Let us therefore go back to the GAT and
{\em define} the function which fulfills the area theorem 
in the general case.
\bdefi[$\gexit$ Function]
\label{def:gexit}
Let $X$ be a vector of length $n$ chosen with
probability $p_X(x)$ from ${\cal X}^n$.
Let the channel from $X$ to $Y$ be memoryless, where
$Y_i$ is the result of passing $X_i$ 
through the smooth
family $\{\text{M}(\cp_i)\}_{\cp_i}$, $\cp_i \in [0, 1]$.
Assume that all individual channels are parameterized
in a smooth (differentiable) way
by a common parameter $\cp$, i.e., $\cp_i=\cp_i(\cp)$, $i \in [n]$.
Let $\Omega$ be a further observation of $X$ so that
$p_{\Omega \mid X, Y}(\omega \mid x, y) = p_{\Omega \mid X}(\omega \mid x)$.
Then the   $i^{\text{th}}$ and the (average) {\em generalized} $\exit$ ($\gexit$) function
are defined by 
\begin{align*}
\gexitfi {}(\cp) & \defas 
\frac{\partial H(X_i | Y, \Omega)}{\partial \cp_i} \frac{\text{d} \cp_i}{\text{d} \cp} \Big|_{\cp}, \\
\gexitf {}(\cp) & \defas \frac{1}{\n}\sum_{i=1}^{\n} \gexitfi {}(\cp).
\end{align*}
\edefi
Discussion: The definition is stated in quite general terms.
First note that if we consider the
integral 
$\int_{\underline{\cp}}^{\overline{\cp}} \gexitf {}(\cp) \text{d}\cp$,
then from Theorem \ref{theo:generalareatheorem} we conclude that
the result is 
$\frac{1}{n}\bigl(H(X \mid Y(\overline{\cp}), \Omega)-H(X \mid Y(\underline{\cp}), \Omega)\bigr)$.
In words, if we smoothly change the individual channel parameters $\cp_i$
as a function of $\cp$,
then the integral of $\gexitfi {}(\cp)$ 
tells us how much the conditional entropy of the system changes
due to the total change of the parameters $\cp_i$. To be concrete,
assume, e.g., that all bits are sent through Gaussian channels. 
We can imagine that we first only change the parameter of the
Gaussian channel through which bit $1$ is sent from its initial
to its final value, then the parameter of the second channel and so
on. Alternatively, we can imagine that all channel parameters are 
changed simultaneously. In the two cases the integrals of the 
individual $\gexit$ functions $\gexitfi {}$ 
differ but their sum is the same and it equals the
total change of the conditional entropy due to the change
of channel parameters. Therefore, $\gexit$ functions can
be considered to be a ``local'' way of measuring 
the change of the conditional entropy of a system.
One should think of the common parameter $\cp$  
as a convenient way of parameterizing the
path through ``channel space'' that we are taking.

In many applications all channels are identical,
and formulas simplify significantly.
In Section \ref{sec:egexit} we will see a case in which the extra degree
of freedom afforded by allowing different channels is important.
The 
additional observation $\Omega$ is useful
if we consider the design or iterative systems and component-wise
$\gexit$ functions. For what follows though we will not need it. Hence,
we will drop $\Omega$ in the sequel.

If we assume that the input is {\em binary}
we obtain a more explicit expression for the $\gexit$ functions.
\blemma[$\gexitf {}$ for $\BM$ Channels]\label{lemma:gexitBM}
Let $X$ be a {\em binary} vector of length $n$ chosen with
probability $p_X(x)$.
Let the channel from $X$ to $Y$ be memoryless, where
$Y_i$ is the result of passing $X_i$ over 
the smooth family $\{\text{BM}(\ent_i)\}_{\ent_i}$, $\ent_i \in [0, 1]$.
Assume that all individual channel families are parameterized
in a smooth (differentiable) way
by a common parameter $\cp$, i.e., $\ent_i=\ent_i(\cp)$, $i \in [n]$.
Then the  $i^{\text{th}}$  and the (average) {\em generalized} $\exit$ ($\gexit$) function
are given by
\begin{align}
\gexitfi {}(\cp) & =
\int_{\phi_i,y_i}
\sum_{x_i}
p(x_i)
p(\phi_i \mid x_i)
\frac{\text{d}\phantom{\ent}}{\text{d} \ent_i} p(y_i|x_i)\cdot 
\label{equ:gexitcompact0} \\
& \phantom{xxxx} \cdot
\log\left\{\sum_{x'_i}\frac{p(x'_i|\phi_i)
p(y_i|x'_i)}{p(x_i|\phi_i)p(y_i|x_i)}
\right\} \frac{\text{d} \ent_i}{\text{d} \cp} \text{d} y_i \text{d} \phi_i, \nonumber \\
\gexitf {}(\cp) & = \frac{1}{n} \sum_{i=1}^{n} \gexitfi {}(\cp) ,
\label{equ:gexitcompact}
\end{align}
where $\phi_i(y_{\sim i})$ and $\Phi_i$ are
defined as in (\ref{equ:phidef}).
\elemma
Discussion: As mentioned above, the derivative
of  $p(y_i|x_i)$ in  Eq.~(\ref{equ:gexitcompact0}) has to be interpreted in 
general as in Eq.~(\ref{eq:FormalDerivative}). Moreover, writing the
same expression as $\gexitfi {}(\cp)  =\int f(y)
\frac{\text{d}\phantom{\ent}}{\text{d} \ent_i} p(y_i|x_i)\de y$,
the existence of such derivative follows from the channel
smootheness and the differentiability of $f(y)$ (if written as a 
function of the log-likelihood $\log\frac{p(y|+1)}{p(y|-1)}$.

\begin{proof}
We proceed as in the proof of Lemma \ref{lem:exitlinear}.
We claim that $X_i \rightarrow (\Phi_i, Y_i) \rightarrow Y$ forms
a Markov chain (equivalently, $(\Phi_i, Y_i)$ constitutes
a sufficient statistic).
To see this, fix $z \in \reals$ and let $y_{\sim i}$ be an element of
$\bigl(\phi_i\bigr)^{-1}(z)$, so that
$z = \phi_i(y_{\sim i})$. Then,
using the fact that $Y_i$ is conditionally independent of $Y_{\sim i}$,
given $X_i=x_i$, we may write
\begin{align*}
p_{X_i \mid Y_i, Y_{\sim i}, \Phi_i}(x_i \mid y_i, y_{\sim i}, z)
=\\ \frac{p_{Y_i\mid X_i}(y_i\mid x_i)
p_{X_i\mid  Y_{\sim i}, \Phi_i}(x_i\mid y_{\sim i},z)}
{\sum_{x_i'\in{\cal X}} 
p_{Y_i\mid X_i}(y_i|x_i')p_{X_i\mid  Y_{\sim i}, \Phi_i}
(x_i'\mid y_{\sim i},z)}
\end{align*}
Since $X_i\to \Phi_i\to Y_{\sim i}$ forms a Markov chain
(as already shown in the proof of Lemma \ref{lem:exitlinear}),
we have $p_{X_i\mid  Y_{\sim i}, \Phi_i}(x_i\mid y_{\sim i},z)=
p_{X_i\mid  \Phi_i}(x_i\mid z)$.
Substituting in the above equation, we get
$p_{X_i \mid Y_i, Y_{\sim i}, \Phi_i}(x_i \mid y_i, y_{\sim i}, z)
= p_{X_i \mid Y_i, \Phi_i}(x_i \mid y_i, z)$, as claimed.

Therefore, we can rewrite $\gexitfi {}(\cp)$ as 
\begin{align*}
\gexitfi {}(\cp) 
& =
\frac{\partial H(X_i | Y)}{\partial \ent_i} \frac{\text{d} \ent_i}{\text{d} \cp} \Big|_{\cp}
= \frac{\partial H(X_i | \Phi^{}_i, Y_i)}{\partial \ent_i} 
\frac{\text{d} \ent_i}{\text{d} \cp} \Big|_{\cp}.
\end{align*}
Expand  $H(X_i | \Phi^{}_i, Y_i)$ as
\begin{align*}
& -\int_{\phi_i, y_i} \sum_{x_i} p(x_i, \phi_i, y_i) 
\logtwo(p(x_i \mid \phi_i, y_i)) \text{d} y_i \text{d} \phi_i \\
&= -\int_{\phi_i, y_i}\sum_{x_i} 
p(x_i) p(\phi_i \mid x_i) p(y_i|x_i)\cdot \\
& \phantom{xxxx} 
\cdot\logtwo\left\{\frac{p(x_i|\phi_i)p(y_i|x_i)}{\sum_{x_i'\in\cX}
p(x'_i|\phi_i)p(y_i|x'_i)}\right\} \text{d} y_i \text{d} \phi_i .\nonumber
\end{align*}
This form has the advantage that the dependence of
$H(X_i | \Phi_i, Y_i)$ upon the channel at position $i$
is completely explicit. Let us therefore differentiate the
above expression with respect to $\ent_i$, the parameter
which governs the transition probability $p(y_i \mid x_i)$. The terms obtained
by differentiating with respect to the channel {\em inside} the $\logtwo$ 
vanish. For instance, when differentiating with respect to the $p(y_i|x_i)$
at the numerator, we get
\begin{align*}
& -\int_{\phi_i, y_i}
\sum_{x_i}
p(x_i) p(\phi_i \mid x_i)
\frac{\text{d}\phantom{\ent}}{\text{d} \ent_i}
p(y_i \mid x_i) \text{d} y_i \text{d} \phi_i \\
= & -\int_{\phi_i}
\sum_{x_i} p(x_i) p(\phi_i \mid x_i)
\frac{\text{d}\phantom{\ent}}{\text{d} \ent_i}\int_{y_i} p(y_i \mid x_i)
\text{d} y_i \text{d} \phi_i
=0\, .
\end{align*}
When differentiating with respect to the {\em outer} $p(y_i|x_i)$
we get the stated result.
\end{proof}
Although the last lemma was stated for the case of binary channels, it poses
no difficulty to generalize it. It is in fact sufficient to replace
$\phi_i(y_{\sim i})$ with any sufficient statistic
of $X_i$, given $Y_{\sim i}=y_{\sim i}$. For instance, one may take
$\phi_i(y_{\sim i}) = \{p_{X_i|Y_{\sim i}}(x_i\mid y_{\sim i}) ;\;
x_{i}\in{\cal X}\}$, which takes value on the 
$(|{\cal X}|-1)$-dimensional simplex, or any parameterization of it.
The log-likelihood can be regarded as a particular parameterization
of the $1$-dimensional simplex. More generally, 
$p_{X_i|Y_{\sim i}}(x_i\mid y_{\sim i})$ is a natural quantity appearing 
in iterative decoding. The proof (as well as the statement) applies verbatimly 
to this case.

We get an even more compact description if we assume that
transmission takes place using a binary {\em proper linear} code
and that the channel is {\em symmetric}.
\blemma[$\gexitf {}$ for Linear Codes and $\BMS$ Channels]
\label{lem:gexitlinear}
Let $X$ be chosen uniformly at random from a proper binary linear code of
length $n$. Let the channel from $X$ to $Y$ be memoryless, where
$Y_i$ is the result of passing $X_i$ over the smooth family 
$\{\BMS(\ent_i)\}_{\ent_i}$.
Assume that all individual channels are parameterized
in a smooth (differentiable) way by a common parameter $\cp$, i.e., $\ent_i=\ent_i(\cp)$, $i \in [n]$.
Let the  $i^{\text{th}}$  channel be characterized by its $L$-density, which by
some abuse of notation we denote by
$\Ldens{c}_{\BMSsmall(\ent_i)}$.
Let $\phi_i$ and $\Phi_i$ be as defined
in (\ref{equ:phidef}) and let $\Ldens{a}_i$ denote the density of 
$\Phi_i$, assuming that the all-one codeword was transmitted.
Then
\begin{align*}
\gexitfi {}(\cp) & = \int_{-\infty}^{\infty} 
\Ldens{a}_i(z) \gexitkl {\Ldens{c}_{\BMSsmall(\ent_i)}} z \;\text{d}z,
\end{align*}
where
\begin{align*}
\gexitkl {\Ldens{c}_{\BMSsmall(\ent_i)}} z & \defas 
\int_{-\infty}^{\infty} 
\frac{\partial \Ldens{c}_{\BMSsmall(\ent_i)}(w)}{\partial \cp }
\logtwo(1+ \mbox{e}^{-z-w}  ) \;\text{d}w.
\end{align*}
\elemma
Discussion: The remarks made after Lemma \ref{lemma:gexitBM}
apply in particular to the present case:
We write 
$\int_{-\infty}^{\infty}
\frac{\partial \Ldens{c}_{\BMSsmall(\ent_i)}(w)}{\partial \cp }
\logtwo(1+ \mbox{e}^{-z-w}  ) \;\text{d}w$ as a proxy 
for $\frac{\partial}{\partial \cp} 
\left\{\int_{-\infty}^{\infty}
\Ldens{c}_{\BMSsmall(\ent_i)}(w)
\logtwo(1+ \mbox{e}^{-z-w}  ) \;\text{d}w \right\}$. The latter
expression exists, since 
$\logtwo(1+ \mbox{e}^{-z-w}  )$ is continuously differentiable
as a function of $w$ and by assumption the channel 
family is smooth. 
Note further that $\gexitkl {\Ldens{c}_{\BMSsmall(\ent_i)}} z$ 
is continuous and non-negative so that $\gexitfi {}(\cp)$ exists as well. 

\begin{proof}
Consider the expression for $\gexitfi {}(\cp)$
as given in (\ref{equ:gexitcompact}). By assumption,
$p(y_i \mid x_i)$ is symmetric for all $i \in [n]$. Further, as
already remarked in the proof of Lemma \ref{lem:exitlinear}, the ``channel''
$p(\phi_i \mid x_i)$ is symmetric as well.
It follows from this and the fact that $p_{X_i}(+1)=p_{X_i}(-1)$ (due to
the assumption that the code is proper and that codewords are chosen
with uniform probability) that the contributions to
$\gexitfi {}(\cp)$ for $x_i=+1$ and $x_i=-1$ are identical.
We can therefore assume without loss of generality that $x_i=+1$.
Recall that the density of $\Phi_i$ assuming that $X_i=1$ 
is equal to the density of $\Phi_i$ assuming that the all-one
codeword was transmitted. The latter is by definition equal to
$\Ldens{a}_i$. As remarked earlier, $\Ldens{a}_i$ is
symmetric. Further, as discussed in the introduction, we can
assume that the  $i^{\text{th}}$  $\BMS$ channel outputs already log-likelihood
ratios. Therefore, $p_{Y_i | X_i}(y_i |+1)=\Ldens{c}_{\BMSsmall(\ent_i)}(y_i)$.
Finally, consider the expression within the $\logtwo$. If $x_i'=+1$
then the numerator and denominator are equal and we get one.
If on the other hand $x_i'=-1$ then we get by the previous remarks
the product of the likelihoods. Putting this all together we get
\begin{align*}
\gexitfi {}(\cp) & =
\int 
\Ldens{a}_i(z)
\frac{\text{d} \Ldens{c}_{\BMSsmall(\ent_i)}(w)}{\text{d} \ent_i}
\logtwo\left(1+\text{e}^{-z-w} \right) \text{d}z \text{d} w.
\end{align*}
The thesis follows by rearranging terms.
\end{proof}

\bex[Alternative Kernel Representations] \label{ex:nonuniquekernel}
Note that because of the symmetry property of $L$-densities we can
write 
\begin{align*}
\gexitf {}(\cp) & = \int_{-\infty}^{\infty}
\Ldens{a}(z) \gexitkl {\Ldens{c}_{\BMSsmall(\ent)}} z \;\text{d}z \\
& = \int_{0}^{\infty}
\absLdens{a}(z) 
\frac{\gexitkl {\Ldens{c}_{\BMSsmall(\ent)}} z + 
e^{-z} \gexitkl {\Ldens{c}_{\BMSsmall(\ent)}} {-z} }{1+e^{-x}}\text{d}z.
\end{align*}
This means that the kernel is uniquely specified on the absolute value domain
$[0, \infty]$, but that for each $z \in [0, \infty]$ 
we can split the weight of the kernel in any desired way between $+z$ and $-z$
so that $\gexitkl {\Ldens{c}_{\BMSsmall(\ent)}} z +
e^{-z} \gexitkl {\Ldens{c}_{\BMSsmall(\ent)}} {-z}$ equals the desired value. 
In the sequel we will use this degree of freedom to bring some
kernels into a more convenient form. Although it constitutes some
abuse of notation we will in the sequel make no notational distinction
between equivalent such kernels even though pointwise they might not
represent the same function.
\eex

As we have already remarked in the discussion
right after Definition \ref{def:gexit}, the $\gexit$ functions
$\gexitfi {}(\cp)$ allow us to ``locally'' measure
the change of the conditional entropy of a system.
This property is particularly apparent in the representation of
Lemma \ref{lem:gexitlinear} where we see that the local measurement has two components:
(i) the kernel which depends on the derivative of the channel
seen at the given position and (ii) the distribution $\Ldens{a}_i$,
which encapsulates all our ignorance
about the code behavior with respect to the $i^{\text{th}}$  position.
This representation is very intuitive. If we improve the observation
of a particular bit (derivative of the channel with respect to the parameter)
then the amount by which the conditional entropy
of the overall system changes clearly depends on how well this particular
bit was already known via the code constraints and
the observations of the other bits (extrinsic posterior density): if
the bit was already perfectly known then the additional observation
afforded will be useless, whereas if nothing was known about the bit
one would expect that the additional reduction in entropy of this bit
fully translates to a reduction of the entropy of the overall system.
We will see some quantitative statements of this nature in Section 
\ref{sec:basicproperties}.
%In particular the density $\Ldens{a}_i$
%is in turn the natural object appearing in message passing
%algorithms and in the density evolution analysis. 

In the next three examples we compute the kernels 
$\gexitkl {\Ldens{c}_{\BMSsmall(\ent_i)}} z$ for the standard
families $\{\BEC(\ent)\}_{\ent}$, $\{\text{BSC}(\ent)\}_{\ent}$, and
$\{\text{BAWGNC}(\ent)\}_{\ent}$. If we consider a single family of
$\BMS$ channels parameterized by the entropy $\ent$ it is
convenient to ``normalize'' the $\gexit$ kernel so that it
measure the ``progress per $\text{d} \ent$''. This means, in
the following examples we compute 
\begin{align}
\gexitkl {\Ldens{c}_{\BMSsmall(\ent)}} z & \defas
\frac{
\int_{-\infty}^{\infty}
\frac{\partial \Ldens{c}_{\BMSsmall(\ent_i)}(w)}{\partial \cp }
\logtwo(1+ \mbox{e}^{-z-w}  ) \;\text{d}w}{
\int_{-\infty}^{\infty}
\frac{\partial \Ldens{c}_{\BMSsmall(\ent_i)}(w)}{\partial \cp }
\logtwo(1+ \mbox{e}^{-w}  ) \;\text{d}w}. \label{kernelcomputation}
\end{align}

\bex[$\gexit$ Kernel, $L$-Domain -- $\{\BEC(\ent)\}_{\ent}$] \label{ex:lbeckernel}
If we take the family $\{\Ldens{c}_{\BECsmall(\ent)}\}_{\ent}$, 
where $\ent=\epsilon$ denotes both, the channel (intrinsic) entropy 
and the cross-over erasure probability, 
then a quick calculation
shows that $\gexitkl {\Ldens{c}_{\BECsmall(\ent)}} z  = 
\log_2(1+e^{-z})=\exitkl z$. In words, the $\gexit$ kernel
with respect to the family 
$\{\BEC(\ent)\}_{\ent}$ is the regular $\exit$ kernel.
\eex

\bex[$\gexit$ Kernel, $L$-Domain -- $\{\text{BSC}(\ent)\}_{\ent}$]
\label{ex:lbsckernel}
Let us now look at the family 
$\{\Ldens{c}_{\BSCsmall(\ent)}\}_{\ent}$. Some calculus shows that
\begin{align*}
\gexitkl {\Ldens{c}_{\BSCsmall(\ent)}} z & = 
\log\left( \frac{1+\frac{1-\cp}{\cp}\mbox{e}^{-z}}{1+\frac{\cp}{1-\cp}\mbox{e}^{-z} }\right)/\log\left(\frac{1-\cp}{\cp} \right),
\end{align*}
where $\cp = h_2^{-1}(\ent)$. For a fixed $z \in \reals$ and
$\ent\to 0$, the kernel converges to $1$ as $1+z/\lognat(\cp)$, whereas the limit when $\ent\to 1$  
is equal to $\frac{2}{1+e^{z}}$.
\eex

\bex[$\gexit$ Kernel, $L$-Domain -- 
$\{\text{BAWGNC}(\ent)\}_{|ent}$]
\label{ex:lbawgnkernel}
Consider now the family
$\{\Ldens{c}_{\BAWGNCsmall(\ent)}\}_{\ent}$, where $\ent$ denotes 
the channel entropy. This family is defined in Example 
\ref{ex:channeldefinition}. 
Recall that the noise is assumed to be Gaussian with 
zero-mean and variance $\sigma^2$.  
A convenient parameterization for this case is 
$\cp\defas2/\sigma^2$. This means that in the following 
$\ent = \entropy(\Ldens{c}_{\text{\tiny BAWGNC}(\sigma^2=2/\cp)})$. 
After some steps of calculus shown 
in Appendix \ref{app:gaussiankernel} and Lemma \ref{lemma:equivkernel}, we get 
\begin{align*}
\gexitkl {\Ldens{c}_{\text{\tiny BAWGNC}(\ent)}} z 
& =
\left({ \int_{-\infty}^{+\infty}\frac{\scriptstyle  \text{e}^{-\frac{(w-\cp)^2}{4 \cp}}    }{ \scriptstyle 1 + \text{e}^{w+z}       }\text{d}w    }\right)/\left( {\int_{-\infty}^{+\infty}\frac{ \scriptstyle \text{e}^{-\frac{(w-\cp)^2}{4 \cp}}    }{ \scriptstyle  1 + \text{e}^{w}  } \text{d}w}\right).
\end{align*}  
In Appendix \ref{app:gaussiankernel} we give alternative  
representations and/or interpretations of
this kernel. In particular we discuss the relationship
to the formulation presented by Guo, Shamai and Verd{\'u} in \cite{GSV04,GSV05}
using a connection to the MSE detector
as well as the formulation by Macris in \cite{Mac05} 
based on the Nishimori identity.
\eex
One convenient feature of standard $\exit$ functions is that 
they are fairly similar for a given code across the whole
range of $\BMS$ channels. Is this still true for $\gexit$ functions?
$\gexit$ functions depend on the channel {\em both} through the kernel as well
as through the extrinsic densities. 
%The densities are the same as
%for the computation of $\exit$ functions. But the kernels
%are now also functions of the channel.
Let us therefore compare the shape of the various kernels. It is most
convenient to compare the kernels not in the $L$-domain but in
the $|D|$-domain. 
A change of variables shows that in general the $L$-domain kernel,
call it $\gexitkl {\Ldens{c}} \cdot$, 
and the associated $|D|$-domain
kernel, denote it by $\gexitkabsd {\Ldens{c}} \cdot$, are linked by
\begin{align}
\label{equ:gexitkernelconversion}
\gexitkabsd {\Ldens{c}} s & =
\frac{1-s}{2} \gexitkl {\Ldens{c}} {\log\frac{1-s}{1+s}} +
\frac{1+s}{2} \gexitkl {\Ldens{c}} {\log \frac{1+s}{1-s}}.
\end{align}
E.g., 
if we apply the above transformation to the previous examples we
get the following results.

\bex[$\gexit$ Kernel, $|D|$-Domain -- $\{\text{BEC}(\ent)\}_{\ent}$] 
\label{ex:absdbeckernel}
We get $\gexitkabsd {\Ldens{c}_{\BECsmall(\ent)}} s=h_2((1+s)/2)$.
\eex

\bex[$\gexit$ Kernel, $|D|$-Domain -- $\{\text{BSC}(\ent)\}_{\ent}$] 
\label{ex:absdbsckernel}
Some calculus shows that
$
\gexitkabsd {\Ldens{c}_{\BSCsmall(\ent(\cp))}} s =1+ \frac{s}{\log((1-\cp)/\cp)}\log\left(\frac{1+2\cp s -s}{1-2\cp s+s}\right).
$
The limiting values are seen to be
%\begin{align*}
$\lim_{\ent\to1}\gexitkabsd {\Ldens{c}_{\BSCsmall(\ent)}} s  = 1-s^2,$ and  
$\lim_{\ent\to0}\gexitkabsd {\Ldens{c}_{\BSCsmall(\ent)}} s  = 1.$
%\end{align*}
\eex

\bex[$\gexit$ Kernel, $|D|$-Domain -- $\{\text{BAWGN}(\ent)\}_{\ent}$]
\label{ex:absdbawgnkernel} 
Using Example \ref{ex:lbawgnkernel} and 
 (\ref{equ:gexitkernelconversion}),  
it is straightforward to write the kernel in the $|D|$-domain as 
\begin{align*}
 \gexitkabsd {\Ldens{c}_{\text{\tiny BAWGNC}(\ent(\cp))}} s 
& =  \sum_{i\in\{-1,+1\} }\frac
{ \int_{-\infty}^{+\infty}\frac{(1-s^2) \text{e}^{-\frac{(w-\cp)^2}{4 \cp}}  }{(1+is)+(1-is)\text{e}^w}  \text{d}w}
{\int_{-\infty}^{+\infty}  \frac{2  \text{e}^{-\frac{(w-\cp)^2}{4 \cp}} }{1+\text{e}^w }\text{d}w }.
\end{align*}
As shown in Appendix \ref{app:kernellimits}, the limiting values are the
same as for the BSC, i.e.,
$\lim_{\ent\to1}\gexitkabsd {\Ldens{c}_{\text{\tiny BAWGNC}(\ent)}} s  = 1-s^2$,
and  
$\lim_{\ent\to0}\gexitkabsd {\Ldens{c}_{\text{\tiny BAWGNC}(\ent)}} s  = 1$.
\eex

\begin{figure}[hbt]
\centering
\setlength{\unitlength}{0.4bp}
\begin{picture}(600,200)
\put(0,0)
{
\put(10,0){\includegraphics[scale=0.4]{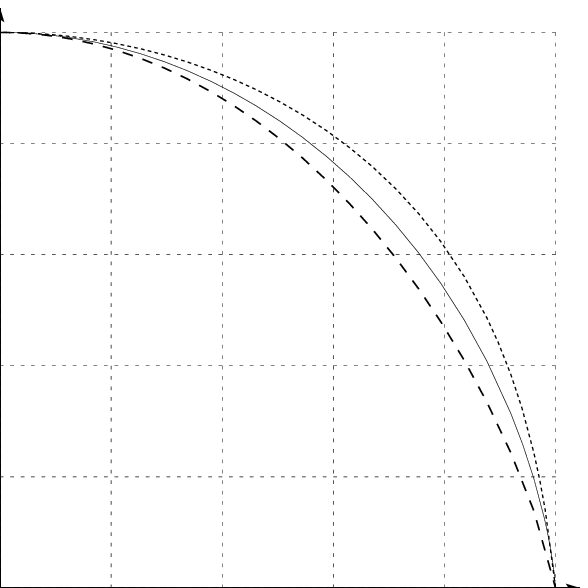}} %length=180bp
{\tiny
%\footnotesize
\multiputlist(20,-2)(32,0)[cb]{$~$,$0.2$,$0.4$,$0.6$,$0.8$,$1.0$}
\multiputlist(18,10)(0,32)[rc]{$~$,$0.2$,$0.4$,$0.6$,$0.8$,$1.0$}
\put(18,-2){\makebox(0,0)[rb]{$0.0$}}
}
}
\put(200,0)
{
\put(10,0){\includegraphics[scale=0.4]{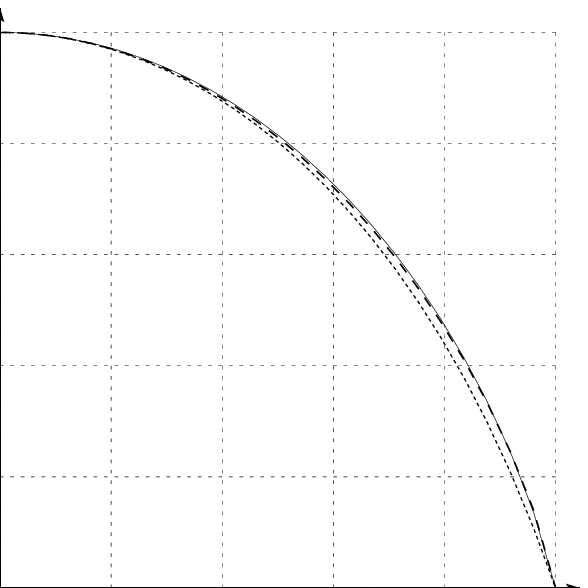}} %length=180bp
{\tiny
%\footnotesize
\multiputlist(20,-2)(32,0)[cb]{$~$,$0.2$,$0.4$,$0.6$,$0.8$,$1.0$}
\multiputlist(18,10)(0,32)[rc]{$~$,$0.2$,$0.4$,$0.6$,$0.8$,$1.0$}
\put(18,-2){\makebox(0,0)[rb]{$0.0$}}
}
}
\put(400,0)
{
\put(10,0){\includegraphics[scale=0.4]{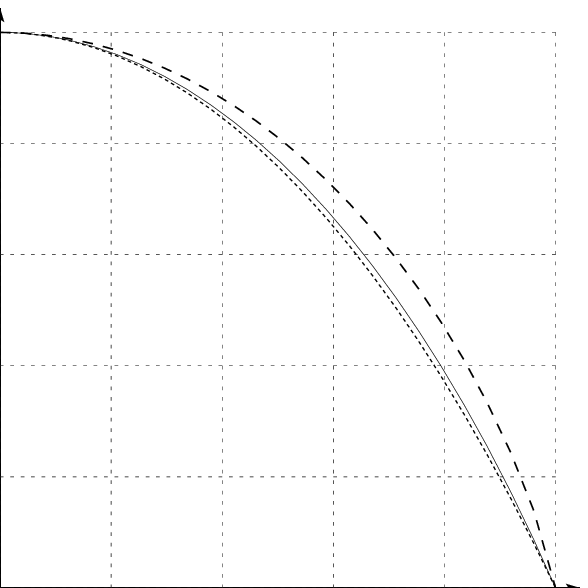}} %length=180bp
{\tiny
%\footnotesize
\multiputlist(20,-2)(32,0)[cb]{$~$,$0.2$,$0.4$,$0.6$,$0.8$,$1.0$}
\multiputlist(18,10)(0,32)[rc]{$~$,$0.2$,$0.4$,$0.6$,$0.8$,$1.0$}
\put(18,-2){\makebox(0,0)[rb]{$0.0$}}
}
}

\end{picture}
\caption{Comparison of the kernels
$\gexitkabsd {\Ldens{c}_{\BECsmall(\ent)}} s$ 
(dashed line) with 
$\gexitkabsd {\Ldens{c}_{\text{\tiny BSC}(\ent)}} s$ 
(dotted line) and
$\gexitkabsd {\Ldens{c}_{\text{\tiny BAWGNC}(\ent)}} s$
(solid line) at channel entropy rate $\ent=0.1$ (left), $\ent=0.5$ (middle) and 
$\ent=0.9$ (right).} 
\label{fig:gexitkernels}
\end{figure}

In Fig.~\ref{fig:gexitkernels} we compare
the $\exit$ kernel (which is also the $\gexit$ kernel for the BEC)
with the $\gexit$ kernels 
for BSC$(\ent)$ and BAWGNC$(\ent)$ in the
$|D|$-domain for several channel parameters. 
Note that these kernels are distinct but quite similar.
In particular, for $\ent=0.5$ the $\gexit$ kernel with respect to BAWGNC$(\ent)$
is hardly distinguishable from the regular $\exit$ kernel.
The $\gexit$ kernel for the $\BSC$ shows more variation.
%For $\ent$ converging to zero the kernel in the $|D|$-domain converges to the 
% constant one for the BAWGN and the $\BSC$, whereas the limit when $\ent$ 
%approaches one is equal to $1-s^2$ as explained in Examples  \ref{ex:absdbsckernel} 
%and  \ref{ex:absdbawgnkernel}. 

\begin{figure}[hbt]
\centering
\setlength{\unitlength}{0.6bp}
\begin{picture}(400,200)
\put(0,0)
{
\put(10,0){\includegraphics[scale=0.6]{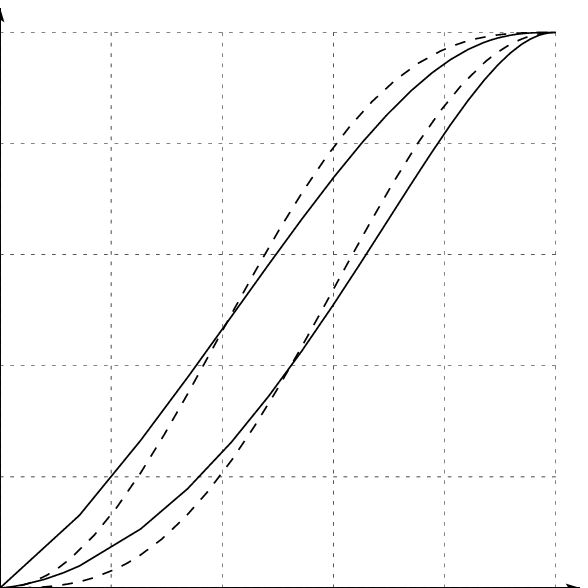}} %length=180bp
{\tiny
%\footnotesize
\multiputlist(20,-2)(32,0)[cb]{$~$,$0.2$,$0.4$,$0.6$,$0.8$,$1.0$}
\multiputlist(18,10)(0,32)[rc]{$~$,$0.2$,$0.4$,$0.6$,$0.8$,$1.0$}
\put(18,-2){\makebox(0,0)[rb]{$0.0$}}
}
\put(181,12){\makebox(0,0)[lb]{$\ih$}}
\put(22,172){\makebox(0,0)[lb]{$\exitf {}$, $\gexitf {}$}}
\put(27,146){\makebox(0,0)[l]{\small $[7,4,3]$ Hamming}}
\put(72,25){\makebox(0,0)[l]{\small $[7,3,4]$ Simplex}}
}
\put(200,0)
{
\put(10,0){\includegraphics[scale=0.6]{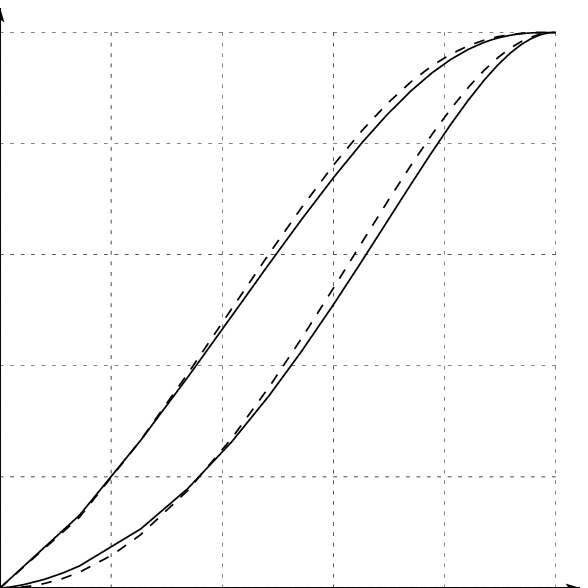}} %length=180bp
{\tiny
%\footnotesize
\multiputlist(20,-2)(32,0)[cb]{$~$,$0.2$,$0.4$,$0.6$,$0.8$,$1.0$}
\multiputlist(18,10)(0,32)[rc]{$~$,$0.2$,$0.4$,$0.6$,$0.8$,$1.0$}
\put(18,-2){\makebox(0,0)[rb]{$0.0$}}
}
\put(181,12){\makebox(0,0)[lb]{$\ih$}}
\put(22,172){\makebox(0,0)[lb]{$\exitf {}$, $\gexitf {}$}}
\put(27,146){\makebox(0,0)[l]{\small $[7,4,3]$ Hamming}}
\put(72,25){\makebox(0,0)[l]{\small $[7,3,4]$ Simplex}}
}
\end{picture}
\caption{Comparison of the $\gexit$ functions for the $[7,4,3]$ Hamming
code and its dual. Left picture: Comparison between $\gexit$ functions 
when transmitting over the BEC (dashed line) 
and over the BSC (solid line). Right picture: Comparison between $\gexit$ (solid line) and $\exit$ (dashed line) functions 
when transmission takes place over the BSC.}
\label{fig:HamBesCgexit}
\end{figure}

\bex[Repetition Code]
Consider the $[\n, 1, \n]$ repetition code.
Let $\{\Ldens{c}_{\ent}\}_{\ent}$ characterize a smooth family of 
$\BMS$ channels. For $\n \in \naturals$, let $\Ldens{c}_{\ent}^{\conv \n}$ denote
the $\n$-fold convolution of $\Ldens{c}_{\ent}$. 
The $\gexit$ function for the $[\n, 1, \n]$ repetition code is then given by
$\gexitf {}(\ent)  = 
\frac{1}{n} \frac{\text{d}\phantom{\ent}}{\text{d} \ent} \entropy(\Ldens{c}_{\ent}^{\conv n})$.
Explicitly, we get $\gexitfc {} {\BECsmall}(\ent) = \ent^n =
\exitfc {}  {\BECsmall}(\ent)$. 
As a further example, $\gexitfc {} {\BSCsmall}$ is given in parametric form
by
\begin{align*}
\Bigl(h_2(\cp), \frac{\sum_{j=\pm1} j\sum_{i=1}^{n} \binom{n}{i}
\cp^i \overline{\cp}^{n-i} \log\bigl(1+(\cp/\overline{\cp})^{n-2i-j} \bigr)}{
n \log\left({\overline{\cp}}/{\cp}\right)} \Bigr),
\end{align*}
with $\overline{\cp}=1-\cp.$
\eex
\bex[Single Parity-Check Code]
Consider the dual code, i.e., 
the $[\n, \n-1, 2]$ parity-check code. Some calculations show that
$\gexitfc {} {\BSCsmall}$ is given in parametric form
by
\begin{align*}
\Bigl(h_2(\cp), 1-(1-2 \cp)^{n-1} 
\frac{\log\bigl( \frac{1+(1-2 \cp)^{n}}{1-(1-2 \cp)^{n}} \bigr)}{\log \bigl(\frac{1-\cp}{\cp} \bigr)} \Bigr). 
\end{align*}
No simple analytic expressions are known for the case of transmission
over the \BAWGNC.
\begin{figure}[hbt]
\centering
\setlength{\unitlength}{0.6bp}
\begin{picture}(400,180)
\put(0,0)
{
\put(10,0){\includegraphics[scale=0.6]{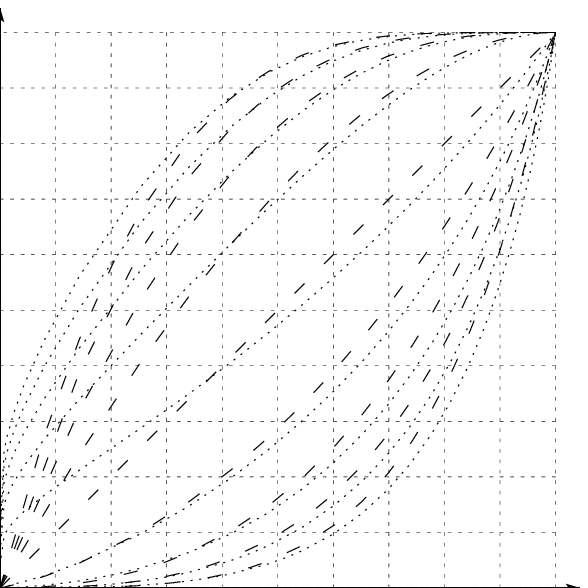}} %length=180bp
{\tiny
%\footnotesize
\multiputlist(20,-2)(32,0)[cb]{$~$,$0.2$,$0.4$,$0.6$,$0.8$,$1.0$}
\multiputlist(18,10)(0,32)[rc]{$~$,$0.2$,$0.4$,$0.6$,$0.8$,$1.0$}
\put(18,-2){\makebox(0,0)[rb]{$0.0$}}
}
\put(181,12){\makebox(0,0)[lb]{$\ih$}}
\put(22,172){\makebox(0,0)[lb]{$\exitf {}$, $\gexitf {}$}}
}
\put(200,0)
{
\put(10,0){\includegraphics[scale=0.6]{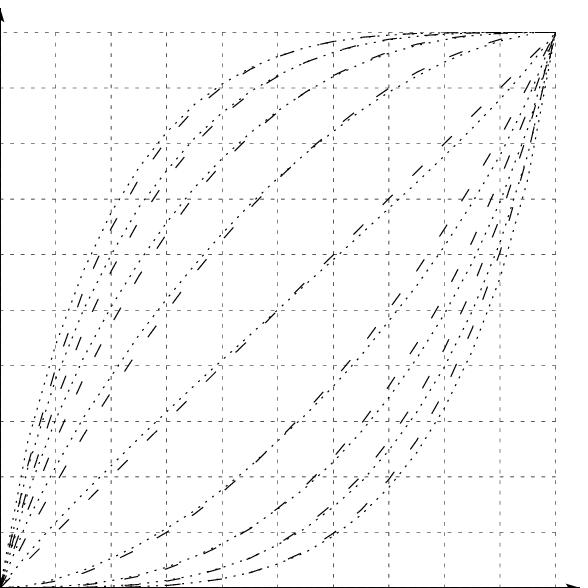}} %length=180bp
{\tiny
%\footnotesize
\multiputlist(20,-2)(32,0)[cb]{$~$,$0.2$,$0.4$,$0.6$,$0.8$,$1.0$}
\multiputlist(18,10)(0,32)[rc]{$~$,$0.2$,$0.4$,$0.6$,$0.8$,$1.0$}
\put(18,-2){\makebox(0,0)[rb]{$0.0$}}
}
\put(181,12){\makebox(0,0)[lb]{$\ih$}}
\put(22,172){\makebox(0,0)[lb]{$\exitf {}$, $\gexitf {}$}}
}
\end{picture}
\caption{\label{fig:exitgexitchartbecexamples} 
The $\exit$ (dashed) and $\gexit$ (dotted) function of
the $[\n, 1, \n]$ repetition code 
and the $[\n, \n-1, 2]$ parity-check code
assuming that transmission takes place over  BSC$(\ent)$ (left picture)
or the BAWGNC$(\ent)$ (right picture), $\n\in\{2,3,4,5,6\}$.
}
\end{figure}
\eex
Fig.~\ref{fig:exitgexitchartbecexamples} compares EXIT to GEXIT curves
for some repetition and some single parity-check codes.
\bex[Hamming Code]
Consider the $[7,4,3]$ Hamming code. 
When transmission takes place over BEC$(\ih)$, 
it is a tedious but conceptually simple exercise to
show that the $\exit$ function is 
$\xh(\ih)=3 \ih^2+4\ih^3-15\ih^4+12\ih^5-3\ih^6$, see, e.g., 
\cite{AKtB04,MeU03}. 
%The lowest degree of this polynomial in $\ih$ 
%is equal to the minimum distance minus 1 of the linear code. Its degree 
%is $\n-1$.
In a similar way, using the derivative of the 
conditional entropy, one can give an analytic expression for
the $\gexit$ function assuming transmission takes place over the \BSC. 
Both expressions are evaluated in 
Fig.~\ref{fig:HamBesCgexit} (left). 
A comparison between $\gexit$ and $\exit$ functions 
for the Hamming code and the BSC is shown in
Fig.~\ref{fig:HamBesCgexit} (right).
\eex
\bex[Simplex Code]
Consider now the dual of the Hamming code, i.e., the $[7,3,4]$ Simplex code. 
%The same discussion as for its dual code holds. 
%The analytic expression 
For transmission over the \BEC\ we have
$\xh(\ih)=4\ih^3-6\ih^5+3\ih^6$.
%which has minimum degree $3=4-1$. 
Fig.~\ref{fig:HamBesCgexit} 
compares $\gexit$ and $\exit$ functions for this code when transmission 
takes place over the BEC and over the BSC.
\eex

%
%***********************************************************
%
\section{Basic Properties of $\gexit$ Functions}
\label{sec:basicproperties}
$\gexit$ functions fulfill the GAT by definition. Let us state a few more
of their properties.

We first show that the $\gexit$ function
preserves the partial order implied by physical degradation.
\begin{lemma}
\label{prop:physicaldegradationgeneral}
Let $X$ be chosen with probability $p_X(x)$ from ${\cal X}^n$. 
Let the channel from $X$ to $Y$ be memoryless, where
$Y_i$ is the result of passing $X_i$ through the smooth and degraded family 
$\{\text{M}(\cp_i)\}_{\cp_i}$, $\cp_i \in I_i$. 
If $X \rightarrow Y_{\sim i} \rightarrow \Phi_i$ forms
a Markov chain then
\begin{align}
\label{eq:generalphysicaldegradation}
\frac{\partial H(X_i \mid Y) }{\partial \cp_i} \le
\frac{\partial H(X_i \mid Y_i, \Phi_i) }{\partial \cp_i}.
\end{align}
\end{lemma}
\begin{proof}
Since the derivatives in Eq.~ (\ref{eq:generalphysicaldegradation})
are known to exist a.e., the above statement is in fact equivalent to saying that,
for any $\cp'_i\ge\cp_i$,
\begin{align*}
H(X_i\mid Y_i(\cp_i'),Y_{\sim i})-
H(X_i\mid Y_i(\cp_i),Y_{\sim i})\le\\
H(X_i\mid Y_i(\cp_i'), \Phi_i)-
H(X_i\mid Y_i(\cp_i), \Phi_i)\, .
\end{align*}
Here, $Y_i(\cp_i)$ and $Y_i(\cp_i')$ are the result of transmitting
$X_i$ through the channels with parameter $\cp_i$ and $\cp'_i$, respectively.
We claim that
\begin{align*}
& X \rightarrow Y_i(\cp_i) \rightarrow Y_i(\cp_i'), \\
& X \rightarrow Y_{\sim i} \rightarrow \Phi_i, \\
& (Y_i(\cp_i), Y_i(\cp_i')) \rightarrow X \rightarrow (Y_{\sim i}, \Phi_i).
\end{align*}
The first claim follows from the assumption
that the channel family is degraded 
and the second claim is also part of the assumption.
Finally, the third claim is true since 
the channel is memoryless.

The thesis is therefore a consequence of 
Lemma \ref{lem:infotheoretic} stated below by making
the following substitutions:
\begin{xalignat*}{3}
Y_i(\cp_i) & \rightarrow Y,  &
Y_i(\cp_i') & \rightarrow Y', \\
Y_{\sim i}  & \rightarrow Z, \;\;&
\Phi_i  & \rightarrow Z'.
\end{xalignat*}
\end{proof}
\begin{lemma}
\label{lem:infotheoretic}
Assume that $X\to Y \to Y'$, $X\to Z\to Z'$, as well as
$(Y,Y') \to X \to (Z,Z')$ form Markov chains.
Then
\begin{align}
H(X\mid Y',Z) - H(X \mid Y,Z) \le  H(X\mid Y',Z') - H(X\mid Y,Z')\, .
\end{align}
\end{lemma}
\begin{proof}
The statement is equivalent to 
$H(X\mid Z,Y',Z') - H(X \mid Y,Z,Y',Z') \le  H(X\mid Y',Z') - 
H(X\mid Y,Y',Z')$.  Let us now condition on a event 
$(Y'=y',Z'=z')$. The proof is completed by showing that
(here the conditioning upon $Y'=y',Z'=z'$ is left implicit for the
sake of simplicity)
\begin{eqnarray}
H(X\mid Y,Z)-H(X\mid Y)-H(X\mid Z)+ H(X)\ge 0\, .
\end{eqnarray}
This inequality can be written in terms of mutual information
as $I(Y;X\mid Z)\le I(Y;X)$. The statement is therefore a well-known
consequence of the data processing inequality, see~\cite[p. 33]{CoT91},
if we can show that, conditioned on $Y'=y',Z'=z'$, 
$Y\to X\to Z$ forms a Markov chain. In formulae, we have to show that 
$p(y, z \mid x, y', z') = p(y \mid x, y', z') p(z \mid x, y', z')$,
which in turn follows if we can show that 
$\frac{p(z \mid x, y', z')}{p(z \mid x, y, y', z')}=1$.
The last equality can be shown by first applying Bayes law, 
then expanding all terms in the order $x, z', y$ and $y'$,
further canceling common terms and, finally,
repeatedly using the conditions
that $X\to Y \to Y'$, $X\to Z\to Z'$, as well as
$(Y,Y') \to X \to (Z,Z')$ form Markov chains. 
\end{proof}

In case of linear codes, and communication over a smooth and degraded family of
$\BMS$ channels, 
Lemma \ref{lem:gexitlinear} provides an explicit representation of the 
$\gexit$ function in terms of $L$-densities. In this case Lemma
\ref{prop:physicaldegradationgeneral}
becomes a statement on the corresponding
kernel. For completeness, let us state the corresponding 
condition explicitly.
\bcor[$ \gexitkl {\Ldens{c}_{\BMSsmall(\ent)}} z$ Preserves Partial Order]
\label{lemma:orderingviaphysicaldegradationexit}
Consider a smooth and degraded family of \BMS\ channels characterized by
the associated family of $L$-densities 
$\{\Ldens{c}_{\BMSsmall(\ent)}\}_{\ent}$.
Let $\Ldens{a}$ and $\Ldens{b}$ denote two symmetric $L$-densities
so that $\Ldens{a} \prec \Ldens{b}$, i.e.,
$\Ldens{b}$ is physically degraded with respect to $\Ldens{a}$.
Then
\begin{align*}
\int_{-\infty}^{\infty} \!\Ldens{a}(z) \, \gexitkl {\Ldens{c}_{\BMSsmall(\ent)}} z \text{d}z \leq
\int_{-\infty}^{\infty} \!\Ldens{b}(z)\,  \gexitkl {\Ldens{c}_{\BMSsmall(\ent)}} z \text{d}z.
\end{align*}
\ecor
An alternative proof of this statement 
is provided
in Appendix \ref{sec:AlternativePhysicalDegradation}.

We continue by examining some limiting cases. In the sequel 
$\perr$ denotes the error-probability operator. In the
$L$-domain it is defined as
$\perr (\Ldens{a}) = \frac12 
\int_{-\infty}^{\infty} \Ldens{a}(z) e^{-(|z/2|+z/2)} \text{d}z$. 
\blemma[Bounds for $\gexit$ Kernel]\label{lem:GeneralBounds}
Let $\gexitkabsd {\Ldens{c}_{\BMSsmall(\ent)}} z$
be the kernel associated to a smooth degraded family of $\BMS$ channels
characterized by their family of $L$-densities
$\{ {\Ldens{c}_{\BMSsmall(\ent)}} \}_{\ent}$.
Then
\begin{align*}
1-z \leq \gexitkabsd {\Ldens{c}_{\BMSsmall(\ent)}} z \leq 1.
\end{align*}
Therefore, if $\Ldens{a}$ is a symmetric $L$-density, we have
\begin{align*}
2 \perr (\Ldens{a}) \leq 
\int_{-\infty}^{\infty} \gexitkl {{\Ldens{c}_{\BMSsmall(\ent)}}} z
\Ldens{a}(z) \text{d}z \leq 1\, .
\end{align*}
\elemma
\begin{proof}
In  Appendix \ref{sec:AlternativePhysicalDegradation}, we show that
$\gexitkabsd {\Ldens{c}_{\BMSsmall(\ent)}} {z}$ is non-increasing and concave.
The upper bound follows from 
$\gexitkabsd {\Ldens{c}_{\BMSsmall(\ent)}} z<
\gexitkabsd {\Ldens{c}_{\BMSsmall(\ent)}} {z=0}=1$.
The lower bound is proved in a similar way by using concavity 
and observing that 
$\gexitkabsd {\Ldens{c}_{\BMSsmall(\ent)}} {z=1}=0$. 
The final claim now follows
from the fact that the $|D|$-domain kernel associated to $\perr$
is equal to $(1-z)/2$.
\end{proof}

\blemma[Further Properties of $\gexit$ Functions]
Let $\gexitf {}(\ent)$ be the $\gexit$ function associated
to a proper binary linear code of minimum distance larger than $1$,
and transmission over a complete smooth family of $\BMS$ channels. Then
\begin{xalignat*}{3}
\gexitf {}(0) & = 0, & \gexitf {}(1) & = 1.
\end{xalignat*}
If the minimum distance of the code is larger than $k$, then
\begin{align*}
\left.
\frac{\text{d}^{k-1} \phantom{\ent}}{\text{d} \ent^{k-1}} 
\gexitf {}(\ent) 
\right|_{\ent=0} = 0.
\end{align*}
Further, $\gexitf {}(\ent)$ is a non-decreasing function in $\ent$.
\elemma
\begin{proof}
Consider the first two assertions. If $\ent=0$,
then the associated $L$-density corresponds to a ``delta at infinity''
(this is an easy consequence of the minimum distance being at least
$2$). On the other
hand, if $\ent=1$ then the corresponding $L$-density is a ``delta at zero.'' 
The claim
in both cases follows now by a direct calculation.

In order to prove the last claim, we use the definition of 
 $\gexitf {}(\ent)$ to write
\begin{align*}
\left.
\frac{\text{d}^{k-1} \phantom{\ent}}{\text{d} \ent^{k-1}} \gexitf {}(\ent) 
\right|_{\ent=0} = 
\frac{1}{n}\left.\frac{\text{d}^{k} \phantom{\ent}}{\text{d} \ent^{k}} 
H(X|Y(\ent)) 
\right|_{\ent=0}\, .
\end{align*}
In order to evaluate the last derivative, we can first assume 
that the $i$-th bit is transmitted through a channel $\BMS(\ent_i)$.
Next we take partial derivatives with respect to $k$ of the entropies
$\{\ent_i\}$. Finally we set $\ent_i=0$ for all bits $i$.
We get therefore (neglecting the factor $1/n$):
\begin{align*}
\left.\sum_{i_1\dots i_{k}}
\frac{\partial^{k} \phantom{\ent}}{\partial \ent_{i_1}\cdots
\partial \ent_{i_k}} 
H(X|Y)\right|_{\ent_i=0}.
\end{align*}
Of course $h_i$ can be set to $0$ right at the beginning for all
the bits that are not differentiated over. This is equivalent
to passing the exact bits $X_i$. We get the expression
\begin{align*}
\sum_{i_1\dots i_{k}}
\frac{\partial^{k} \phantom{\ent}}{\partial \ent_{i_1}\cdots
\partial \ent_{i_k}} 
H(X|Y_{i_1}(\ent_{i_1})\dots Y_{i_k}(\ent_{i_{k}}), X_{\sim i_1\dots i_k} ) 
\end{align*}
to be evaluated at $\ent_{i_1}=\cdots=\ent_{i_k}=0$. If the
code has minimum distance larger than $k$, then any $n-k$ bits determine the
whole codeword and 
$H(X|Y_{i_1}(\ent_{i_1})\dots Y_{i_k}(\ent_{i_{k}}), 
X_{\sim i_1\dots i_k} )=0$. This finishes the proof.
\end{proof}

So far we have used the compact notation $\gexitf {}(\ent)$
for the GEXIT function. In some circumstance it is more convenient to
use a notation that makes the dependence of the functional on the
involved densities more explicit.
\bdefi[Alternative Notation for GEXIT Functional]
Consider a binary linear code and transmission over a smooth
family of BMS channels characterized by the associated family of $L$-densities 
$\{\Ldens{c}_{\cp}\}_{\cp}$.
Let $\{\Ldens{a}_{\cp}\}_{\cp}$ denote the associated family of average extrinsic
MAP densities (which we assume smooth).
Define
\begin{align*}
\gentropy(\Ldens{c}_{\cp}, \Ldens{a}_{\cp}) & \defas
\int_{-\infty}^{\infty} \Ldens{a}_{\cp}(z) \gexitkl {\Ldens{c}_{\cp}} z \text{d}z,
\end{align*}
where
\begin{align*}
\gexitkl {\Ldens{c}_{\cp}} z
& =
\frac{\int_{-\infty}^{\infty} \frac{\text{d} \Ldens{c}_{\cp}(w)}{\text{d} \cp}
\lognat(1+e^{-z-w}) \text{d}w}{
\int_{-\infty}^{\infty} \frac{\text{d} \Ldens{c}_{\cp}(w)}{\text{d} \cp}
\lognat(1+e^{-w}) \text{d}w}.
\end{align*}
\edefi
\blemma[GEXIT and Dual GEXIT Function]
\label{lem:dualgexit}
Consider a binary code $C$ and transmission over a complete and smooth
family of BMS channels characterized by the associated family of $L$-densities
$\{\Ldens{c}_\cp\}_{\cp}$. Let $\{\Ldens{a}_{\cp}\}_{\cp}$ denote
the corresponding family of (average) extrinsic \MAP\ densities.
Then the standard GEXIT curve is given in parametric form by
$\{\entropy(\Ldens{c}_{\cp}), \gentropy(\Ldens{c}_{\cp}, \Ldens{a}_{\cp})\}$.
The {\em dual}
GEXIT curve is defined by
$\{\gentropy(\Ldens{a}_{\cp}, \Ldens{c}_{\cp}), \entropy(\Ldens{a}_{\cp})\}$.
Both, standard and dual GEXIT curve have an area equal to
$r(C)$, the rate of the code.
\elemma
Discussion:
Note that both curves are ``comparable'' in that the first component measures
the channel $\Ldens{c}$ and the second argument measure the \MAP\ density
$\Ldens{a}$. The difference between the two
lies in the choice of measure which is applied to each component.

\begin{proof}
The statement that $\{\entropy(\Ldens{c}_{\cp}), 
\gentropy(\Ldens{c}_{\cp}, \Ldens{a}_{\cp})\}$ 
represents the standard GEXIT function follows by unwinding the 
corresponding definitions.
The only statement that requires a proof is the one concerning the
area under the ``dual GEXIT'' curve. We proceed as follows:
Consider the entropy
$\entropy(\Ldens{c}_{\cp} \conv \Ldens{a}_{\cp})$. We have
\begin{align*}
\entropy(\Ldens{c}_{\cp} \conv \Ldens{a}_{\cp}) & =
\int_{-\infty}^{\infty} \Bigl(\int_{-\infty}^{\infty}
\Ldens{c}_{\cp}(w) \Ldens{a}_{\cp}(v-w) \text{d}w \Bigr) \log(1+e^{-v}) \text{d}v \\
& = \int_{-\infty}^{\infty} \int_{-\infty}^{\infty}
\Ldens{c}_{\cp}(w) \Ldens{a}_{\cp}(z) \log(1+e^{-w-z}) \text{d}w \text{d}z.
\end{align*}
Consider now $\frac{\text{d} \entropy(\Ldens{c}_{\cp} \conv \Ldens{a}_{\cp})}{\text{d} \cp}$.
Using the previous representation we get
\begin{align*}
\frac{\text{d} \entropy(\Ldens{c}_{\cp} \conv \Ldens{a}_{\cp})}{\text{d} \cp} & =
\int_{-\infty}^{\infty} \int_{-\infty}^{\infty}
\frac{\text{d}\Ldens{c}_{\cp}(w)}{\text{d} \cp} \Ldens{a}_{\cp}(z) \log(1+e^{-w-z}) \text{d}w \text{d}z + \\
& \phantom{=} \int_{-\infty}^{\infty} \int_{-\infty}^{\infty}
\Ldens{c}_{\cp}(w) \frac{\text{d} \Ldens{a}_{\cp}(z)}{\text{d} \cp} \log(1+e^{-w-z}) \text{d}w \text{d}z.
\end{align*}
The first expression can be identified with the standard GEXIT curve
except that it is parameterized by a generic parameter $\cp$.
The second expression is essentially the same, but the roles of
the two densities are exchanged.
Integrate now this relationship over the whole range of $\cp$ and
assume that this range goes from ``perfect'' (channel) to ``useless''.
The integral on the left clearly equals 1. To perform the integrals
on the right, reparameterize the first expression with respect to
$\ent \defas \int_{\infty}^{\infty} \Ldens{c}_{\cp}(w) \lognat(1+e^{-w}) \text{d} w$
so that the integral is equal to the area under the standard GEXIT curve
given by
$\{\entropy(\Ldens{c}_{\cp}), \gentropy(\Ldens{c}_{\cp}, \Ldens{a}_{\cp})\}$.
In the same manner, reparameterize the second expression by
$\ent \defas \int_{\infty}^{\infty} \Ldens{a}_{\cp}(w) \lognat(1+e^{-w}) \text{d} w$.
Therefore the value of second expression is equal the area
under the curve given by
$\{\entropy(\Ldens{a}_{\cp}), \gentropy(\Ldens{a}_{\cp}, \Ldens{c}_{\cp})\}$.
Since the sum of the two areas equals one and the area under the
standard GEXIT curve equals $r(C)$, it follows that the area under
the second curve equals $1-r(C)$. Finally, note that if we consider the inverse
of the second curve by exchanging the two coordinates, i.e., if we consider the
curve
$\{\gentropy(\Ldens{a}_{\cp}, \Ldens{c}_{\cp}), \entropy(\Ldens{a}_{\cp})\}$,
then the area under this curve is equal to $1-(1-r(C))=r(C)$, as claimed.
\end{proof}
\begin{example}[GEXIT Versus Dual GEXIT]
Fig.~\ref{fig:gexitanddualgexit} shows the standard
GEXIT function and the dual GEXIT function for the $[5, 4, 2]$ code
and transmission over the $\BSC$. Although the two curves have quite
distinct shapes, the area under the two curves is the same.
\begin{figure}[htp]
\setlength{\unitlength}{0.75bp}%
\begin{center}
\begin{picture}(120,120)
\put(0,0)
{
	\put(0,0){\includegraphics[scale=0.75]{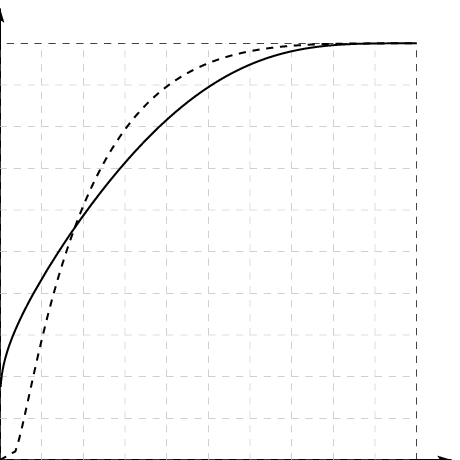}}
	\put(60, -2){\makebox(0,0)[t]{\small $\entropy(\Ldens{c}_{\ent})$, $\gentropy(\Ldens{a}_{\ent}, \Ldens{c}_{\ent})$}}
	\put(-2, 60){\makebox(0,0)[r]{\small \rotatebox{90}{$\gentropy(\Ldens{c}_{\ent}, \Ldens{a}_{\ent})$,$\entropy(\Ldens{a}_{\ent})$}}}
	\put(60, 30){\makebox(0,0)[t]{\small both GEXIT}}
	\put(-2, -2){\makebox(0,0)[rt]{\small $0$}}
	\put(120,-2){\makebox(0,0)[t]{\small $1$}}
	\put(-2,120){\makebox(0,0)[r]{\small $1$}}
}
\end{picture}
\end{center}
\caption{\label{fig:gexitanddualgexit} Standard and dual GEXIT function of $[5, 4, 2]$
code and transmission over the $\BSC$.}
\end{figure}
\end{example}

%
%************************************************************************
%
\section{Ensembles: Concentration and Asymptotic Setting}
\label{sec:asymptotic}
For simple codes, like, e.g.,
single parity-check codes or repetition codes, $\exitf {}$
and $\gexitf {}$ are relatively easy to compute.
In general though it is not a trivial matter to
determine the density of $\Phi^{}_i$ required
for the calculation.
What we {\em can} typically compute are the extrinsic estimates
if we use the $\BP$ decoder instead of the $\MAP$ decoder. It is therefore
natural to look at the equivalent of $\exit$ and $\gexit$ functions
if we substitute the extrinsic $\MAP$ estimates by their equivalent extrinsic
$\BP$ estimates. 
Although most of the subsequent definitions and statements
can be as easily derived for $\exit$ as for $\gexit$ functions, 
we 
focus on the latter. After all, these are the natural objects
to study as suggested by the GAT.
\bdefi[$\gexitf {\BPsmall}$ for Linear Codes and $\BMS$ Channels]
\label{def:bpgexit}
Let $X$ be chosen uniformly at random from a proper binary linear code.
Let the channel from $X$ to $Y$ be memoryless, where
$Y_i$ is the result of passing $X_i$ through the
smooth family $\{\BMS(\ent_i)\}_{\ent_i}$, $\ent_i \in [0, 1]$.
Assume that all individual channels are parameterized in a smooth way
by a common parameter $\cp$, i.e., $\ent_i=\ent_i(\cp)$, $i \in [n]$.
Let $\Phi^{\BPsmall, \itersmall}_i$ denote the extrinsic estimate of the
 $i^{\text{th}}$  bit at the $\iter^{\text{th}}$ round of $\BP$ decoding, assuming
an arbitrary  but fixed representation of the code by a Tanner graph
as well as an arbitrary but fixed schedule of the decoder.
Then the $\BP$ $\gexit$ function is defined as
\begin{align*}
\gexitfi {\BPsmall, \itersmall}(\cp)
& \defas
\frac{\partial H(X_i | \Phi^{\BPsmall, \itersmall}_i, Y_i)}{\partial \ent_i}
\frac{\text{d} \ent_i}{\text{d} \cp} \Big|_{\cp}.
\end{align*}
\edefi
The following statement, which is a direct consequence of the previous
definition and Lemma \ref{prop:physicaldegradationgeneral}, confirms
the intuitive fact that the BP GEXIT function (which is associated
to the suboptimal BP decoder) is at least as large as the the GEXIT
function itself, assuming only that the channel family is degraded.
\bcor[GEXIT Versus BP GEXIT]
\label{cor:gexitversusbpgexit}
Let $X$ be chosen uniformly at random from a proper binary linear code.
Let the channel from $X$ to $Y$ be memoryless, where
$Y_i$ is the result of passing $X_i$ through a
smooth and degraded family $\{\BMS(\ent_i)\}_{\ent_i}$, $\ent_i \in [0, 1]$.
Assume that all individual channels are parameterized in a smooth (differentiable) way
by a common parameter $\cp$, i.e., $\ent_i=\ent_i(\cp)$, $i \in [n]$.
Let $\gexitfi{}(\cp)$ and $\gexitfi {\BPsmall, \itersmall}(\cp)$ be as defined in Definitions \ref{def:gexit} and \ref{def:bpgexit}.
Then
\begin{align*}
\gexitfi{}(\cp) \leq \gexitfi {\BPsmall, \itersmall}(\cp).
\end{align*}
\ecor
\bdefi[Asymptotic $\BP$ $\exit$ and $\gexit$ Functions]
\label{def:asymptoticbpgexit}
Consider a \ddp $(\ledge, \redge)$ and the corresponding sequence of
ensembles $\eldpc n \ledge \redge$. Further consider a smooth
and degraded family $\{\BMS(\ent)\}_{\ent}$. 
Assume that all bits of $X$ are
sent through the channel $\BMS(\ent)$. 
For $\graph \in \eldpc n \ledge \redge$ and $i \in [n]$, let 
$\gexitfi {}(\graph, \cp)$ and
$\gexitfi {\BPsmall, \itersmall}(\graph, \cp)$
denote the $i^{\text{th}}$ $\MAP$ and $\BP$
$\gexit$ function associated to code $\graph$. 
By some abuse of notation, define
the asymptotic (and average) quantities
\begin{align*}
\gexitf {}(\ent) & \defas
\limsup_{n \rightarrow \infty} \expectation_{\graph} 
\Bigl[\frac{1}{n} \sum_{i \in [n]}\gexitfi {}(\graph, \ent) \Bigr], \\
\gexitf {\BPsmall, \itersmall}(\ent) & \defas
\lim_{n \rightarrow \infty} \expectation_{\graph}
\Bigl[
\frac{1}{n} \sum_{i \in [n]} \gexitfi {\BPsmall, \itersmall}(\graph, \ent)
\Bigr], \\
\gexitf {\BPsmall}(\ent) & \defas
\lim_{\iter \rightarrow \infty} \gexitf {\BPsmall, \itersmall}(\ent).
\end{align*}
For notational simplicity we suppress the dependence of the 
above quantities on the \ddp and the channel family $\{\BMS(\ent)\}_{\ent}$.
\edefi

In the above definitions we have taken the average of the individual curves
over the ensemble. Let us now justify this approach by showing that the quantities
are concentrated. The proof of the following statement, 
which asserts the concentration of the conditional entropy,
can be found in \cite{MMU05}.
\btheo[Concentration of Conditional Entropy] 
\label{theo:concentrationml}
Let $\graph(n)$ be chosen uniformly at random from $\eldpc n \ledge \redge$.
Assume that $\graph(n)$ is used to transmit over a $\BMS(\ent)$ channel.
By some abuse of notation, let
$H_{\graph(n)} =H_{\graph(n)}(X \mid Y)$ be the associated conditional entropy.
Then for any $\xi > 0$
\begin{align*}
\Pr \left\{|H_{\graph(n)} - \expectation_{\graph(n)}[H_{\graph(n)}]|>\n \xi\right\} &\leq 2\,
\text{e}^{-\n B \xi^2},\\
\end{align*}
where $B= 1/(2 (\drmax+1)^2(1-r))$ and $\drmax$ is the maximal check-node degree.
\etheo

\btheo[Concentration of $\gexitf {\BPsmall, \iter}$]
\label{theo:concentrationexitbp}
Consider the sequence of ensembles 
$\eldpc n \ledge \redge$, where $(\ledge, \redge)$ is fixed
and $n$ tends to infinity. Then the limits
$\gexitf {\BPsmall, \itersmall}(\ent) = 
\lim_{n \rightarrow \infty} n^{-1}\expectation_{\graph}[
\sum_{i \in [n]} \gexitfi {\BPsmall, \itersmall}(\graph, \ent)]$ and
$\gexitf {\BPsmall}(\ent) = 
\lim_{\iter \rightarrow \infty} \gexitf {\BPsmall, \itersmall}(\ent)$
exist.
Further, let $\graph(n)$ be chosen uniformly at random from 
$\eldpc n \ledge \redge$.
Assume that $\graph(n)$ is used to transmit over a $\BMS(\ent)$ channel.
Then, for all  $\xi>0$, there exists $\alpha_\xi>0$, such that,
for $n$ large enough
\begin{multline}
\Pr\Bigl\{\Bigl| \gexitf {\BPsmall, \itersmall}(\graph(n), \ent)-
\gexitf {\BPsmall, \itersmall}(\ent) 
\Bigr|> n \xi\Bigr\}
\leq \text{e}^{-\alpha_\xi \n}.
\end{multline}
\etheo
\begin{proof}
Note that for a fixed iteration number $\iter$, the distribution 
of $\Phi^{\BPsmall}_i$ (with $i$ a uniformly random node), assuming
that the all-one codeword was sent, converges (at a speed of $1/n$)
to the corresponding
distribution of density evolution, 
denote it by $\Ldens{a}_{\iter}$.
The result now follows by 
noting that $\gexitf {\BPsmall, \itersmall}$ is the result of
applying a bounded linear operator to this distribution
$\Ldens{a}_{\iter}$.
The proof of concentration is almost verbatimly the same as
the proof in \cite{RiU05},
which shows the concentration of the probability
of error under $\BP$ decoding, or the proof in \cite{MMU05},
which relates to the
concentration of the $\BP$ $\exit$ function. We will therefore skip the details.
\end{proof}
\btheo[Concentration of $\gexitf {}$]
\label{theo:concentrationexitml}
Let $\graph$ be chosen uniformly at random from $\ldpc(\n,\ledge,\redge)$
and consider the smooth and degraded family $\{\BMS(\ent)\}_{\ent}$, $\ent \in [0, 1]$.
Assume that $\graph$ is used to transmit over the $\BMS(\ent)$ channel.
Let $H_{\graph(n)} =H_{\graph(n)}(X \mid Y)$ be the associated conditional entropy, $\gexitf {}(\graph(n),\ent)$ the corresponding $\MAP$ $\gexit$ function,
and $\gexitf {}_n(\ent)=\expectation\,\gexitf {}(\graph(n),\ent)$.
Let $J \subseteq [0, 1]$ be an interval on which
$\lim_{n\to\infty} \frac{1}{n}\expectation \left[H_{\graph(n)}\right]$
exists and is differentiable with respect
to $\ent$. Then, for any $\cp\in J$ and $\xi>0$ there exist an
$\alpha_{\xi}>0$ such that, for $n$ large enough
\begin{align*}
\Pr \left\{|\gexitf {}(\graph(n),\ent)-\gexitf {}_n(\ent)|>\n \xi\right\} &\leq
\text{e}^{-\n \alpha_{\xi}}.
\end{align*}
Furthermore, if
$\lim_{n\to\infty} \frac{1}{n}\expectation \left[H_{\graph(n)}\right]$
is twice differentiable with respect
to $\ent \in J$, there exists a 
strictly positive constant $A$ such that $\alpha_\xi>A\xi^4$.
\etheo
The proof of this statement can be found in \cite{MMU05}.

Let us summarize. We have seen that all the quantities which we introduced
in Definition \ref{def:asymptoticbpgexit} are concentrated 
and that the $\BP$
quantities $\gexitf {\BPsmall, \itersmall}$ and $\gexitf {\BPsmall}$ exist.
Unfortunately, we had to use $\limsup$ for the definition of
$\gexitf {}$ since to prove the existence of the limit seems 
to be difficult.
As discussed in \cite{MMU05}, even in the case of transmission 
over the $\BEC$
the existence of the corresponding limit is not known in general but only follows
from the explicit construction of the Maxwell decoder in all those cases where
the Maxwell construction can be shown to result in $\MAP$ performance.

Note that $\gexitf {\BPsmall, \itersmall}$ and $\gexitf {\BPsmall}$
have again a convenient representation in terms of the asymptotic $\BP$ densities.
More precisely, we have
\begin{align*}
\gexitf {\BPsmall, \itersmall} (\ent) & = 
\int_{-\infty}^{\infty} \Ldens{a}^{\BPsmall, \itersmall} (z) 
\gexitkl {\Ldens{c}_{\BMSsmall(\ent)}} z
\;\text{d}z, \\
\gexitf {\BPsmall} (\ent) & =
\int_{-\infty}^{\infty} \Ldens{a}^{\BPsmall} (z) 
\gexitkl {\Ldens{c}_{\BMSsmall(\ent)}} z
\;\text{d}z,
\end{align*}
where $\Ldens{a}^{\BPsmall, \itersmall}$ is the limiting density 
of $\Phi_i^{\BPsmall, \iter}$ (with $i$ a uniformly random node) under the 
all-one codeword assumption as $n$ tends to infinity 
associated to the \ddp $(\ledge, \redge)$. This density can easily be computed
by density evolution. In a similar manner, $\Ldens{a}^{\BPsmall}$ is the 
corresponding fixed-point density of density evolution.

\begin{figure}[hbt]
\centering
\setlength{\unitlength}{0.6bp}
\begin{picture}(400,200)
\put(0,0)
{
\put(10,0){\includegraphics[scale=0.6]{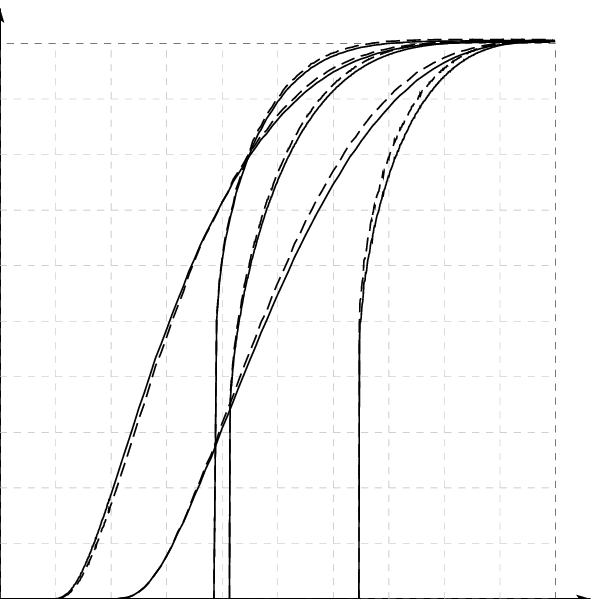}} %length=180bp
{\tiny
%\footnotesize
\multiputlist(20,-2)(32,0)[cb]{$~$,$0.2$,$0.4$,$0.6$,$0.8$,$1.0$}
\multiputlist(18,10)(0,32)[rc]{$~$,$0.2$,$0.4$,$0.6$,$0.8$,$1.0$}
\put(18,-2){\makebox(0,0)[rb]{$0.0$}}
}
\put(181,12){\makebox(0,0)[lb]{$\ih$}}
\put(130,25){\makebox(0,0)[lb]{\footnotesize{$(3, 4)$}}}
\put(90,25){\makebox(0,0)[lb]{\footnotesize{$(3, 6)$}}}
\put(85,90){\makebox(0,0)[rb]{\footnotesize{$(4, 8)$}}}
\put(52,35){\makebox(0,0)[lb]{\footnotesize{$(2, 4)$}}}
\put(45,25){\makebox(0,0)[rb]{\footnotesize{$(2, 6)$}}}
}
\put(200,0)
{
\put(10,0){\includegraphics[scale=0.6]{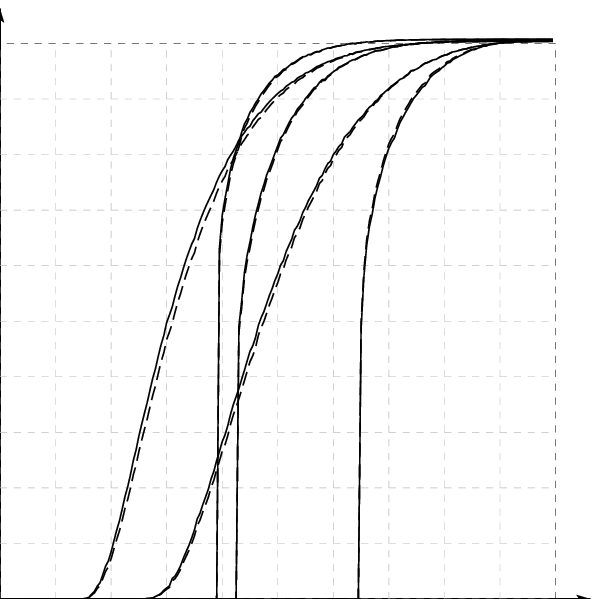}} %length=180bp
{\tiny
%\footnotesize
\multiputlist(20,-2)(32,0)[cb]{$~$,$0.2$,$0.4$,$0.6$,$0.8$,$1.0$}
\multiputlist(18,10)(0,32)[rc]{$~$,$0.2$,$0.4$,$0.6$,$0.8$,$1.0$}
\put(18,-2){\makebox(0,0)[rb]{$0.0$}}
}
\put(181,12){\makebox(0,0)[lb]{$\ih$}}
\put(130,25){\makebox(0,0)[lb]{\footnotesize{$(3, 4)$}}}
\put(90,25){\makebox(0,0)[lb]{\footnotesize{$(3, 6)$}}}
\put(85,90){\makebox(0,0)[rb]{\footnotesize{$(4, 8)$}}}
\put(52,35){\makebox(0,0)[lb]{\footnotesize{$(2, 4)$}}}
\put(45,25){\makebox(0,0)[rb]{\footnotesize{$(2, 6)$}}}
}
\end{picture}
\newcaption{$\BP$ $\gexit$ (solid curves)
versus $\BP$ $\exit$ (dashed curves) for several
regular LDPC ensembles for the $\BSC$ (left picture) and the 
$\BAWGNC$ (right picture).}
\label{fig:gexitbsc}
\end{figure}

In Fig.~\ref{fig:gexitbsc} we plot the $\BP$ $\gexit$ function
$\gexitf {\BPsmall}$ for a
few regular LDPC ensembles 
and we compare them
with the corresponding $\BP$ $\exit$ functions, 
which we denote by $\exitf {\BPsmall}$. We see
that the curves are quite similar.

\blemma[$\gexitf {} \leq \gexitf {\BPsmall}$]
\label{lemma:gexitmapversusgexitbp}
Consider a  \ddp $(\ledge, \redge)$ and transmission over the 
smooth and degraded family 
$\{\BMS(\ent)\}_{\ent}$.
Let  $\gexitf {}(\ent)$ and $\gexitf {\BPsmall}(\ent)$ denote
respectively the corresponding asymptotic $\MAP$ and $\BP$ $\gexit$ functions
as defined in Definition \ref{def:asymptoticbpgexit}
when the code is chosen uniformly at random from the ensemble
$\eldpc n \ledge \redge$.
Then 
\begin{align*}
\gexitf {}(\ent) \leq \gexitf {\BPsmall} (\ent).
% \leq  \gexitf {\BPsmall, \iter} (\ent)\, .
%
\end{align*}
\elemma
\begin{proof}
Using Corollary \ref{cor:gexitversusbpgexit}, we know that for any
$\graph \in \eldpc n \ledge \redge$ and $\ell \in \naturals$
\begin{align*}
{\gexitf {}}_{\graph}(\cp)  \leq {\gexitf {}}^{\BPsmall, \ell}_{\graph}(\cp).
\end{align*}
If we take first the expectation over the elements of the ensemble,
then the $\limsup$ on both sides with respect to $n$, and finally
the limit $\ell \rightarrow \infty$, we get the desired result.

\end{proof}

\section{An Upper Bound on the MAP Threshold}
\label{sec:upperbound}
One important consequence of the area theorem is that it gives
rise to an easy to compute upper bound on the threshold
of $\MAP$ decoding. 
\bdefi[$\MAP$ Threshold]
Consider a \ddp $(\ledge, \redge)$ and a smooth and degraded family
$\{\BMS(\ent)\}_{\ent}$. The {\em threshold} $\ent^{\MAPsmall}$ is {\em defined} as
\begin{align*}
\ent^{\MAPsmall} & \defas
\min\{\ent \in [0, 1]: \liminf_{n \rightarrow \infty} 
\expectation_{\graph}[H(X \mid Y(\ent))]/n > 0\}.
\end{align*}
\edefi
%Discussion: Let us consider the operational meaning of the above
%definition. Let $\ent < \ent^{\MAPsmall}$. Then for any $\delta > 0$,
%there exists a finite $n = n(\delta)$ so that 
%$\expectation_{\graph}[H(X \mid Y(\ent))]/n < \delta$.
%By Theorem \ref{theo:concentrationml} it follows that if
%transmission takes place over the channel with parameter $\ent$
%then all sufficiently
%long codes (of properly chosen blocklength)
%have a normalized conditional entropy less than any fixed constant.
%Since $\delta$ can be chosen arbitrarily small, this means in
%essence that arbitrarily reliable communications is possible with
%most codes at that channel parameter.
%
%On the other hand, assume that  $\ent > \ent^{\MAPsmall}$. In this case 
%the average (over the ensemble) 
%normalized conditional entropy stays bounded away from 
%zero by a strictly positive constant for an infinite sequence of
%blocklengths. 
%Again, by Theorem \ref{theo:concentrationml} 
%this is not only true for the average over the ensemble but
%for most elements from the ensemble. It follows that
%(at least for those blocklengths) reliable transmission is not possible.

Discussion: Let us consider the operational meaning of the above
definition. Let $\ent < \ent^{\MAPsmall}$. Then by definition of 
the threshold,
there exists a sequence of blocklengths $n_1, n_2, n_3, \cdots,$ so
that the normalized (divided by the blocklength $n$)
{\em average} conditional entropy converges to zero. 
By Theorem \ref{theo:concentrationml} it follows that most of the codes
in the corresponding ensembles
have a normalized conditional entropy less than any fixed constant.
%This means that for this sequence of blocklengths and the 
%conditional entropy behaves sublinearly for at least some
%elements in the ensemble. 
For sufficiently
large blocklengths, a conditional entropy which 
grows sublinearly implies
that the receiver can limit the set of hypothesis to a subexponential 
list which with high probability contains the correct codeword.
Therefore, in this sense reliable communication is possible.

On the other hand, assume that  $\ent > \ent^{\MAPsmall}$. In this case 
the normalized conditional entropy stays bounded away from 
zero by a strictly positive constant for all sufficiently
large blocklengths. By Theorem \ref{theo:concentrationml} 
this is not only true for the average over the ensemble but
for most elements from the ensemble. It follows that
with most elements from the ensemble 
reliable communication is not possible.

\btheo[Upper Bound on $\MAP$ Threshold]\label{theo:UBMAP}
Consider a \ddp $(\ledge, \redge)$ whose 
asymptotic rate converges to the design rate $r(\ledge, \redge)$, see \cite[Lemma 7]{MMU05}.
Assume further that transmission takes place over a smooth and degraded family 
$\{\BMS(\ent)\}_{\ent}$. Let $\gexitf {\BPsmall}(\ent)$
denote the associated $\BP$ $\gexit$ function. 
Then 
\begin{align}
 \liminf_{n \to \infty} 
\expectation_{\graph}[H(X \mid Y(\ent))]/n\ge r(\ledge, \redge)-\int_{\ent}^1 
\gexitf {\BPsmall}(\ent') \;\text{d}\ent'\, .\label{eq:LowerBoundEntropyGen}
\end{align}
Furthermore, if
$\overline{\ent}$ denotes the largest positive number so that
\begin{align*}
\int_{\overline{\ent}}^{1} \gexitf {\BPsmall}(\ent) \;\text{d}\ent=r(\ledge,
\redge),
\end{align*}
then $\ent^{\MAPsmall} \leq \overline{\ent}$, where
$\ent^{\MAPsmall}$ denotes the $\MAP$ threshold. 
\etheo
\begin{proof}
Let $\graph$ be chosen uniformly at random
from the ensemble $\eldpc n \ledge \redge$. By the GAT
\begin{align*}
r(\ledge, \redge)-& \liminf_{n \to \infty} 
\expectation_{\graph}[H(X \mid Y(\ent))]/n = \\
&=\limsup_{n\to\infty} \frac{1}{n}\,
\expectation_{\graph}[H(X \mid Y(1))-H(X \mid Y(\ent))]=\\
&=\limsup_{n \to \infty}\expectation_{\graph} \Bigl[\int_{\ent}^1 
\gexitf {}(\graph,\ent') \;\text{d}\ent' \Bigr].
\end{align*}
We can exchange the expectation and the integral by Fubini's theorem:
in fact $\gexitf {}(\graph,\ent')$ is measurable
and $\gexitf {}(\graph,\ent')\in [0,1]$. 
We can furthermore exchange the limit and the integral by the 
Fatou-Lebesgue lemma.
We get
\begin{align*}
\liminf_{n \to \infty} 
\expectation_{\graph}[H(X \mid Y(\ent))]/n  \geq
r(\ledge, \redge)-\! \int_{\ent}^1 
\gexitf {}(\ent') \;\text{d}\ent'\, .
\end{align*}
Equation (\ref{eq:LowerBoundEntropyGen}) is proved by applying Lemma
\ref{lemma:gexitmapversusgexitbp}.

The upper bound on the $\MAP$ threshold follows from the
observation that the r.h.s. of  Eq.~(\ref{eq:LowerBoundEntropyGen})
is non-decreasing in $h$. Therefore  $\limsup_{n \to \infty} 
\expectation_{\graph}[H(X \mid Y(\ent))]/n$ is bounded away from $0$
for any $\ent>\overline{\ent}$ and the thesis follows from the definition
of $\ent^{\MAPsmall}$.
\end{proof}

\bex 
The following table presents the upper bounds on the $\MAP$ threshold
for transmission over the $\BAWGNC(\ent)$
as derived from Theorem \ref{theo:UBMAP} for a few regular ensembles:
$\ledge(x) = x^{\ldegree-1}$, $\redge(x) = x^{\rdegree-1}$.
The same threshold were first computed using the (non-rigorous)
replica method from 
statistical physics~\cite{andrea}. In~\cite{Mon04}, they were shown to
be upper bounds for $\rdegree$ even, using an interpolation technique. 
The present proof applies also to the case of odd $\rdegree$. 
It can be proved that the three characterizations of the threshold are
indeed equivalent, i.e., they give {\em exactly} the same value.
\tablespace
\begin{center}
\small{
\begin{tabular}{cccccc}
$\ldegree$  & $\rdegree$ & $\ih^\BP$ & $\overline{\ih}$ & $\overline{\ih} ($\cite{WiS05,BKLM02}) & $\ih^\Sh$  \\ \hline
$3$     &   $4$   &  $0.6507(5)$ &   $0.7417(1)$   & $0.743231$ & $3/4$   \\
$3$     &   $5$   &  $0.5113(5)$ &   $0.5800(3)$   & $0.583578$ & $3/5$  \\
$3$     &   $6$   &  $0.4160(5)$ &   $0.4721(5)$   & $0.476728$ & $1/2$  \\
$4$     &   $6$   &  $0.5203(5)$ &   $0.6636(2)$   & $0.663679$ & $1/3$  \\ 
\end{tabular}
}
\end{center}
\tablespace
Also shown is the result of the information theoretic upper bound given in
\cite{WiS05}, which in turn is an improved version of the bound
developed in \cite{BKLM02}. For the specific case of transmission over the \BSC\ and
regular codes it is given by $h_2(\overline{\cp})$, where 
$\overline{\cp}$ is the unique positive root
of the equation 
$\rdegree h_2(\cp)= \ldegree h_2((1-(1-2\cp)^\rdegree)/2)$.
\eex
%
%*****************************************************************
%
\section{The Extended BP $\gexit$ Curve}
\label{sec:egexit}
\subsection{Extended $\BP$ $\gexit$ Curve}
As discussed in detail in \cite{MMU05} for 
the case of transmission over the $\BEC$, the fundamental relationship
which appears in the limit of large blocklengths between the $\MAP$
and the $\BP$ decoder is best described in terms of the {\em extended}
$\exit$ curve. For the $\BEC$ this is the curve with parametric
description
$\left( \frac{\xl}{\ledge(1-\redge(1-\xl))}, \lnode(1-\redge(1-\xl)) \right)$,
where $\xl$ takes values in the subset $J\subseteq [0, 1]$
such that $\xl\le\ledge(1-\redge(1-\xl))$ 
($J$ is in fact the union of a finite number of intervals). 
 Note that the families
$\{\Ldens{f}_\xl\}_\xl \defas \{\BEC(\xl)\}_\xl$ and 
$\{\Ldens{c}_\xl\}_\xl \defas \{\BEC(\frac{\xl}{\ledge(1-\redge(1-\xl)))}\}_\xl$,
$\xl \in J$, have the following property: For each $\xl \in J$,
$\Ldens{f}_\xl$ constitutes a fixed-point density (of density evolution) 
for the channel
$\Ldens{c}_\xl$. Furthermore both channel families are {\em smooth}
and satisfy $H(\Ldens{f}_\xl)=\xl$.
Finally if $J=[0,1]$ (a necessary condition for this to happen is 
$\ledge'(0)\redge'(1)\ge 1$)
the families are said to be {\em complete}. 
\bdefi[Complete Fixed-Point Family, $\gexitf \EBPsmall$ and $\gexitf \BPsmall$]
Consider a degree distribution pair $(\ledge, \redge)$.
We say that the families $\{\Ldens{f}_\xl\}_\xl$
and $\{\Ldens{c}_\xl\}_\xl$, $\xl \in [0, 1]$, form a {\em complete fixed-point family} 
for $(\ledge, \redge)$ if 
\begin{itemize}
\item[(i)] there exists a complete and degraded family $\{\BMS(\ent)\}_{\ent}$
such that for each $\xl \in [0, 1]$, $\Ldens{c}_\xl \in \{\BMS(\ent)\}_{\ent}$
\item[(ii)] for each $\xl \in [0, 1]$, $\Ldens{f}_\xl$ is a fixed-point density
with respect to the degree distribution $(\ledge, \redge)$ and the
channel $\Ldens{c}_\xl$; this means that for each $\xl \in[0, 1]$,
$\Ldens{f}_\xl =\Ldens{c}_\xl \conv \ledge(\redge(\Ldens{f}_x))$
\item[(iii)] $\{\Ldens{f}_\xl\}_\xl$ and $\{\Ldens{c}_\xl\}_\xl$ are 
smooth with respect to $\xl$
\item[(iv)] $\entropy(\Ldens{f}_\xl)=\xl$
\end{itemize}
Let $\Ldens{a}_\xl(y) \defas \lnode(\redge(\Ldens{f}_\xl))$.
The {\em extended} $\BP$ ($\EBP$) $\gexit$ curve, 
call it $\gexitf {\EBPsmall}(\xl)$, is then given in parametric form
by $(H(\xl), \gexitf {\EBPsmall}(\xl))$, where
\begin{align*}
\gexitf {\EBPsmall}(\xl) & \defas 
\int_{-\infty}^{\infty} \Ldens{a}_\xl(y) \gexitkl {\Ldens{c}_\xl} y \;\text{d}y.
\end{align*}
Finally, the $\BP$ $\gexit$ curve, call it $\gexitf \BPsmall$,
is the ``envelope''
of the $\gexitf {\EBPsmall}$ curve.
\edefi
Discussion: Contrary to our usual notation, we have used $\xl$ to parameterize
the channel families and the function $\gexitf {\EBPsmall}(\xl)$ 
and we have assumed that $\entropy(\Ldens{f}_\xl)=\xl$ (rather
than $\entropy(\Ldens{c}_\xl)=\xl$). This has the following reason: in general,
the $\EBP$ $\gexit$ function is not a single-valued function of the
{\em channel} entropy but it is a single-valued function of the 
{\em fixed-point} entropy (see Fig.~\ref{fig:multijump}). We prefer
to use the parameter $\xl$ instead of the usual parameter $\ent$,
to remind ourselves that the channel $\Ldens{c}_{\xl}$ is the
channel which belongs to the fixed-point density $\Ldens{f}_{\xl}$
(and not the channel $\Ldens{c}_{\ent}$, which by our previous
notational convention has entropy $\ent$). {\em Complete} fixed-point
families do not always exist. 
If, for instance, $\ledge_2=0$, 
then $\xl$ cannot be chosen arbitrarily close to $0$.
This is easily seen for transmission over the $\BEC$.
In this case $\xl\ge \underline{\xl}$ with $\underline{\xl}$ 
the smallest (non-vanishing) root of the
equation $\ledge(1-\rho(1-\xl))=\xl$.

From the definition it is not immediately obvious
that for a given degree distribution pair $(\ledge, \redge)$ 
and a complete and degraded family $\{\BMS(\ent)\}_{\ent}$,
such a (complete or incomplete) fixed-point family always exists,
or that it is unique.
For the $\BEC$ we have an explicit formula for the family, but in the general
case the existence is far from trivial.
We will get back to this point in the next section.

One of the important applications of the $\EBP$ $\gexit$ curve is that 
it encodes very clearly the connection between $\MAP$ and $\BP$ decoding.
As mentioned above, the $\BP$ $\gexit$ function is obtained 
as the `envelope' of the  $\EBP$ curve. More precisely,
one has to choose, for each value of the channel entropy $\ent$,
the branch of the $\EBP$ curve whose $\gexit$ value is the largest.
As pointed out in the introduction when discussing Fig.~\ref{fig:multijump},
a different single valued function can be obtained by applying the Maxwell
construction, described in detail in \cite{MMU05}, to the $\EBP$ $\gexit$ 
curve. Motivated by the GAT as well as by the $\BEC$ case, we formulate
the following
\begin{conj}
\label{con:theconjecture}
The ($\MAP$) 
$\gexit$ function $\gexitf{}(\ent)$ is obtained by applying the Maxwell
construction to the extended $\BP$ $\gexit$ curve  $(H(\xl), 
\gexitf {\EBPsmall}(\xl))$.
\end{conj} 

\begin{figure}[hbt]
\centering
\setlength{\unitlength}{0.6bp}
\begin{picture}(200,200)
\put(0,0)
{
\put(10,0){\includegraphics[scale=0.6]{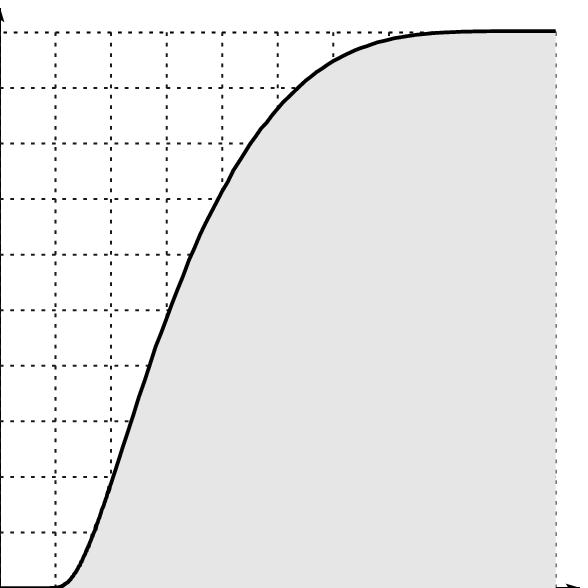}} %length=180bp
{\tiny
%\footnotesize
\multiputlist(20,-2)(32,0)[cb]{$~$,$0.2$,$0.4$,$0.6$,$0.8$,$1.0$}
\multiputlist(18,10)(0,32)[rc]{$~$,$0.2$,$0.4$,$0.6$,$0.8$,$1.0$}
\put(18,-2){\makebox(0,0)[rb]{$0.0$}}
}
\put(181,12){\makebox(0,0)[lb]{$\ih$}}
\put(22,172){\makebox(0,0)[lb]{$\gexitf{}$}}
\put(90,100){\makebox(0,0)[lb]{{$\gexitf{\BPsmall}(\ih)$}}}
\put(90,80){\makebox(0,0)[lb]{{$=\gexitf{\EBPsmall}(\ih)$}}}
\put(90,60){\makebox(0,0)[lb]{{$=\gexitf{}(\ih)$}}}
\put(46,12){\makebox(0,0)[lb]{$\ih^\BPsmall=\ih^\BPsmall$}}
}
\end{picture}
\caption{EBP GEXIT curve for the cycle-code ensemble with \ddp $(x,x^5)$. 
 The EBP GEXIT curve, BP GEXIT curve and MAP GEXIT curve coincide. }
\label{fig:26ebpmapgexit}
\end{figure}

Let us 
consider a few typical examples. In each of the following
cases the complete fixed-point family was computed by a {\em numerical}
procedure, which will be explained in the next section.
\bex[LDPC($x,x^5$) -- BSC] \label{ex:CycleEx}
Consider the \ddp $(\ledge,\redge)=(x, x^5)$ and the
corresponding $\ldpc$ ensemble with design rate $\drate=2/3$.
We assume that transmission takes place over the family
$\{\text{BSC}(\cp)\}$. Recall that for this code the
$\BP$ threshold is given by the stability condition.
From Fig.~\ref{fig:26ebpmapgexit} we see that, according to
the numerical calculation, the $\EBP$ $\gexit$ curve is a monotone
function. Assuming this is true, it follows that
the $\EBP$ $\gexit$ is equal to the $\BP$ $\gexit$ curve for
this example. 
For any value of the channel parameter a single fixed point 
density (apart from the `delta at infinity') is found. 
Also: a single fixed point density exists for each value of the 
density entropy $\xl$. The Maxwell construction is trivial in this
case and yields a $\MAP~ \gexit$ equal to 
the $\BP~ \gexit$ curve.  
\eex

\begin{figure}[hbt]
\centering
\setlength{\unitlength}{0.6bp}
\begin{picture}(400,200)
\put(0,0)
{
\put(10,0){\includegraphics[scale=0.6]{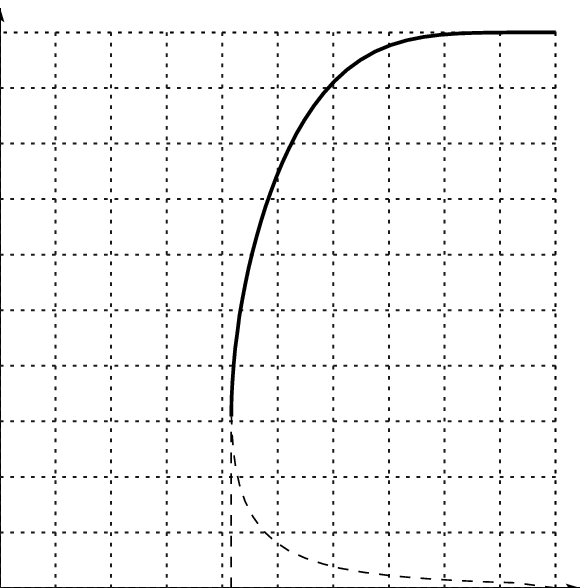}} %length=180bp
{\tiny
%\footnotesize
\multiputlist(20,-2)(32,0)[cb]{$~$,$0.2$,$0.4$,$0.6$,$0.8$,$1.0$}
\multiputlist(18,10)(0,32)[rc]{$~$,$0.2$,$0.4$,$0.6$,$0.8$,$1.0$}
\put(18,-2){\makebox(0,0)[rb]{$0.0$}}
}
\put(181,12){\makebox(0,0)[lb]{$\ih$}}
\put(22,172){\makebox(0,0)[lb]{$\gexitf{}$}}
\put(50,110){\makebox(0,0)[lb]{{$\gexitf{\BPsmall}(\ih)$}}}
\put(66,12){\makebox(0,0)[lb]{$\ih^\BPsmall$}}
}
\put(200,0)
{
\put(10,0){\includegraphics[scale=0.6]{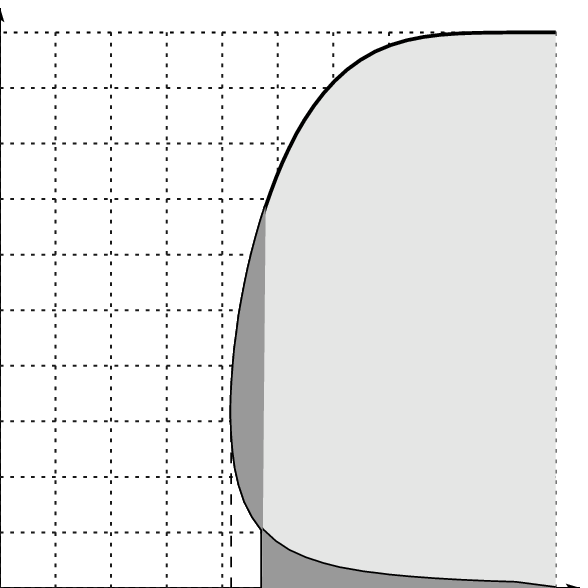}} %length=180bp
{\tiny
%\footnotesize
\multiputlist(20,-2)(32,0)[cb]{$~$,$0.2$,$0.4$,$0.6$,$0.8$,$1.0$}
\multiputlist(18,10)(0,32)[rc]{$~$,$0.2$,$0.4$,$0.6$,$0.8$,$1.0$}
\put(18,-2){\makebox(0,0)[rb]{$0.0$}}
}
\put(181,12){\makebox(0,0)[lb]{$\ih$}}
\put(22,172){\makebox(0,0)[lb]{$\gexitf {}$}}
\put(105,110){\makebox(0,0)[lb]{{$\gexitf{}(\ih)$}}}
\put(66,12){\makebox(0,0)[lb]{$\ih^\BPsmall$}}
\put(94,12){\makebox(0,0)[lb]{$\ih^\MAPsmall$}}
}
\end{picture}
\caption{EBP GEXIT curve for the $(3,6)$ ensemble. Left: EBP GEXIT curve and corresponding BP GEXIT curve. 
Right: The conjectured MAP GEXIT curve according to the Maxwell construction. }
\label{fig:36ebpbpmapgexit}
\end{figure}

\bex[(3,6) LDPC Ensemble -- BSC]\label{example:36}
Consider the \ddp $(\ledge,\redge)=(x^2, x^5)$ and the
corresponding $\ldpc$ ensemble with design rate $\drate=1/2$.
We assume that transmission takes place over the family
$\{\text{BSC}(\cp)\}$.
Fig.~\ref{fig:36ebpbpmapgexit} shows on the left the $\EBP$ $\gexit$ curve
and the corresponding $\BP$ $\gexit$ curve, which has one jump.
The picture on the right shows the conjectured
$\MAP$ $\gexit$ curve according to the Maxwell construction.
 For this ensemble, we have $\ih^\BPsmall\approx0.416$.
The $\MAP$ threshold implied by the Maxwell construction coincides with 
the one of Theorem \ref{theo:UBMAP}:
$\overline{\ih}^\MAPsmall\approx 0.472$.

\begin{figure}[hbt]
\centering
\setlength{\unitlength}{0.6bp}
\begin{picture}(400,200)
\put(0,0)
{
\put(10,0){\includegraphics[scale=0.6]{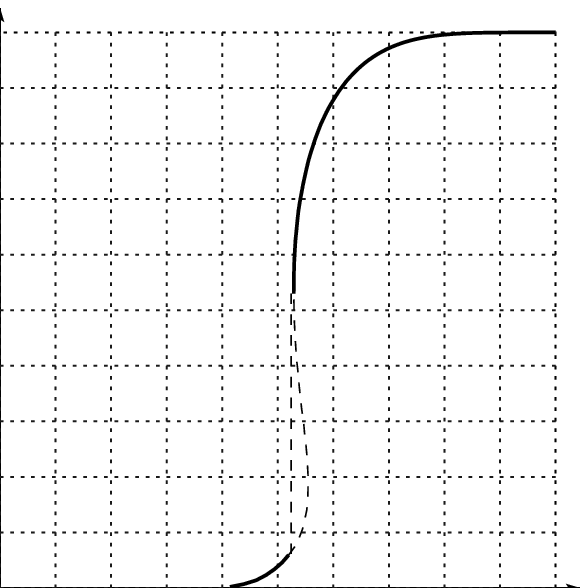}} %length=180bp
{\tiny
%\footnotesize
\multiputlist(20,-2)(32,0)[cb]{$~$,$0.2$,$0.4$,$0.6$,$0.8$,$1.0$}
\multiputlist(18,10)(0,32)[rc]{$~$,$0.2$,$0.4$,$0.6$,$0.8$,$1.0$}
\put(18,-2){\makebox(0,0)[rb]{$0.0$}}
}
\put(181,12){\makebox(0,0)[lb]{$\ih$}}
\put(22,172){\makebox(0,0)[lb]{$\gexitf{}$}}
\put(50,110){\makebox(0,0)[lb]{{$\gexitf{\BPsmall}(\ih)$}}}
\put(66,12){\makebox(0,0)[lb]{$\ih^\BPsmall$}}
}
\put(200,0)
{
\put(10,0){\includegraphics[scale=0.6]{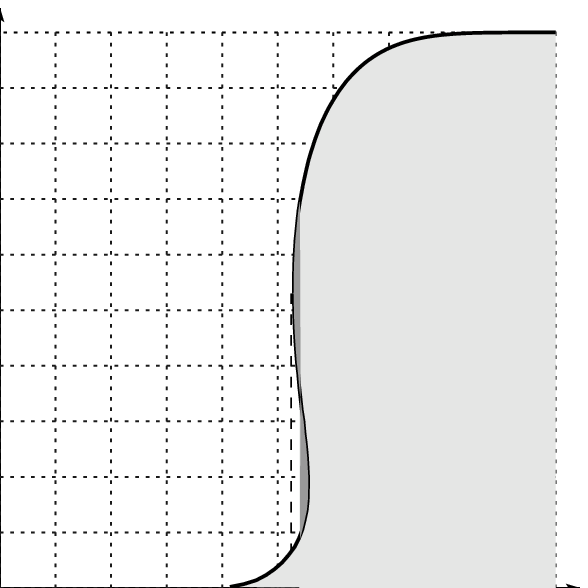}} %length=180bp
{\tiny
%\footnotesize
\multiputlist(20,-2)(32,0)[cb]{$~$,$0.2$,$0.4$,$0.6$,$0.8$,$1.0$}
\multiputlist(18,10)(0,32)[rc]{$~$,$0.2$,$0.4$,$0.6$,$0.8$,$1.0$}
\put(18,-2){\makebox(0,0)[rb]{$0.0$}}
}
\put(181,12){\makebox(0,0)[lb]{$\ih$}}
\put(22,172){\makebox(0,0)[lb]{$\gexitf {}$}}
\put(110,110){\makebox(0,0)[lb]{{$\gexitf{}(\ih)$}}}
\put(66,12){\makebox(0,0)[lb]{$\ih^\MAPsmall$}}
}
\end{picture}
\caption{EBP GEXIT curve for the $(\ledge,\redge)=(2/5 x + 3/5 x^5,x^5)$ ensemble. Left: EBP GEXIT curve and corresponding BP GEXIT curve. 
Right: The conjectured MAP GEXIT curve according to the Maxwell construction. }
\label{fig:stabJmapmaxgexit}
\end{figure}

\eex
\bex[LDPC$(2/5 x + 3/5 x^5,x^5)$ -- BSC]\label{exampleMult1}
Consider the \ddp $(\ledge,\redge)=(2/5 x + 3/5 x^5,x^5)$ and the
corresponding $\ldpc$ ensemble with design rate $\drate=4/9$.
We assume that transmission takes place over the family 
$\{\text{BSC}(\cp)\}$.
Fig.~\ref{fig:stabJmapmaxgexit}
 shows on the left the $\EBP$ $\gexit$ curve
and the corresponding $\BP$ $\gexit$ curve, which has one jump. 
The picture on the right shows the conjectured 
$\MAP$ $\gexit$ curve according to the Maxwell construction.
The $\BP$ threshold is given by the stability condition. 
As a consequence of this and Conjecture \ref{con:theconjecture},
$\ih^\BPsmall=\ih^\MAPsmall$ . 

\eex 

\begin{figure}[hbt]
\centering
\setlength{\unitlength}{0.6bp}
\begin{picture}(400,200)
\put(0,0)
{
\put(10,0){\includegraphics[scale=0.6]{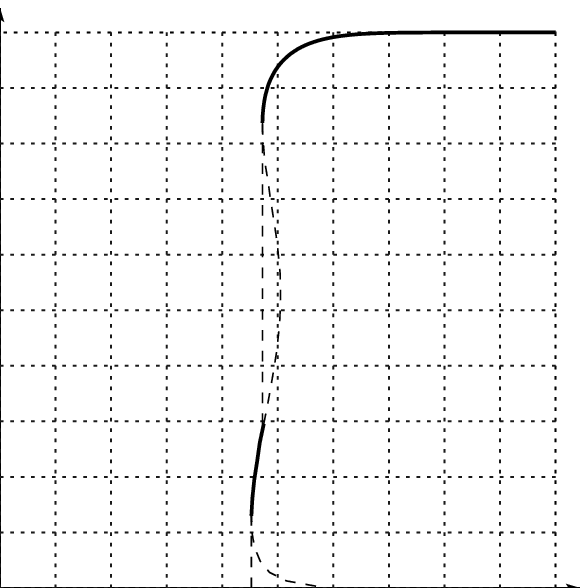}} %length=180bp
{\tiny
%\footnotesize
\multiputlist(20,-2)(32,0)[cb]{$~$,$0.2$,$0.4$,$0.6$,$0.8$,$1.0$}
\multiputlist(18,10)(0,32)[rc]{$~$,$0.2$,$0.4$,$0.6$,$0.8$,$1.0$}
\put(18,-2){\makebox(0,0)[rb]{$0.0$}}
}
\put(181,12){\makebox(0,0)[lb]{$\ih$}}
\put(22,172){\makebox(0,0)[lb]{$\gexitf{}$}}
\put(50,110){\makebox(0,0)[lb]{{$\gexitf{\BPsmall}(\ih)$}}}
\put(68,12){\makebox(0,0)[lb]{$\ih^\BPsmall$}}
\put(98,44){\makebox(0,0)[lb]{$\ih^{\BPsmall,2}$}}
}
\put(200,0)
{
\put(10,0){\includegraphics[scale=0.6]{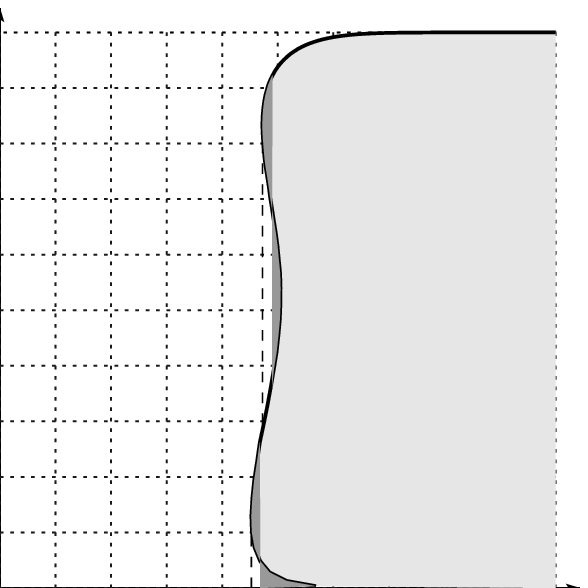}} %length=180bp
{\tiny
%\footnotesize
\multiputlist(20,-2)(32,0)[cb]{$~$,$0.2$,$0.4$,$0.6$,$0.8$,$1.0$}
\multiputlist(18,10)(0,32)[rc]{$~$,$0.2$,$0.4$,$0.6$,$0.8$,$1.0$}
\put(18,-2){\makebox(0,0)[rb]{$0.0$}}
}
\put(181,12){\makebox(0,0)[lb]{$\ih$}}
\put(22,172){\makebox(0,0)[lb]{$\gexitf {}$}}
\put(105,110){\makebox(0,0)[lb]{{$\gexitf{}(\ih)$}}}
\put(66,12){\makebox(0,0)[lb]{$\ih^\BPsmall$}}
\put(96,18){\makebox(0,0)[lb]{$\ih^\MAPsmall$}}
\put(100,58){\makebox(0,0)[lb]{$\ih^{\MAPsmall,2}$}}
}
\end{picture}
\caption{EBP GEXIT curve for the \ddp ($\frac{3x+6x^2+11x^{17}}{20},x^9$). Left: EBP GEXIT curve and corresponding BP GEXIT curve. 
Right: The conjectured MAP GEXIT curve according to the Maxwell construction. }
\label{fig:2Jebpbpgexit}
\end{figure}

\bex[LDPC($\frac{3x+6x^2+11x^{17}}{20},x^9$) -- BSC]
Consider the
 \ddp ($\frac{3x+6x^2+11x^{17}}{20},x^9$).
We assume that transmission takes place over the family
$\{\text{BSC}(\cp)\}$.
Fig.~\ref{fig:2Jebpbpgexit} shows on the left the $\EBP$ $\gexit$ curve
and the corresponding $\BP$ $\gexit$ curve, which has two jumps.
The picture on the right shows the conjectured
$\MAP$ $\gexit$ curve according to the Maxwell construction: 
This curve has also 2 jumps. 
\eex

\begin{figure}[hbt]
\centering
\setlength{\unitlength}{0.6bp}
\begin{picture}(400,200)
\put(0,0)
{
\put(10,0){\includegraphics[scale=0.6]{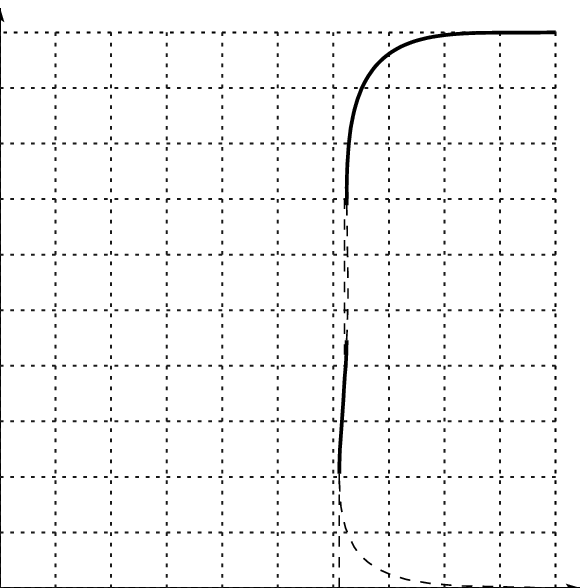}} %length=180bp
{\tiny
%\footnotesize
\multiputlist(20,-2)(32,0)[cb]{$~$,$0.2$,$0.4$,$0.6$,$0.8$,$1.0$}
\multiputlist(18,10)(0,32)[rc]{$~$,$0.2$,$0.4$,$0.6$,$0.8$,$1.0$}
\put(18,-2){\makebox(0,0)[rb]{$0.0$}}
}
\put(181,12){\makebox(0,0)[lb]{$\ih$}}
\put(22,172){\makebox(0,0)[lb]{$\gexitf{}$}}
\put(50,110){\makebox(0,0)[lb]{{$\gexitf{\BPsmall}(\ih)$}}}
\put(92,12){\makebox(0,0)[lb]{$\ih^\BPsmall$}}
\put(128,74){\makebox(0,0)[lb]{$\ih^{\BPsmall,2}$}}
}
\put(200,0)
{
\put(10,0){\includegraphics[scale=0.6]{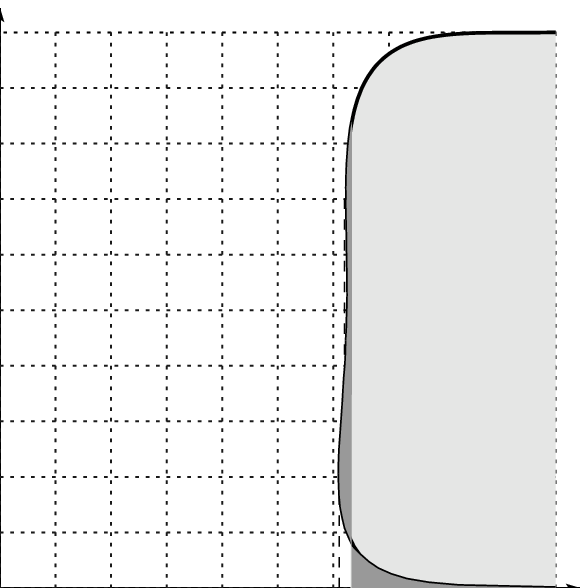}} %length=180bp
{\tiny
%\footnotesize
\multiputlist(20,-2)(32,0)[cb]{$~$,$0.2$,$0.4$,$0.6$,$0.8$,$1.0$}
\multiputlist(18,10)(0,32)[rc]{$~$,$0.2$,$0.4$,$0.6$,$0.8$,$1.0$}
\put(18,-2){\makebox(0,0)[rb]{$0.0$}}
}
\put(181,12){\makebox(0,0)[lb]{$\ih$}}
\put(22,172){\makebox(0,0)[lb]{$\gexitf {}$}}
\put(128,110){\makebox(0,0)[lb]{{$\gexitf{}(\ih)$}}}
\put(92,12){\makebox(0,0)[lb]{$\ih^\BPsmall$}}
\put(130,18){\makebox(0,0)[lb]{$\ih^\MAPsmall$}}
}
\end{picture}
\caption{EBP GEXIT curve for the \ddp $(\frac{x+2x^2+2x^{13}}{5},x^5)$. Left: EBP GEXIT curve and corresponding BP GEXIT curve. 
Right: The conjectured MAP GEXIT curve according to the Maxwell construction. }
\label{fig:2BP1MAPmapmaxgexit}
\end{figure}

\bex[$(\frac{x+2x^2+2x^{13}}{5},x^5)$ -- BSC] \label{exampleMult3}
Consider the \ddp 
$(\frac{x+2x^2+2x^{13}}{5},x^5)$ and the
corresponding $\ldpc$ ensemble. .
We assume that transmission takes place over the family
$\{\text{BSC}(\cp)\}$.
Fig.~\ref{fig:2BP1MAPmapmaxgexit} shows on the left the $\EBP$ $\gexit$ curve
and the corresponding $\BP$ $\gexit$ curve, which has two jumps.
The picture on the right shows the conjectured
$\MAP$ $\gexit$ curve according to the Maxwell construction.
This example shows that a \ddp can have more $\BP$ jumps than
$\MAP$ jumps.

\eex

%*************************************************************************
%
\section{How to Compute
$\EBP$ $\gexit$ Curves: Basic Properties and Area Theorem}
\label{sec:HowToEBP}

In the previous pages we presented examples
of $\EBP$ $\gexit$ curves for several $\ldpc$ ensembles.
In this section we explain how these curves have  been computed
and we derive some of their basic properties, 
including the EBP Area Theorem.

We start by noticing that ordinary density evolution cannot be
applied to the present case because of two reasons. First,
$\EBP$ curves include `unstable branches'. We refer by such a term to
branches along which the $\gexit$ curve is a decreasing
function of the channel entropy. Such branches are expected to
correspond to fixed point densities which are locally unstable under
density evolution (whence the name). This expectation can be confirmed
analytically for the $\BEC$ case, and numerically for a general
$\BMS$ channel. As a consequence, these fixed points cannot be
approximated by iterating density evolution with a generic initial
condition.

The second problem is related to values of the channel parameter
for which multiple locally stable fixed point densities coexist.
This is the case for instance in the Examples \ref{exampleMult1}
to \ref{exampleMult3} above. In this case different initial conditions
are required to achieve each of these densities by density evolution.
A systematic way for constructing all such initial conditions is however not available.

The crucial observation for overcoming both these problems consists
in noticing that $\EBP$ $\gexit$ curves are naturally parameterized
by the {\em entropy of the fixed point density}. More precisely,
consider a smooth and degraded family $\{\BMS(\ih)\}$ and
$\xl\in [0,1]$. Then, we expect that there exists at most one
value of the channel parameter $\ih = \ih(\xl)$ and one density
$\Ldens{f}_{\xl}$, such that $H(\Ldens{f}_{\xl})=\xl$ and
$(\Ldens{c}_{\xl} \defas \BMS(\ih(\xl)),\Ldens{f}_{\xl})$ forms a fixed
point pair.

This naturally suggests to run density evolution {\em at fixed
density entropy}. Let us denote by $\Tc_{\ih}$ the ordinary density evolution
operator at fixed channel $\BMS(\ih)$. Formally
\begin{eqnarray}
\Tc_{\ih}(\Ldens{a})  \defas \Ldens{c} \conv \ledge(\redge(\Ldens{a}))\, .
\end{eqnarray}
where $\Ldens{c}$ is the density associated to the channel $\BMS(\ih)$.
For any $\xl\in[0,1]$, we define the density evolution operator at fixed
entropy $\xl$, $\Td_{\xl}$ as
\begin{eqnarray}
\Td_{\xl}(\Ldens{a}) \defas \Tc_{\ih(\Ldens{a}, \xl)}(\Ldens{a})
\end{eqnarray}
where $\ih(\Ldens{a}, \xl)$ is the solution of $H(\Tc_{\ih}(\Ldens{a})) =\xl$.
Whenever no such value of $\ih$ exists, $\Td_{\xl}(\Ldens{a})$
is left undefined.
Since, for a given $\Ldens{a}$, the family
$\Tc_{\ih}(\Ldens{a})$ is ordered by physical degradation,
$H(\Tc_{\ih}(\Ldens{a}))$ is a non decreasing function of $\ih$.
As a consequence the equation  $H(\Tc_{\ih}(\Ldens{a})) =\xl$
cannot have more than a single solution.
Furthermore, by the smoothness of the channel family
$\BMS(\ih)$,  $H(\Tc_{\ih}(\Ldens{a}))$ is continuous.
Notice  that
$H(\Tc_{0}(\Ldens{a})) = 0$: if the channel is noiseless the
output density at a variable nodes is noiseless as well.
Therefore, a necessary
and sufficient condition for a solution $\ih(\Ldens{a}, \xl)$
to exist (when the family $\{\BMS(\ent)\}_{\ent}$ is complete) is that
$H(\Tc_{1}(\Ldens{a}))= H(\ledge(\redge(\Ldens{a}))) \ge\xl$.

Any fixed point of the above transformation $\Td_{\xl}$, i.e. any
$\Ldens{f}$ such that
$\Ldens{f} = \Td_{\xl}(\Ldens{f})$, is also a fixed point of
ordinary density evolution for the channel $\BMS(\ih)$ with
$\ih = \ih(\Ldens{f}, \xl)$, and corresponds to
a point on the $\EBP$ $\gexit$ curve.
Furthermore if a sequence of densities such that $\Ldens{a}_{\ell+1} =
\Td_{\xl}(\Ldens{a}_{\ell})$ converges (weakly)
to  a density $\Ldens{f}$, then $\Ldens{f}$ is a fixed point of
$\Td_{\xl}$, with entropy $\xl$.

This motivates the following numerical procedure which has been used to
determine the $\gexit$ curves plotted in the previous section. $(i)$
Set the initial condition  $\Ldens{a}_0 = \BMS(\xl)$.
$(ii)$ For $\ell\ge 0$ compute  $\Ldens{a}_{\ell+1} =
\Td_{\xl}(\Ldens{a}_{\ell})$.  In practice the convolutions are evaluated
numerically either by sampling or, via Fourier transforms as in ordinary
density evolution. Due to the monotonicity
of $H(T_{\ih}(\Ldens{a}_{\ell}))$ in $\ih$, the value of
$\ih(\Ldens{a}_{\ell}, \xl)$ can be efficiently found by bisection.
$(iii)$  The current estimate of
the $\gexit$ function is given by 
$({\ih}_{\ell},\gexitf{\EBPsmall}_{\ell})$.
Here $\ih_\ell\defas\ih(\Ldens{a}_{\ell}, \xl)$ 
is the current estimate of the channel entropy, and
\begin{eqnarray}
\gexitf{\EBPsmall}_\ell & \defas
\int_{-\infty}^{\infty} \Ldens{b}_{\ell}(y)\; \gexitkl {\BMS(\ih_{\ell})}
y \;\text{d}y.
\end{eqnarray}
with $\Ldens{b}_{\ell}\defas \Lambda(\rho(\Ldens{a}_{\ell}))$.
$(iv)$ Halt when some convergence criterion is met and return the
current estimate $({\ih}_{\ell},\gexitf{\EBPsmall}_{\ell})$. In
practice one can require that (a properly defined)
distance between $\Ldens{a}_\ell$ and $\Ldens{a}_{\ell+1}$
becomes smaller than a threshold.

In all the examples discussed in the previous section, we found
that this procedure converges rapidly, and that the limit point
is (within numerical precision) independent of the initial condition
$\Ldens{a}_0$. Proving these statements seems a challenging task
(notice that unlike in ordinary density evolution, the
sequence $\{\Ldens{a}_{\ell}\}$ is in general not ordered by physical
degradation). However it is easy to show that, if $\xl$ is such that
$\Td_{\xl}$ is `well defined', then this procedure
has at least one fixed point.
\btheo\label{theo:ExistenceEntro}
Let $(\ledge,\redge)$ be a \ddp, $\xl\in [0,1]$, and $\Td_{\xl}$
the corresponding density evolution operator at fixed density entropy 
defined as above,
for the smooth, complete and degraded family $\{\BMS(\ih)\}_{\ih}$.
If $H(\ledge(\redge(\Ldens{a}))) \ge\xl$ for any density $\Ldens{a}$ with
$H(\Ldens{a}) = \xl$, then there exists at least one density
$\Ldens{f}$ such that $\Td_{\xl}(\Ldens{f}) = \Ldens{f}$.
Equivalently, $H(\Ldens{f})=\xl$ and there exists 
$\ent\in[0,1]$ such that $\Ldens{f}$ is a fixed point of
density evolution  for the channel $\BMS(\ih)$.
\etheo
\bproof
Consider the space ${\sf S}_{\xl}$ of $L$-densities $\Ldens{a}$
such that $H(\Ldens{a}) = \xl$. Any element
in ${\sf S}_{\xl}$ is a probability measure on the completed real line,
satisfying
the symmetry condition (formally $\Ldens{a}(-x) = e^{-x}\Ldens{a}(x)$).
Vice versa, any such probability measure (to be denoted formally
by its `density' $\Ldens{a}$) with $\E [\log(1+e^{-x})] = \xl$) corresponds
to a unique element of   ${\sf S}_{\xl}$.
Notice that the completed linear line $\reals_{\infty}$
is a compact metric space (we can for instance identify
it with $[-1,1]$ through the mapping $x\mapsto \tanh(x/2)$ and
use the euclidean metric on $[-1,1]$).
Therefore, the space of probability measure on
$\reals_{\infty}$ is sub-sequentially compact under the weak topology
by Prohorov's theorem \cite{Shir96}.
Both the  symmetry condition and  $H(\Ldens{a}) = \xl$ are
closed under the same topology, and therefore ${\sf S}_{\xl}$ is compact
as well.

Let ${\sf BL}$ be the space of bounded Lipshitz function
on $\reals_{\infty}$ (as above, we identify  $\reals_{\infty}$
with $[-1,1]$ and consider the Lipschitz condition with respect
to the induced distance) with the corresponding norm $||\cdot ||_{\sf BL}$.
The space of probability measures on $\reals_\infty$
can be viewed as a convex subset of
the dual space ${\sf BL}^*$, and the topology
induced by the dual norm $||\cdot ||_{\sf BL}^*$ coincides with the
weak topology (cf. \cite[Chapter III, \S 7]{Shir96}).
As a consequence ${\sf S}_{\xl}$ is a compact convex subspace of a normed
linear space.

By hypothesis the mapping $\Ldens{a}\mapsto \Td_{\xl}(\Ldens{a})$,
is well defined for any $\Ldens{a}\in {\sf S}_{\xl}$, and
maps ${\sf S}_{\xl}$ into itself. Furthermore, it is easily seen to
be continuous with respect to the weak topology.
This is a consequence of the Lipschitz continuity of the
functions $(x_1,\dots,x_{\ldegree})\to (x_1+\dots+x_{\ldegree})$
and $(x_1,\dots,x_{\rdegree-1})\to 2\, {\rm atanh}(\tanh(x_1/2)\cdots
\tanh(x_{\rdegree-1}/2))$. Therefore $\Td_{\xl}$ is compact and,
by Schauder's fixed point theorem (cf. \cite{Bro93}, Chapter 4) it has
at least one fixed point.
\eproof
Notice that the above procedure, as well as Theorem \ref{theo:ExistenceEntro},
holds unchanged if the entropy functional $H(\,\cdot\,)$ is substituted
by any continuous linear functional which preserves physical degradation.

In checking the hypothesis of Theorem \ref{theo:ExistenceEntro},
as well as in applications, it is important to prove bounds on the
entropy of fixed point pairs $(\Ldens{f},\Ldens{c})$.
We start by recalling upper and lower bounds on the entropy 
of $\Tc_{\ent}(\Ldens{a})$  which follows
straightforwardly from \cite{SSZ03,LHHH03,HuH03,LHHH05}.
\blemma[Lower Bound]\label{lemmaLBH}
Consider a  \ddp $(\ledge, \redge)$ and transmission over the
channel $\BMS(\ent)$. Let
\begin{align*}
\underline{l}(\xl) \defas  \ledge(\xl),  \;\;\;\;\;
\underline{r}(\xl) \defas
\sum_{i} \redge_i
h_2\Bigl(\frac{1-(1-2 \cp(\xl))^{i-1}}{2} \Bigr)\, ,
\end{align*}
where $\cp(\xl) \defas h_2^{-1}(\xl)$.
If $\Ldens{a}$ is an $L$-density with $H(\Ldens{a})=\xl$, then
\begin{align*}
H(\Tc_{\ent}(\Ldens{a}))\ge \ent\; \underline{l}(\underline{r}(\xl))\, .
\end{align*}
\elemma
\bproof
Following Refs.~\cite{SSZ03,LHHH03,HuH03,LHHH05},
for fixed $H(\Ldens{a})$ and $H(\Ldens{b})$,
$\Ldens{a}\conv \Ldens{b}$ has minimum entropy 
if $\Ldens{a}$ and $\Ldens{b}$ are the densities corresponding to a $\BEC$.
On the other hand,
for the convolution at a parity-check node the minimum is achieved
when the input densities correspond to a $\BSC$.
The lemma follows by applying these bounds to random variable
and check nodes with degree distributions given by 
$\lambda$ and $\rho$.
\eproof
This result can be used to check the hypotheses of
Theorem \ref{theo:ExistenceEntro}.
We deduce that, if $\underline{l}(\underline{r}(\xl))\ge \xl$
for some $\xl\in [0,1]$, 
then there exists a fixed point pair $(\Ldens{f},\Ldens{c})$
with $H(\Ldens{f}) = \xl$ and $\Ldens{c}=\BMS(\ent)$ for some $\ent$.
For instance, for cycle codes (i.e., for $\lambda(x) =x$) this 
implies that such a fixed point pair $(\Ldens{f},\Ldens{c})$
exists for any $H(\Ldens{f})= \xl\in[0,1]$.

\blemma[Upper Bound]
\label{lemmaUBH}
Consider a  \ddp $(\ledge, \redge)$ and transmission
over the channel $\BMS(\ent)$. Let
\begin{align*}
\overline{l}(\ent, \xl) \defas & \sum_{i} \ledge_i f_{i-1}(\ent, \xl)\, ,
\;\;\;\;\;\;\;\overline{r}(\xl) \defas & 1 - \redge(1-\xl)
\end{align*}
where  
\begin{align*}
f_i(\ent, \xl) \defas
\sum_{k \in \{ \pm 1\}}
\sum_{j=0}^{i} \binom{i}{j}
(1-\cp(\xl))^{j}  \cp(\xl)^{i-j} a_k(\ent)\\
 \cdot
\logtwo \Bigl( 1+
\frac{ \cp(\xl)^{2j-i} a_{-k}(\ent)}{(1-\cp(\xl))^{2j-i} a_k(\ent)}
\Bigr)\, ,
\end{align*}
$a_{+1}(\ent) \defas 1- \cp(\ent)$,
$a_{-1}(\ent) \defas \cp(\ent)$, and $\cp(\ent) \defas h_2^{-1}(\ent)$ as
above.
If $\Ldens{a}$ is an $L$-density with $H(\Ldens{a})=\xl$, then
\begin{align*}
H(\Tc_{\ent}(\Ldens{a}))\le \overline{l}(\ent,\overline{r}(\xl))\, .
\end{align*}
\elemma
\bproof
Apply the upper bounds of~\cite{SSZ03,LHHH03,HuH03,LHHH05} 
(simply interchange $\BEC$ and $\BSC$).
\eproof

\btheo[Bounds on $\exit$ Function]
Consider a \ddp $(\ledge, \redge)$ and transmission
over the degraded family $\{\BMS(\ent)\}_{\ent}$. Define the functions
\begin{align*}
\underline{L}(\xl) \defas   \lnode(\xl)\, , \;\;\;\;\;
\overline{L}(\xl) \defas  \sum_{i} \lnode_i f_i(1,\xl),
\end{align*}
and 
$f(\xl,\xl') \defas  \max \{\ent: \overline{l}(\ent,\xl')=\xl \}$
(with the convention $f(\xl,\xl')=0$, if the set is empty).
Let $\Ldens{f}$ denote any fixed point of density evolution,
i.e., $\Ldens{f} = \Tc_{\ent}(\Ldens{f})$.
If $\entropy(\Ldens{f})=\xl$ then
\begin{align*}
f(\xl,\overline{r}(\xl))
\leq \ent \leq
\xl/\underline{l}(\underline{r}(\xl)), \\
\underline{L}(\underline{r}(\xl))
\leq \exitf {\EBPsmall} \leq
\overline{L}(\overline{r}(\xl)).
\end{align*}

In words, the entropy parameters of any fixed points of density evolution, and
so in particular the function $\exitf {\EBPsmall}$, are contained in
the union of rectangles as given above.
\etheo
\begin{proof}
The first two inequality follow from Lemma \ref{lemmaLBH} and \ref{lemmaUBH}.
From Lemma \ref{lemmaLBH} we get 
$\xl = H(\Ldens{f}) = H(\Tc_{\ent}(\Ldens{f}))\ge \ent\,
\underline{l}(\underline{r}(\xl))$ which gives the upper bound
on $\ent$. Analogously,  Lemma \ref{lemmaUBH} implies
$\xl\ge \overline{l}(\ent,\overline{r}(\xl))$. Since
$\overline{l}(\ent,\overline{r}(\xl))$ is monotonically 
increasing in $\ent$, this relation can be inverted as
in the thesis of the theorem.

Given the fixed point $\Ldens{f}$, the corresponding
$\exit$ entropy at variable nodes is 
$\exitf {\EBPsmall} = H(L(\rho(\Ldens{f})))$.
The bounds are obtained as in the proofs of Lemmas \ref{lemmaLBH}
and \ref{lemmaUBH}.
\end{proof}
Discussion: The bounds given above are by no means
best possible. First, the given bounds are ``universal'' in
the sense that the are valid for {\em all} channel distributions.
Better bounds for any specific channel family can be derived
by taking the actual input distribution into account.
Even in the universal case slightly better bounds can be given by
taking into account that at the variable node before convolution
with the channel, the incoming message density can not be of arbitrary shape but
that it is already the convolution
of several message densities.
Second, tighter bounds on the extremes of information combining
have been derived in \cite{SSZ05} and can be translated to giver tighter
bounds on $\exit$ functions, albeit at the prize of
more complex expressions. Finally, by using a similar techniques
one can also give bounds on the entropy versus $\gexit$ parameter of
any fixed point with respect to any smooth channel family.

\begin{figure}[hbt]
\centering
\setlength{\unitlength}{0.6bp}
\begin{picture}(400,200)
\put(0,0)
{
\put(10,0){\includegraphics[scale=0.6]{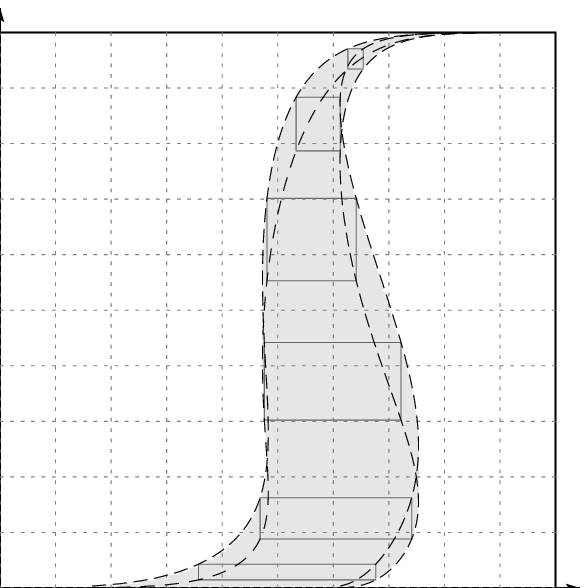}} %length=180bp
{\tiny
%\footnotesize
\multiputlist(20,-2)(32,0)[cb]{$~$,$0.2$,$0.4$,$0.6$,$0.8$,$1.0$}
\multiputlist(18,10)(0,32)[rc]{$~$,$0.2$,$0.4$,$0.6$,$0.8$,$1.0$}
\put(18,-2){\makebox(0,0)[rb]{$0.0$}}
}
\put(181,12){\makebox(0,0)[lb]{$\ih$}}
}
\put(220,0)
{
{\tiny
%\footnotesize
\put(10,0){\includegraphics[scale=0.6]{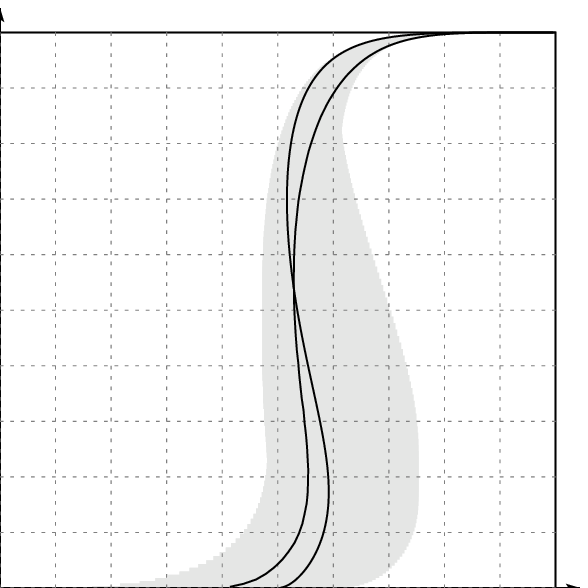}} %length=180bp
\multiputlist(20,-2)(32,0)[cb]{$~$,$0.2$,$0.4$,$0.6$,$0.8$,$1.0$}
\multiputlist(18,10)(0,32)[rc]{$~$,$0.2$,$0.4$,$0.6$,$0.8$,$1.0$}
\put(18,-2){\makebox(0,0)[rb]{$0.0$}}
}
\put(181,12){\makebox(0,0)[lb]{$\ih$}}
\put(50,110){\makebox(0,0)[lb]{{$\exitf{\BPsmall}(\ih)$}}}
}
\end{picture}
\caption{
Left: Construction of bounding region for all
$\EBP$ $\exit$ curves for the \ddp $(\ledge,\redge)=(2/5 x + 3/5 x^5,x^5)$.
Right: The $\EBP$ $\exit$ curves for transmission over the $\BSC$ and
the $\BEC$ families.
\label{fig:exitbound}}
\end{figure}

\bex[LDPC$(2/5 x + 3/5 x^5,x^5)$]
Consider again the \ddp $(\ledge,\redge)=(2/5 x + 3/5 x^5,x^5)$.
Fig.~\ref{fig:exitbound}
shows on the left the construction of the bounded region (union
of rectangles) which contains
all $\EBP$ $\gexit$ curves.
The dashed lines represent the individual curves traced out by the
corner points of the rectangles.
On the right this is compared to the actual
$\EBP$ $\gexit$ curves
for transmission over the $\BSC$ and
the $\BEC$ families (solid lines).
\eex
\btheo[EPP Area Theorem]
\label{theo:ebpareatheorem}
Consider the \ddp $(\ledge, \redge)$ and transmission
over the smooth and degraded family $\{\BMS(\ent)\}$. 
Let $\gexitf {\EBPsmall}$
denote the corresponding $\EBP$ $\gexit$ function.
Assume that the corresponding $\{\Ldens{f}_\xl\}_\xl$
and $\{\Ldens{c}_\xl\}_\xl$, $\xl \in [0, 1]$,
form a {\em complete fixed-point family}.
Then
\begin{align*}
\int_{0}^1 \gexitf {\EBPsmall}(\xl) \text{d}\xl  = 1 -\frac{\int \redge}{\int \ledge}.
\end{align*}
\etheo
\bproof
%\par
\begin{figure}[hbt]
\centering
\setlength{\unitlength}{0.5bp}
\begin{picture}(160,160) % size of small graph
\put(20,0){\includegraphics[scale=0.5]{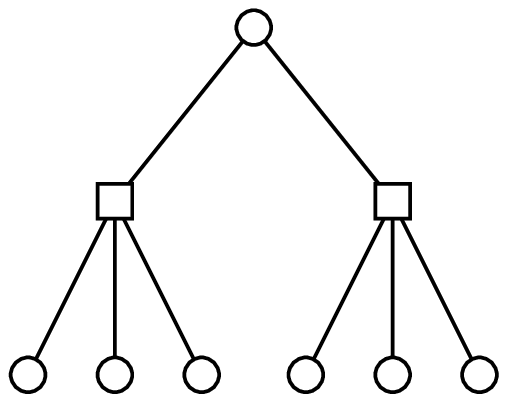}}
\put(100,10){\makebox(0,0){\small{leaves}}}
\put(100,150){\makebox(0,0){\small{root}}}
\end{picture}
\caption{
Computation tree of depth one for the $(2,4)$-regular LDPC ensemble.}
\vspace{7bp}
\label{fig:smallgraph}
\end{figure}
First, let us assume that the ensemble is $(\ldegree, \rdegree)$-regular. Consider a
variable node and the corresponding computation tree of depth one as shown in
Fig.~\ref{fig:smallgraph}.
Let us assume that the bit associated to the
root node is passed through the channel characterized by $\Ldens{c}_\xl$, while the ones
associated to the leaf nodes are  passed through a channel
characterized by $\Ldens{f}_\xl$.
Apply the GAT:
let $X=(X_1,\dots,X_{1+\ldegree\times(\rdegree-1)})$
be the transmitted codeword  chosen uniformly at random
from the tree code and $Y(\xl)$ be the result of
passing the bits of $X$ through their respective channels
with parameter $\xl$.
Note that $H(X \mid Y(\xl=1))-H(X \mid Y(\xl=0)) =H(X)$. This
follows since by assumption the fixed-point family is complete.
In particular this implies that the channel for $\xl=0$ is
the ``noiseless'' channel so that $H(X \mid Y(\xl=0))=0$.
By the GAT, this difference is equal to the sum of the integrals
of the individual $\gexitfi {}$ curves, where the integral extends from $\xl=0$ to $\xl=1$.
There are two types of individual $\gexitfi {}$ curves, namely the one
associated to the root node, call it $\gexitfroot {}$,
and the $\ldegree (\rdegree-1)$ ones associated to the leaf nodes,
call them $\gexitfleaf {}$. To summarize, the GAT
states
\begin{align*}
H(X) & = \int_0^1 \gexitfroot {} (\xl)\, \text{d}\xl
+\ldegree (\rdegree-1) \int_0^1\gexitfleaf {} (\xl)\text{d}\xl.
\end{align*}
Note that $H(X) = 1+\ldegree (\rdegree-1)-\ldegree=1-\ldegree (\rdegree-2)$ since the computati
on tree
contains $1+\ldegree(\rdegree-1)$ variable nodes and $\ldegree$ check nodes.
Moreover, $\int_0^1 \gexitfleaf {} (\xl)\text{d}\xl=\int_0^1 1-\redge(1-\xl)\text{d}\xl=
({\rdegree-1})/{\rdegree}$. This follows by applying the GAT once
again to a $[\rdegree, 1, \rdegree-1]$ single parity check code.
Collecting these observations and solving for
$\int_{0}^{1} \gexitfroot {} (\xl) \;\text{d}\xl$, we get
\begin{align*}
\int_{0}^{1} \gexitfroot {}(\xl) \;\text{d}\xl = 1- \ldegree/\rdegree=\drate,
\end{align*}
as claimed since $\gexitfroot{} =\gexitf {\EBPsmall}$.

The irregular case follows in the same manner: we consider the ensemble of computation
trees of depth one where the degree of the root note is chosen according to the
node degree distribution $\lnode$ and each edge emanating from this root node
is connected to a check node whose degree is chosen according to the edge degree distribution
$\redge$. As before, leaf nodes experience the channel characterized by $\Ldens{f}_x$,
whereas the root node experiences the channel characterized
by $\Ldens{c}_x$. We apply the GAT to each such choice and
average with the respective probabilities.
\end{proof}

This result imposes some strong constraint on $\BP$ $\gexit$ functions 
and their relation to $\MAP$ $\gexit$  functions. Here is an example.
\begin{corollary}\label{CoroCycle}
Consider communication over the smooth and degraded family 
$\{\BMS(\ent)\}_{\ent}$, $\ent\in[0,1]$ using uniformly
random codes from the ensemble $\eldpc n \ledge \redge$ and assume that the
rate of this ensemble converges to the design rate, see \cite[Lemma 7]{MMU05}.
Assume that the $\BP$ fixed point family
$\{\BMS(\ent),\Ldens{a}_{\ent}\}$, is smooth and complete. 
Then ($\MAP$) $\gexit$ function and $\BP$ $\gexit$ function
coincide: $\gexitf{}(\ent)=\gexitf{\BPsmall}(\ent)$ for almost every
$\ent\in[0,1]$.
\end{corollary}
\begin{proof}
By hypothesis we can apply Theorem \ref{theo:ebpareatheorem} to 
the $\BP$ $\gexit$ function. We get
\begin{eqnarray*}
\int_0^1\gexitf{\BPsmall}(\ent)\, \de\ent = r\, .
\end{eqnarray*}
Further, by the GAT (and applying Fubini theorem and 
Fatou's lemma as in the proof of Theorem \ref{theo:UBMAP})
\begin{eqnarray*}
\int_0^1\gexitf{}(\ent)\, \de\ent = r\, .
\end{eqnarray*}
The proof if completed by noticing that, because of Lemma 
\ref{lemma:gexitmapversusgexitbp}, $\gexitf {}(\ent) 
\leq \gexitf {\BPsmall}(\ent)$ for every $\ent\in[0,1]$.
\end{proof}
Proving that the hypotheses of this Corollary hold for some
\ddp $(\ledge,\redge)$ is a challenging task
(see also next section). On the other hand, numerical computations
show very clearly that this is the case, for instance, for cycle ensembles,
cf. Example \ref{ex:CycleEx}. 
%
%***********************************************************************
%
\section{Regularity of Extended BP $\gexit$ Curves}
\label{sec:Regularity}

Theorem \ref{theo:ExistenceEntro} ensures (for many $\ldpc$ 
ensembles) the existence of a fixed point pair 
$(\Ldens{f}_{\xl},\Ldens{c}_{\xl})$ for each value of
$\xl = H(\Ldens{f}_{\xl})$. However, for applying the extended Area
Theorem \ref{theo:ebpareatheorem} the resulting family has to be smooth with
respect to the parameter $\xl$. That this is indeed the case
is strongly suggested by the numerical computation of the $\EBP$ curve, cf. Sec.~\ref{sec:egexit}. 
We provide here some partial analytic results in this direction.

Throughout this section, we denote by $\batta(\Ldens{a})$ the Battacharyya 
parameter for the $L$-density $\Ldens{a}$. Furthermore, when assuming 
communication through the channel $\BMS(\ih)$, we denote by 
$B_{\ih}$ the Battacharyya parameter of the channel.
\begin{lemma}
Assume communication over the degraded family $\{\BMS(\ih)\}_{\ih}$ channel using the 
\ddp $(\lambda,\rho)$. Then, for any $\ih$, there exists at most a unique fixed point
density $\Ldens{f}_{\ih}$ such that
\begin{eqnarray}
B_{\ih}\lambda'(1)\rho''(1-\batta(\Ldens{f}_{\ih})^2)<1
\, .\label{eq:UniquenessCond}
\end{eqnarray}
Furthermore, if such a density $\Ldens{f}_{\ih}$ exists, it coincides with 
the  BP fixed point. Finally,
$\batta(\Ldens{f}_{\ih})$ is Lipschitz continuous with respect to $B_\ih$. 
More precisely, if the two fixed points
$\Ldens{f}_{\ih_1}$, $\Ldens{f}_{\ih_2}$ satisfy the condition
$B_{\ih_i}\lambda'(1)\rho''(1-\batta(\Ldens{f}_{\ih_i})^2)\le 1-\delta$
for some $\delta>0$, then there exists $C = C(\delta,\lambda,\rho)$, 
such that 
\begin{eqnarray*}
|\batta(\Ldens{f}_{\ih_1})-\batta(\Ldens{f}_{\ih_2})|\le C\,|B_{\ih_1}-B_{\ih_2}|\, .\label{eq:Lipschitz}
\end{eqnarray*}
\end{lemma}
\begin{proof}
Consider two channel parameters $\ih_1\le\ih_2$ and 
two $L$-densities $\Ldens{a}_1$ and
$\Ldens{a}_2$ satisfying  the condition
$B_{\ih_i}\lambda'(1)\rho''(1-\batta(\Ldens{f}_{\ih_i})^2)<1-\delta$
for some $\delta>0$.
Assume that $\Ldens{a}_2$ is physically degraded with 
respect to $\Ldens{a}_1$. We prove in Appendix \ref{sec:ProofContraction} 
that there exists a constant $\alpha=\alpha(\lambda,\rho,\delta)<1$ 
on $\delta$, the channel family and the degree distribution, such that 
\begin{align}
|\batta(\Tc_{\ih_1}(\Ldens{a}_{1}))-&\batta(\Tc_{\ih_2}(\Ldens{a}_{2}))|
\le \label{eq:Contraction}\\ &\alpha\, 
|\batta(\Ldens{a}_{1})-\batta(\Ldens{a}_{2})|
+|B_{\ih_1}-B_{\ih_2}|\, .\nonumber
\end{align}

Let us show that  this result implies the thesis.
Denote by $\Ldens{f}_{\ih}$ the BP fixed point for the channel 
$\BMS(\ih)$ and notice that any other fixed point $\Ldens{f}'_{\ih}$
for the same channel is necessarily physically upgraded with respect to 
$\Ldens{f}_{\ih}$. Using the standard notation 
$\Ldens{f}_{\ih}\succ \Ldens{f}'_{\ih}$. 
In fact $\Delta_0\succ\Ldens{f}'_{\ih}$. By applying the 
density evolution operator, we deduce that 
$\Ldens{a}^{\BPsmall,\ell}_{\ih}\succ\Ldens{f}'_{\ih}$,
where $\Ldens{a}^{\BPsmall,\ell}$ is the density after $\ell$ iterations of BP.
By taking the limit $\ell\to\infty$ we get $\Ldens{f}_{\ih}\succ 
\Ldens{f}'_{\ih}$.

Next notice that, if $\Ldens{f}_\ih$ satisfies Eq.~(\ref{eq:UniquenessCond})
there cannot be a distinct fixed point, physically upgraded with respect
to $\Ldens{f}_{\ih}$, also satisfying
Eq.~(\ref{eq:UniquenessCond}). If such a density $\Ldens{f}_{\ih}'$ existed,
we could apply (\ref{eq:Contraction}) to get
\begin{align*}
|\batta(T_{\ih}(\Ldens{f}_{\ih}))-\batta(T_{\ih}(\Ldens{f}'_{\ih}))|
\le \alpha\, 
|\batta(\Ldens{f}_{\ih})-\batta(\Ldens{f}'_{\ih})|\, ,
\end{align*}
with $\alpha<1$. But, since 
$T_{\ih}(\Ldens{f}_{\ih}) = \Ldens{f}_{\ih}$ and
$T_{\ih}(\Ldens{f}'_{\ih}) = \Ldens{f}'_{\ih}$, this would imply
$\batta(\Ldens{f}_{\ih})=\batta(\Ldens{f}'_{\ih})$ which is
impossible because  $\Ldens{f}_{\ih}\succ 
\Ldens{f}'_{\ih}$.

Let us finally prove Lipschitz continuity, cf. Eq.~(\ref{eq:Lipschitz}). 
Under our hypotheses,
the two fixed points $\Ldens{f}_{\ih}$, $\Ldens{f}_{\ih'}$ are 
the BP fixed points for channels $\BMS(\ent)$ and $\BMS(\ent')$. 
Consider therefore the BP sequences
$\{\Ldens{a}^{\BPsmall,\ell}_{\ih}\}_{\ell\ge 0}$, 
$\{\Ldens{a}^{\BPsmall,\ell}_{\ih'}\}_{\ell\ge 0}$. For each $\ell$,
$\Ldens{a}^{\BPsmall,\ell}_{\ih}$ (respectively $\Ldens{a}^{\BPsmall,\ell}_{\ih'}$) 
is physically degraded with respect to $\Ldens{f}_{\ih}$ (respectively
$\Ldens{f}_{\ih'}$), 
and therefore satisfies the  condition (\ref{eq:UniquenessCond}),
since the latter does. 
Furthermore, assuming without loss of generality 
$\ent'>\ent$, we have $\Ldens{a}^{\BPsmall,\ell}_{\ih'}\succ 
\Ldens{a}^{\BPsmall,\ell}_{\ih}$.
Let $\delta_{\ell} \defas
|\batta(\Ldens{a}^{(\ell)}_{\ih}-\batta(\Ldens{a}^{(\ell)}_{\ih'})|$.
Clearly $\delta_0=0$. By applying Eq.~(\ref{eq:Contraction}), we
get
$\delta_{\ell+1}\le \alpha\, \delta_{\ell}
+|B_{\ih_1}-B_{\ih_2}|$, and therefore
\begin{align*}
\delta_{\ell}\le (\alpha+\alpha^2+\cdots+\alpha^{\ell})\, 
|B_{\ih_1}-B_{\ih_2}|\le \frac{\alpha}{1-\alpha}\, |B_{\ih_1}-B_{\ih_2}|\, .
\end{align*}
The thesis follows by taking the $\ell\to\infty$ limit.
\end{proof}
It is worth mentioning that the Lipschitz condition  Eq.~(\ref{eq:Lipschitz})
implies analogous regularity properties for other functionals
of the density $\Ldens{a}_{\ih}$. For instance, it is easy to show that
$|H(\Ldens{f}_{\ih_1})-H(\Ldens{f}_{\ih_2})|\le A \,
|\batta(\Ldens{f}_{\ih_1})-\batta(\Ldens{f}_{\ih_2})|$, for some
universal constant $A$. Also, the Battacharyya parameter is, for most 
channel families, a smooth function of the channel parameter. Regularity
with respect to $B_{\ih}$ translates therefore immediately into regularity
with respect to $\ih$.

In applying the above result, it is helpful to have bounds on the 
Battacharyya parameter of the fixed point densities.
\begin{lemma}\label{lemma:LB_Bhatta}
Assume communication over the channel $\BMS(\ih)$ using random
codes from the $(\lambda,\rho)$ ensemble. If $\Ldens{f}$ is a fixed point 
density with Battacharyya parameter $b=\batta(\Ldens{f})$, then
\begin{align*}
b\ge B_{\ih}\lambda(\tilde{b})\, ,\;\;\;\;\;
\tilde{b} \defas \sum_{\rdegree}\rho_{\rdegree}\sqrt{1-(1-b^2)^{\rdegree-1}}\,.
\end{align*}
\end{lemma}
\begin{proof}
First notice that $b =  B_{\ih}\lambda(\tilde{b}')$ where 
$\tilde{b} \defas \sum_{\rdegree}\rho_{\rdegree}
\batta(\Ldens{f}^{\boxast (\rdegree-1)})$, and $\boxast$ denotes the 
convolution at check nodes. It
is convenient to write densities in terms of the variable
$u\defas\sqrt{1-\tanh^2(x/2)}$. With a slight abuse of notation, 
we use the same symbol $\Ldens{f}$ to denote the density with respect to 
$u$. We get
\begin{align*}
\batta(\Ldens{f}^{\boxast i})\!=\!\!
\int \!\sqrt{1-(1-u^2_1)\cdots(1-u_i)^2}\,
\Ldens{f}(u_1)\de u_1\cdots\Ldens{f}(u_i)\de u_i\, .
\end{align*}
The proof is completed by using convexity with respect to the 
$u_1$,\dots $u_i$ together with the fact that 
$b=\int \!u   \Ldens{f}(u)\,\de u$.
\end{proof}

\begin{figure}[hbt]
\centering
\setlength{\unitlength}{0.5bp}
\begin{picture}(170,170)
\put(0,0){\includegraphics[scale=0.5]{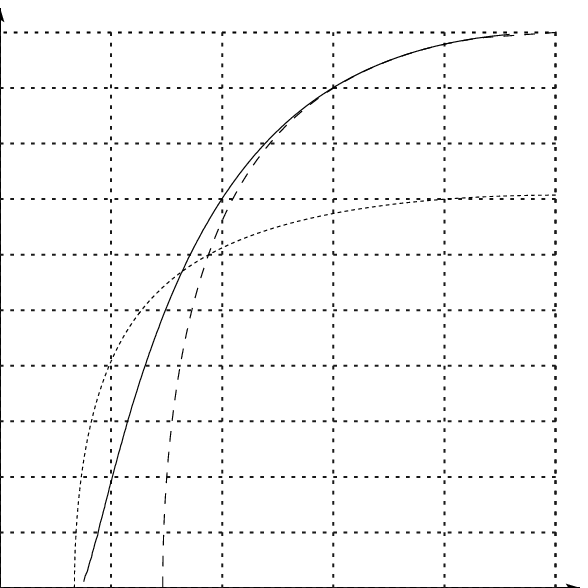}}
{\footnotesize
\multiputlist(0,-4)(32,0)[tc]{$0.0$,$0.1$,$0.2$,$0.3$,$0.4$, $0.5$}
\put(170,0){\makebox(0,0)[lb]{$\epsilon$}}
}
\end{picture}
\caption{\label{fig:23example}
The solid line shows the Battacharyya constant of the fixed point
density as a function of the channel parameter of the \BSC.
The dotted line corresponds to the bound stated in (\ref{eq:Uniq23}),
whereas the dashed curve corresponds to the bound $\batta(\Ldens{f})\ge \sqrt{2-B(\epsilon)^{-2}}$.}
\end{figure}

\bex
Consider the $(2,3)$ ensemble and communication over the $\BSC(\epsilon)$.
In this case Eq.~(\ref{eq:UniquenessCond}) is equivalent to 
\begin{eqnarray}
\batta(\Ldens{f})>\sqrt{1-\frac{1}{2B(\epsilon)}}\, ,\label{eq:Uniq23}
\end{eqnarray}
where $B(\epsilon)$ denotes the channel Battacharyya parameter as a 
function of the flip probability $B(\epsilon) = \sqrt{4\epsilon(1-\epsilon)}$.
Lemma \ref{lemma:LB_Bhatta} implies (if we neglect the 
case $\batta(\Ldens{f})=0$ which corresponds to a no-error
fixed point) $\batta(\Ldens{f})\ge \sqrt{2-B(\epsilon)^{-2}}$.
This lower bound lies in the region described by equation (\ref{eq:Uniq23})
as soon as $B(\epsilon)\ge (\sqrt{17}-1)/4$, i.e., $\epsilon>\epsilon_*$
with
\begin{eqnarray*}
\epsilon_* = \frac{1}{2}-\sqrt{\frac{\sqrt{17}-1}{32}}\, ,
\end{eqnarray*}
which yields $\epsilon_*\approx 0.18759473$. The above results imply that 
a unique fixed point density (apart from the no-error one) exists
for any $\epsilon>\epsilon_*$. On the other hand numerical computations 
suggest this to be the case for all values above the local stability 
threshold $\epsilon_{\rm ls} = (2-\sqrt{2})/4\approx 0.066987298$.
Fig.~\ref{fig:23example} shows the Battacharyya constant of the fixed point
density as a function of the channel parameter of the \BSC\ (solid line),
the bound stated in (\ref{eq:Uniq23}) (dotted line), as well as
the bound $\batta(\Ldens{f})\ge \sqrt{2-B(\epsilon)^{-2}}$ (dashed line).
\eex

As stressed in the previous section, $\EBP$ curves are expected to be
single valued smooth functions of the  entropy $H(\Ldens{f})$ of the 
fixed point density. The same expectation holds, if entropy is replaced by 
any linear functional which preserves physical degradation.
The following result confirms that better regularity properties can 
indeed be obtained by taking this point of view.
\begin{lemma}
Let  $\{\BMS(\ent)\}_{\ent}$ be a degraded family.
Assume $\Ldens{f}_1$ and $\Ldens{f}_2$ to be fixed point densities 
for the channel parameters $\ih_1$, $\ih_2$, and that 
$\Ldens{f}_1$ is physically degraded with respect to $\Ldens{f}_2$.
If $\batta(\Ldens{f}_1)\ge\delta>0$, then there exists a constant
$C=C(\lambda,\rho,\delta)$ such that
\begin{eqnarray*}
|B_{\ih_1}-B_{\ih_2}|\le C\, |\batta(\Ldens{f}_1)-\batta(\Ldens{f}_2)|\, .
\end{eqnarray*}
In words, the channel is a Lipschitz continuous function
of the Battacharyya parameter of the fixed point density.
\end{lemma}
\begin{proof}
Proceeding as in the proof of Eq.~(\ref{eq:Contraction}), 
cf.~Appendix \ref{sec:ProofContraction} it is easy to show
that, if $\Ldens{a}_1\succ\Ldens{a}_2$, then
\begin{align*}
|\batta(\Tc_{1}(\Ldens{a}_{1}))-\batta(\Tc_1(\Ldens{a}_{2}))|
\le \lambda'(1)\rho''(1)|\batta(\Ldens{a}_{1})-\batta(\Ldens{a}_{2})|\, .
\end{align*}
Furthermore, if $\batta(\Ldens{a})\ge\delta>0$, then 
$\batta(\Tc_1(\Ldens{a}))\ge\delta'$ for some $\delta'>0$.

Consider now the difference $|\batta(\Tc_{\ih_1}(\Ldens{f}_1))-
\batta(\Tc_{\ih_2}(\Ldens{f}_2))|$. Since 
$\Ldens{f}_{1/2}$ are density evolution fixed points, this is equal
to $|\batta(\Ldens{f}_1)-\batta(\Ldens{f}_2)|$.
We get therefore
\begin{align*}
|\batta&(\Ldens{f}_1)-\batta(\Ldens{f}_2)| =
|B_{\ih_1}\batta(\Tc_{1}(\Ldens{f}_1))-B_{\ih_2}\batta(\Tc_{1}(\Ldens{f}_2))|\\
\ge & |B_{\ih_1}-B_{\ih_2}|\batta(\Tc_{1}(\Ldens{f}_2))-
B_{\ih_1}|\batta(\Tc_{1}(\Ldens{f}_1))-\batta(\Tc_{1}(\Ldens{f}_2))|\\
\ge & |B_{\ih_1}-B_{\ih_2}|\delta'-
\lambda'(1)\rho''(1)|\batta(\Ldens{f}_{1})-\batta(\Ldens{f}_{2})|\,,
\end{align*}
which implies the thesis after solving for $|B_{\ih_1}-B_{\ih_2}|$.
\end{proof}
%

%
%***********************************************************************
%
\section{$\MAP$ Versus $\BP$ Marginals}
\label{sec:mapversusbp}

As we saw in Sections \ref{sec:egexit} and \ref{sec:HowToEBP}, the 
$\MAP$ and $\BP$ $\gexit$ curves are strictly related for 
LDPC codes in the large blocklength limit.
We conjectured that they can be connected through the
Maxwell construction.
In particular, this would imply that they are asymptotically equal above the 
$\MAP$ threshold for a large family of ensembles, cf. for instance
Example \ref{example:36}.
 
Does the coincidence of $\gexit$ curves mean that
$\BP$ and $\MAP$ decoding in fact coincide {\em bit by bit}?
More precisely, belief propagation can be
regarded as a low complexity (approximate) algorithm for computing
the marginal distributions $p_{X_i\mid Y}(x_i\mid y)$. It is well established
\cite{RiU01}, that the $\BP$ estimate is asymptotically correct in the
low noise regime $\ent<\ent_{\BP}$. We wonder whether the same is true
whenever the two $\gexit$ functions coincide.

Perhaps surprising, the answer is positive. In order to proceed, it is
convenient to introduce some notations. For the sake of simplicity
we consider the case of a binary channel. Rather than the
marginal distributions $p_{X_i|Y}(x_i\mid y)$, it is convenient to
focus on the extrinsic soft bits
\begin{align*}
\mu_{i}(y) \equiv \E[X_i\mid Y_{\sim i} = y_{\sim i}]\, .
\end{align*}
We will further denote by $\mu_{i}^{\BPsmall,\ell}(y)$,
the estimate of this quantity
provided by $\BP$, after $\ell$ iterations.
Notice that $\mu_{i}(y) = \tanh\phi_i^{}(y_{\sim i})$,
and $\mu_{i}^{\BPsmall,\ell}(y) = \tanh\phi_i^{\BPsmall,\ell}(y_{\sim i})$.

A meaningful measure of how much `incorrect' is $\BP$, is the mean square
error
\begin{align*}
\Delta^{(\ell)}(y) \equiv\frac{1}{n}\sum_{i=1}^n
\left|\mu_i^{\BPsmall,\ell}(y)-\mu_i(y)\right|^2\, .
\end{align*}
Let us stress that $\Delta^{(\ell)}(y)$ implies a rather strict notion
of correctness. We are not just requiring the hard decision
reached by $\BP$ to be (approximatively) the same that would be provided
by a $\MAP$ decoder. Rather, $\BP$ should be able to reconstruct the
full information about $X_i$, given the received message.

Our main result is presented below (here we refer to the Tanner graph
associated to the code parity  check matrix, which is naturally related to
belief propagation).
\begin{theorem}\label{thm:BPcorrect}
Consider communication using a linear code over a smooth channel $\BMS(\ent)$,
and let $Y$ be the channel output if the input is uniformly random codeword
$X$.
Let $\gexitkabsd{}{\cdot}$ denote the $\gexit$ kernel in the $|D|$-domain
and  $K \equiv -\sup\left\{\frac{\text{d}^2\gexitkabsd{}{x}}{\text{d} x^2} :\, x\in[0,1]\right\}>0$.
Assume that, for a uniformly
random variable node $i$ in the Tanner graph, the shortest loop
through $i$ has length larger than $2\ell$ with probability at least
$1-\delta$. Then
\begin{eqnarray*}
\E\, \Delta^{(\ell)}(Y)\le \frac{2}{K}[\gexitf{\BPsmall,\ell}(\ent)-
\gexitf{}(\ent)] +4\delta\, .
\end{eqnarray*}
\end{theorem}
Let us stress that this result holds, not just for random elements
of an $\eldpc n \ledge \redge$ ensemble, but for any code with the prescribed 
sparseness properties.

The proof makes use of a technical lemma, which we state below, and prove
in Appendix~\ref{app:BPcorrect}.
\blemma\label{lem:GEXITSquareError}
Consider  a random variable $X$ taking values in $\{+1,-1\}$ and
assume that $X\to Y\to Z$ forms a Markov chain. Let
$k:[0,1]\to {\mathbb R}$ be twice differentiable with
$k'(0)\le 0$, and  $k''(x)\le -K<0$ for any $x\in [0,1]$.
If we denote $\mu_{{\sf Y}}(y) = \E[X\mid Y=y]$ and
$\mu_{{\sf Z}}(z) = \E[X\mid Z=z]$,
then
\begin{align*}
\E[k(|\mu_{{\sf Y}}(Y)|)] \le\E[k(|\mu_{{\sf Z}}(Z)|)]
-\frac{1}{2}\,K\,\E[|\mu_{{\sf Y}}(Y)-\mu_{{\sf Z}}(Z)|^2]\, .
\end{align*}
\elemma

\begin{proof}[Theorem \ref{thm:BPcorrect}]
The $\MAP$ $\gexit$ function can be written as
\begin{eqnarray*}
\gexitf{}(\ent) = \frac{1}{n}\sum_{i=1}^n\,
\E[\gexitkabsd{}{|\mu_i(Y)|}]\, .
\end{eqnarray*}
An analogous expression holds for the $\BP$   $\gexit$ function if we replace
$\mu_i(Y)$ with $\mu^{\BPsmall,\ell}_i(Y)$. We claim that,
if the shortest loop through $i$ in the Tanner graph is longer than
$2\ell$, then
\begin{align}
\E[\gexitkabsd{}{|\mu_i(Y)|}]\le&\E
[\gexitkabsd{}{|\mu^{\BPsmall,\ell}_i(Y)|}]-\label{eq:LocalErr}\\
&-\frac{1}{2}\,K\, \E[(\mu_i(Y)-\mu^{\BPsmall,\ell}_i(Y))^2]\, .\nonumber
\end{align}
The thesis follows by rearranging the terms, using the trivial
bound $(\mu_i(Y)-\mu^{\BPsmall,\ell}_i(Y))^2\le 4$ whenever the
shortest loop through $i$ is not longer than $2\ell$ and summing over $i$.

In order to prove the above claim, let $Y_{\sim i}^{(\ell)}$ denote the subset
of received signals within a distance $\ell$ from the variable node $i$
on the Tanner graph. Notice that $X_i\to Y_{\sim i}\to Y_{\sim i}^{(\ell)}$
is a Markov chain, and that $\mu_i(Y) = \E[X_i|Y_{\sim i}]$,
 $\mu^{(\ell)}_i(Y) = \E[X_i|Y_{\sim i}^{(\ell)}]$.
We can therefore apply Lemma~\ref{lem:GEXITSquareError}, with
$k(x) =\gexitkabsd{}{x}$. This yields Eq.~(\ref{eq:LocalErr}), and thus
concludes the proof.
\end{proof}
One may wonder whether the distortion measure  $\Delta^{(\ell)}(y)$
is appropriate. One could, for instance consider the
actual soft bits, rather than the extrinsic ones. If we let
$\tilde{\mu}_i(y) = \E[X_i|Y=y]$, and denote as
$\tilde{\mu}^{\BPsmall,\ell}_i(y)$ the corresponding $\BP$
estimate, we may define
\begin{eqnarray*}
\widetilde{\Delta}^{(\ell)}(y) = \frac{1}{n}
\sum_{i=1}^n\left|\tilde{\mu}^{\BPsmall,\ell}_i(y)-\tilde{\mu}_i(y)
\right|^2\, .
\end{eqnarray*}
Recall that hard decoding decisions are taken in terms of $\tilde{\mu}_i(y)$,
rather than $\mu_i(y)$. We are therefore interested in knowing whether
$\widetilde{\Delta}^{(\ell)}(y)$ can be much larger than
$\Delta^{(\ell)}(y)$. The answer is generically negative, as shown by the
lemma below.
\begin{lemma}\label{lem:ExtrinsicDistortion}
Assume communication over a $\BMS$ channel with $L$-density
$\Ldens{c}(l)$. Then
\begin{eqnarray*}
\E\,\widetilde{\Delta}^{(\ell)}(Y)\le C\, \E\,\Delta^{(\ell)}(Y)
\end{eqnarray*}
where  $C \equiv \int\! e^{2|l|}\Ldens{c}(l)  \, \text{d}l$.
\end{lemma}
The proof is deferred to Appendix~\ref{app:BPcorrect}.

Theorem \ref{thm:BPcorrect} obviously imply that belief propagation
is `asymptotically correct' every time the $\BP$ and $\MAP$ $\gexit$ functions
asymptotically coincide. We conjectured in Section \ref{sec:egexit} that
the $\MAP$ $\gexit$ function can be obtained from the $\EBP$ one through
the Maxwell construction. This construction allows therefore to determine in
which domain of $\ent$  $\BP$ and $\MAP$ $\gexit$ functions do coincide.
It is worth stating the final result explicitly for
a few simple cases cases.
\begin{corollary}
Consider communication over degraded, smooth and complete family 
$\{\BMS(\ent)\}_{\ent}$,  using uniformly
random codes from the ensemble $\eldpc n \ledge \redge$ and assume that the
rate of this ensemble converges to the design rate. 
Assume that the $\BP$ fixed point family
$\{\BMS(\ent),\Ldens{a}_{\ent}\}$, is smooth and complete. 
Then, for almost every $\ent\in [0,1]$
\begin{align*}
\lim_{\ell\to\infty}\lim_{n\to\infty}\expectation\,\Delta^{(\ell)}(Y) = 0\, .
\end{align*}
\end{corollary}
The proof follows easily from Corollary \ref{CoroCycle}.

A somewhat more general statement is the following.
\begin{corollary}
Consider communication over over degraded, smooth and complete family 
$\{\BMS(\ent)\}_{\ent}$, using uniformly
random codes from the ensemble $\eldpc n \ledge \redge$ and assume that the
rate of this ensemble converges to the design rate. 
Assume that the upper bound in on the $\MAP$ threshold in
Theorem \ref{theo:UBMAP} is tight: $\ent^{\MAPsmall}= \overline{\ent}$. 
Then, for almost any $\ent\in [\overline{\ent},1]$,
\begin{align*}
\lim_{\ell\to\infty}\lim_{n\to\infty}\expectation\,\Delta^{(\ell)}(Y) = 0\, .
\end{align*}
\end{corollary}
\begin{proof}
Proceeding as in the proof of Theorem \ref{theo:UBMAP},
one obtain that
\begin{align*}
\int_{\overline{\ent}}^1\, \gexitf {}(\ent) \, \de\ent=
\int_{\overline{\ent}}^1\, \gexitf {\BPsmall}(\ent) \, \de\ent = r\, .
\end{align*}
Since $\gexitf {}(\ent)\le \gexitf {\BPsmall}(\ent)$ for all $\ent$,
we have necessarily $\gexitf {}(\ent)= \gexitf {\BPsmall}(\ent)$
for almost any $\ent\in[\overline{\ent},1]$. The thesis follows by
applying Theorem \ref{thm:BPcorrect}.
\end{proof}
%
%**************************************************************************

\section{Why We Can Not Surpass Capacity: The Matching Condition}
\label{sec:Matching}

The upper bound $\overline{\ent}$ on the MAP threshold,
cf. Theorem \ref{theo:UBMAP} cannot be larger than the 
Shannon threshold $1-r$. This follows by noticing that the $\gexit$
kernel is not larger than $1$, and implies that iterative coding systems
do not allow to communicate reliably above capacity.
Of course, this result is also a straightforward  consequence of Shannon's 
channel coding theorem. In this section we shall provide yet another
proof of this basic fact. The interest of the new proof is three-fold:
$(i)$ it does not assume communication over a smooth channel family;
$(ii)$ it uses only quantities appearing in density evolution (and not
just fixed points); $(iii)$ component codes (and their `matching')
play a crucial role.

For general BMS channels,
and motivated by the
geometric statement observed for the
BEC and the relationship between the derivative of the
mutual information and the MSE introduced by \cite{GSV04,GSV05}, 
a similar chart,
called MSE chart was constructed by Bhattad and Narayanan \cite{BaN04}.
Assuming that the input densities to the component codes are Gaussian, this
chart again fulfills the Area Theorem. 
In order to apply the MSE chart in the context of iterative coding
the authors proposed to approximate the intermediate densities which
appear in density evolution by ``equivalent'' Gaussian densities.
This was an important first step in generalizing the matching condition
to the whole class of BMS channels.
In the following we show how to overcome the need for making the
Gaussian approximation by using GEXIT functions. 

\begin{figure}[hbt]
\centering
\setlength{\unitlength}{1.0bp}%
\begin{picture}(110,110)
\put(0,0){\includegraphics[scale=1.0]{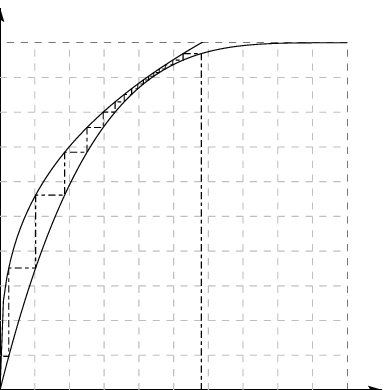}}
%\put(100,-3){\makebox(0,0)[t]{\small $1$}}
%\put(-5,102){\makebox(0,0){\small $1$}}
\put(112, 0){\makebox(0,0)[l]{\small $x$}}
%\put(18, 93){\makebox(0,0){\small $\ledge^{-1}(x/\ent)$}}
\put(40, 60){\makebox(0,0){\small $c(x)$}}
\put(16, 89){\makebox(0,0){\small $v^{-1}_{\ent}(x)$}}
\put(58, 102){\makebox(0,0)[b]{{\small $\ent=0.58$}}}
\put(100, -2){\makebox(0,0)[tr]{$h_{\text{out-variable}}=h_{\text{in-check}}$}}
\put(-4,100){\makebox(0,0)[tr]{\rotatebox{90}{$h_{\text{out-check}}=h_{\text{in-variable
}}$}}}
\end{picture}
\caption{\label{fig:becmatching} The EXIT chart method for
the degree distribution $(\ledge(x)=x^3, \redge(x)=x^4)$ and
transmission over the $\BEC(\ent = 0.58)$.}
\end{figure}

To start, let us review the case of transmission over the $\BEC(\ent)$
using a degree distribution pair $(\ledge, \redge)$.
In this case density evolution is equivalent to the EXIT chart
approach and the condition for successful decoding under \BP\ reads
\begin{align*}
c(x) \defas 1-\redge(1-x) \leq \ledge^{-1}(x/\ent) \defas v^{-1}_{\ent}(x).
\end{align*}
This is shown in Fig.~\ref{fig:becmatching} for the degree distribution pair
$(\ledge(x)=x^3, \redge(x)=x^4)$.
The area under the curve $c(x)$ equals $1-\int \!\redge$ and the
area to the left of the curve $v^{-1}_{\ent}(x)$ is equal to
$\ent \int \!\ledge$. By the previous remarks, a necessary condition
for successful \BP\ decoding
is that these two areas do not overlap.
Since the total area equals $1$ we get the necessary condition
$\ent \int \ledge+1-\int \redge\leq 1$. Rearranging terms, this
is equivalent to the condition
\begin{align*}
1-C_{\Shsmall} = \ent \leq \frac{\int \redge}{\int \ledge}= 1 - r(\ledge, \redge).
\end{align*}
In words, the design rate $r(\ledge, \redge)$ of any LDPC ensemble which, for
increasing block lengths, allows successful
decoding over the $\BEC(\ent)$, can not surpass the Shannon limit
$1-\ent$.
An argument very similar to the above was introduced
by Shokrollahi and Oswald \cite{Sho00,OsS01} (albeit not using the language and geometric
interpretation of EXIT functions and applying a slightly different range of integration).
It was the first bound on the performance of iterative systems in which the Shannon capacity
appeared explicitly using only quantities of density evolution.
A substantially more general version of this bound can be found in 
\cite{AKtB02a,AKTB02,AKtB04}. The extension to parallel turbo schemes is addressed  
in \cite{MeU02jccc,MeU03a}. See also \cite{For05}.

Although the final result (namely that transmission above capacity
is not possible) is trivial, the method of proof is well worth the effort
since it shows how capacity enters in the calculation of the performance
of iterative coding systems. By turning this bound around, we
can find conditions under which iterative systems achieve capacity:
In particular it shows that the two component-wise
EXIT curves have to be matched perfectly. Indeed, all currently known
capacity achieving degree-distributions for the BEC can be derived
by starting with this perfect matching condition and working backwards.
Let us now show that, by using component-wise GEXIT functions, the perfect
matching condition holds in the general case. This might in the future serve 
as a starting point to find capacity-achieving degree distributions for
general BMS channels. We need one preliminary definition.

\bdefi[Interpolating Channel Families]
\label{def:interpolation}
Consider a degree distribution pair $(\ledge, \redge)$
and transmission over the BMS channel characterized by its
$L$-density $\Ldens{c}$. Let $\Ldens{a}_{-1}=\Delta_0$
and $\Ldens{a}_0=\Ldens{c}$ and set $\Ldens{a}_{\alpha}$,
$\alpha \in [-1, 0]$, to
$\Ldens{a}_{\alpha}=-\alpha \Ldens{a}_{-1} + (1+\alpha) \Ldens{a}_0$.
The {\em interpolating density evolution families}
$\{\Ldens{a}_{\alpha}\}_{\alpha=-1}^{\infty}$
and $\{\Ldens{b}_{\alpha}\}_{\alpha=0}^{\infty}$ are then defined as follows:
\begin{align*}
\Ldens{b}_{\alpha} & = \sum_{i} \redge_i \Ldens{a}_{\alpha-1}^{\boxast (i-1)},
\;\;\;\;\; \alpha \geq 0,\\
\Ldens{a}_{\alpha} & =
\sum_{i} \ledge_i \Ldens{c} \conv \Ldens{b}_{\alpha}^{\conv (i-1)},
\;\;\;\;\;\alpha \geq 0,
\end{align*}
where $\conv$ denotes the standard convolution of densities and
$\Ldens{a} \boxast \Ldens{b}$ denotes the density at the output of
a check node, assuming that the input densities are $\Ldens{a}$ and $\Ldens{b}$,
respectively.
\edefi
Discussion: First note that $\Ldens{a}_{\ell}$ ($\Ldens{b}_{\ell}$),
$\ell \in \naturals$,
represents the sequence of $L$-densities of density evolution
emitted by the variable (check) nodes in the $\ell$-th iteration.
By starting density evolution not only with $\Ldens{a}_{0}=\Ldens{c}$
but with all possible convex combinations of $\Delta_0$ and
$\Ldens{c}$, this discrete sequence of densities is completed to
form a continuous family of densities ordered by physical degradation.
The fact that the densities are ordered by physical degradation
can be seen as follows: note that the computation tree for $\Ldens{a}_{\alpha}$
can be constructed by taking
the standard computation tree of $\Ldens{a}_{\lceil \alpha \rceil}$
and independently erasing the observation associated to each variable leaf node with probability
$\lceil \alpha \rceil-\alpha$. It follows that we can convert the computation tree of
$\Ldens{a}_{\alpha}$ to that of $\Ldens{a}_{\alpha-1}$ by erasing all
observations at the leaf nodes and by independently erasing
each observation in the second (from the bottom) row of variable nodes
with probability $\lceil \alpha \rceil-\alpha$.
The same statement is true for $\Ldens{b}_{\alpha}$.
If $\lim_{\ell \rightarrow \infty} \entropy(\Ldens{a}_{\ell})=0$, i.e.,
if \BP\
decoding is successful in the limit of large blocklengths, then
the families are both complete.
\begin{example}[Density Evolution and Interpolation]
Consider transmission over the $\BSC(\epsilon = 0.07)$ using a
$(3, 6)$-regular ensemble. Fig.~\ref{fig:debsc} depicts
the density evolution process for this case.
\begin{figure}[htp]
\setlength{\unitlength}{0.5bp}%
\begin{center}
\begin{picture}(480,380)
\put(0,210)
{
\put(0,0){\includegraphics[scale=0.5]{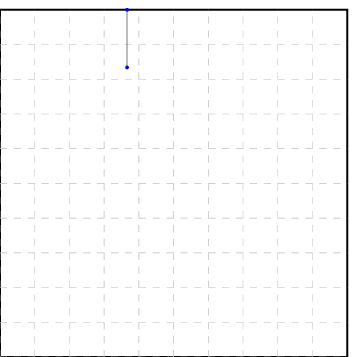}}
\put(120,50){\includegraphics[scale=0.5]{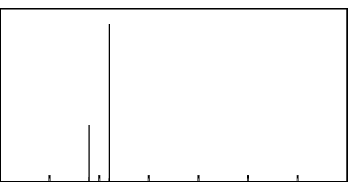}}
\put(0,120){\includegraphics[scale=0.5]{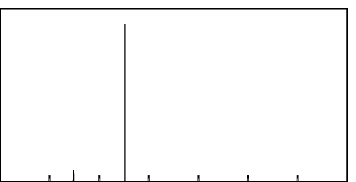}}
\put(50, 110){\makebox(0,0)[c]{\tiny $a_{0}$}}
\put(170, 40){\makebox(0,0)[c]{\tiny $b_{1}$}}
\put(50, -2){\makebox(0,0)[t]{\tiny $\entropy(\Ldens{a})$}}
\put(-2, 50){\makebox(0,0)[r]{\tiny \rotatebox{90}{$\entropy(\Ldens{b})$}}}
}
\put(260,210)
{
\put(0,0){\includegraphics[scale=0.5]{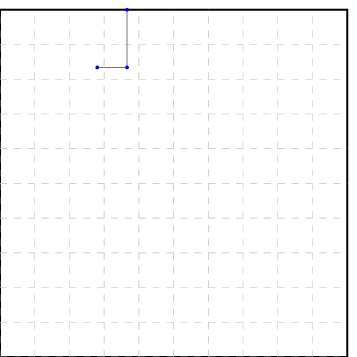}}
\put(120,50){\includegraphics[scale=0.5]{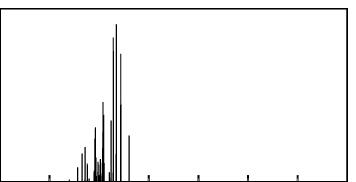}}
\put(0,120){\includegraphics[scale=0.5]{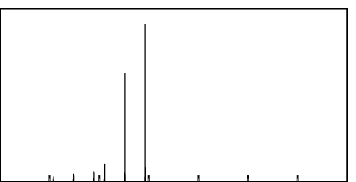}}
\put(50, 110){\makebox(0,0)[c]{\tiny $a_{1}$}}
\put(170, 40){\makebox(0,0)[c]{\tiny $b_{2}$}}
\put(50, -2){\makebox(0,0)[t]{\tiny $\entropy(\Ldens{a})$}}
\put(-2, 50){\makebox(0,0)[r]{\tiny \rotatebox{90}{$\entropy(\Ldens{b})$}}}
}
\put(0,0)
{
\put(0,0){\includegraphics[scale=0.5]{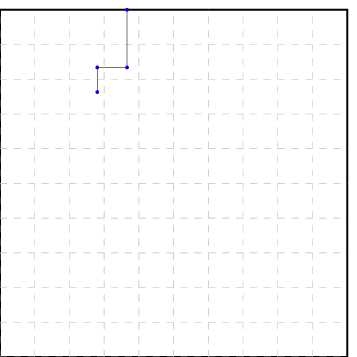}}
\put(120,50){\includegraphics[scale=0.5]{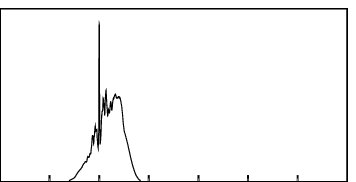}}
\put(0,120){\includegraphics[scale=0.5]{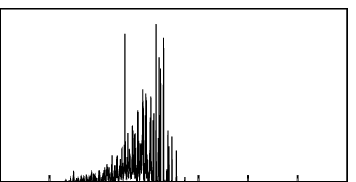}}
\put(50, 110){\makebox(0,0)[c]{\tiny $a_{2}$}}
\put(170, 40){\makebox(0,0)[c]{\tiny $b_{3}$}}
\put(50, -2){\makebox(0,0)[t]{\tiny $\entropy(\Ldens{a})$}}
\put(-2, 50){\makebox(0,0)[r]{\tiny \rotatebox{90}{$\entropy(\Ldens{b})$}}}
}
\put(260,0)
{
\put(0,0){\includegraphics[scale=0.5]{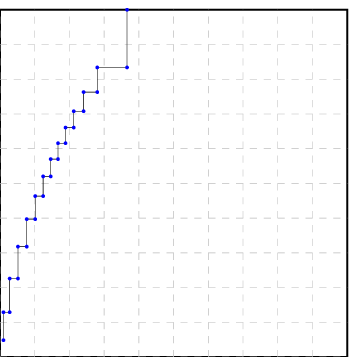}}
\put(120,50){\includegraphics[scale=0.5]{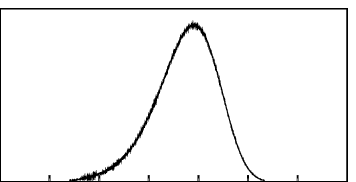}}
\put(0,120){\includegraphics[scale=0.5]{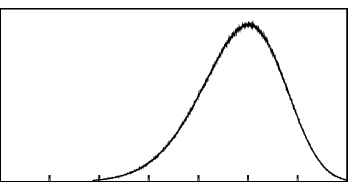}}
\put(50, 110){\makebox(0,0)[c]{\tiny $a_{12}$}}
\put(170, 40){\makebox(0,0)[c]{\tiny $b_{13}$}}
\put(50, -2){\makebox(0,0)[t]{\tiny $\entropy(\Ldens{a})$}}
\put(-2, 50){\makebox(0,0)[r]{\tiny \rotatebox{90}{$\entropy(\Ldens{b})$}}}
}
\end{picture}
\end{center}
\caption{\label{fig:debsc} Density evolution for $(3, 6)$-regular ensemble over $\BSC(0.07)$.}
\end{figure}
This process gives rise to the sequences of densities $\{\Ldens{a}_{\ell}\}_{\ell =0}^{\infty}$,
and $\{ \Ldens{b}_{\ell}\}_{\ell=1}^{\infty}$. Fig.~\ref{fig:interpolation} shows
the interpolation of these sequences for the choices $\alpha=1.0, 0.95, 0.9$ and $0.8$
and the complete such family.
\begin{figure}[htp]
\setlength{\unitlength}{0.6bp}%
\begin{center}
\begin{picture}(360,250)
\put(0, 150){
\put(0,0){\includegraphics[scale=0.6]{de25}}
\put(50, 102){\makebox(0,0)[b]{\tiny $\alpha=1.0$}}
\put(50, -2){\makebox(0,0)[t]{\tiny $\entropy(\Ldens{a})$}}
\put(-2, 50){\makebox(0,0)[r]{\tiny \rotatebox{90}{$\entropy(\Ldens{b})$}}}
\put(130,0){\includegraphics[scale=0.6]{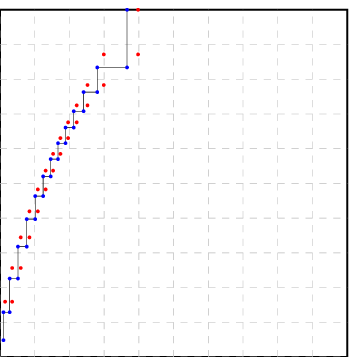}}
\put(180, 102){\makebox(0,0)[b]{\tiny $\alpha=0.95$}}
\put(180, -2){\makebox(0,0)[t]{\tiny $\entropy(\Ldens{a})$}}
\put(108, 50){\makebox(0,0)[r]{\tiny \rotatebox{90}{$\entropy(\Ldens{b})$}}}
\put(260,0){\includegraphics[scale=0.6]{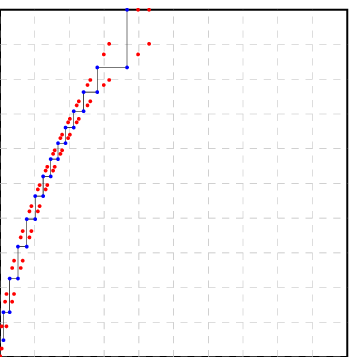}}
\put(310, 102){\makebox(0,0)[b]{\tiny $\alpha=0.9$}}
\put(310, -2){\makebox(0,0)[t]{\tiny $\entropy(\Ldens{a})$}}
\put(258, 50){\makebox(0,0)[r]{\tiny \rotatebox{90}{$\entropy(\Ldens{b})$}}}
}
\put(-390,0)
{
\put(390,0){\includegraphics[scale=0.6]{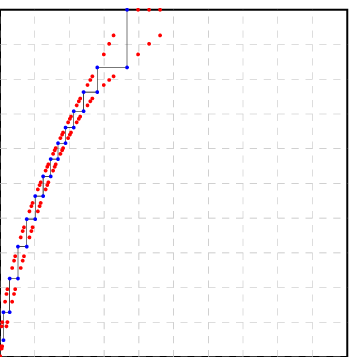}}
\put(440, 102){\makebox(0,0)[b]{\tiny $\alpha=0.8$}}
\put(440, -2){\makebox(0,0)[t]{\tiny $\entropy(\Ldens{a})$}}
\put(388, 50){\makebox(0,0)[r]{\tiny \rotatebox{90}{$\entropy(\Ldens{b})$}}}
\put(520,0){\includegraphics[scale=0.6]{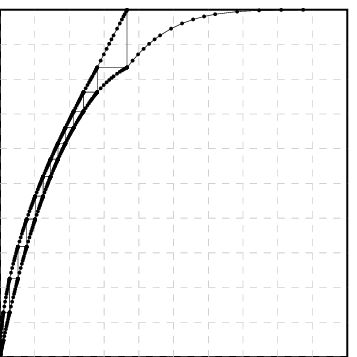}}
\put(570, -2){\makebox(0,0)[t]{\tiny $\entropy(\Ldens{a})$}}
\put(518, 50){\makebox(0,0)[r]{\tiny \rotatebox{90}{$\entropy(\Ldens{b})$}}}
}
\end{picture}
\end{center}
\caption{\label{fig:interpolation} Interpolation of densities.}
\end{figure}
\end{example}

\blemma
Consider a degree distribution pair $(\ledge, \redge)$
and transmission over an BMS channel characterized by its
$L$-density $\Ldens{c}$ so that density evolution converges to
$\Delta_{\infty}$.
Let $\{\Ldens{a}_{\alpha}\}_{\alpha=-1}^{\infty}$
and $\{\Ldens{b}_{\alpha}\}_{\alpha=0}^{\infty}$ denote the interpolated
families as defined in Definition \ref{def:interpolation}.

Then the two GEXIT curves parameterized by
\begin{align*}
\{ \entropy(\Ldens{a}_{\alpha}),
\gentropy(\Ldens{a}_{\alpha}, \Ldens{b}_{\alpha+1}) \}, \tag*{GEXIT of check nodes} \\
\{ \entropy(\Ldens{a}_{\alpha}),
\gentropy(\Ldens{a}_{\alpha}, \Ldens{b}_{\alpha}) \}, \tag*{inverse of dual GEXIT of variable nodes}
\end{align*}
do not cross and faithfully represent density evolution.
Further, the area under the ``check-node'' GEXIT function
is equal to $1-\int \!\redge$ and the area to the left of the
``inverse dual variable node'' GEXIT function is equal to $\entropy(\Ldens{c}) \int \!\ledge$.
It follows that $r(\ledge, \redge) \leq 1-\entropy(\Ldens{c})$, i.e.,
the design rate can not exceed the Shannon limit.
\elemma
\begin{proof}
First note that $\{ \entropy(\Ldens{a}_{\alpha}),
\gentropy(\Ldens{a}_{\alpha}, \Ldens{b}_{\alpha+1}) \}$
is the standard GEXIT curve representing the action
of the check nodes: $\Ldens{a}_{\alpha}$ corresponds to
the density of the messages {\em entering} the check nodes and
$\Ldens{b}_{\alpha+1}$ represents the density of the corresponding
output messages.
On the other hand,
$\{ \entropy(\Ldens{a}_{\alpha}),
\gentropy(\Ldens{a}_{\alpha}, \Ldens{b}_{\alpha}) \}$
is the inverse of the dual GEXIT curve
corresponding to the action at the variable nodes:
now the input density to the check nodes is
$\Ldens{b}_{\alpha}$ and $\Ldens{a}_{\alpha}$ denotes the
corresponding output density.

The fact that the two curves do not cross can be seen as follows.
Fix an entropy value. This entropy value corresponds to a
density $\Ldens{a}_{\alpha}$ for a unique value of $\alpha$.
The fact that
$G(\Ldens{a}_{\alpha}, \Ldens{b}_{\alpha}) \geq
G(\Ldens{a}_{\alpha}, \Ldens{b}_{\alpha+1})$ now follows from
the fact that $\Ldens{b}_{\alpha+1} \prec \Ldens{b}_{\alpha}$ and
that for any symmetric $\Ldens{a}_{\alpha}$ this relationship
stays preserved by applying the GEXIT functional according to
Corollary \ref{lemma:orderingviaphysicaldegradationexit}.

The statements regarding the areas of the two curves
follow in a straightforward manner from the GAT and Lemma \ref{lem:dualgexit}.
The bound on the achievable rate follows in the same manner as for
the BEC: the total area of the GEXIT box equals one and the two curves do not
overlap and have areas $1-\int \redge$ and $\entropy(\Ldens{c})$.
It follows that
$1-\int \!\redge + \entropy(\Ldens{c}) \int \!\ledge \leq 1$,
which is equivalent to the claim $r(\ledge, \redge) \leq 1-\entropy(\Ldens{c})$.
\end{proof}

\begin{figure}[hbt]
\centering
\setlength{\unitlength}{1.0bp}
\begin{picture}(230,230)
\put(0,130){
\put(0,0){\includegraphics[scale=1.0]{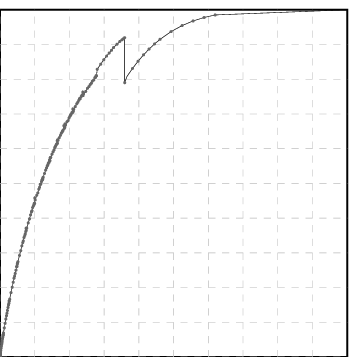}}
\put(50, -2){\makebox(0,0)[t]{\small $\entropy(\Ldens{a}_{\alpha})$}}
\put(102, 50){\makebox(0,0)[l]{\small \rotatebox{90}{$\gentropy(\Ldens{a}_{\alpha}, \Ldens{b}_{\alpha+1})$}}}
\put(50, 40){\makebox(0,0)[t]{\small GEXIT: checks}}
\put(50, 30){\makebox(0,0)[t]{\small $\text{area}=\frac56$}}
\put(50, 10){\makebox(0,0)[c]{$\Ldens{b}_{\alpha+1} = \sum_{i} \redge_i \Ldens{a}_{\alpha}^{\boxast (i-1)} $}}
}
\put(130, 130)
{
\put(0,0){\includegraphics[scale=1.0]{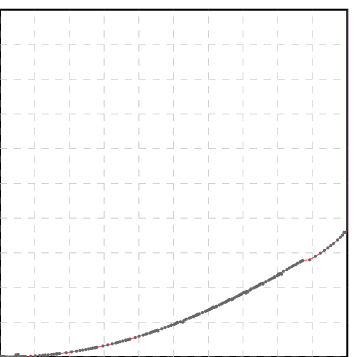}}
\put(50, -2){\makebox(0,0)[t]{\small {$\gentropy(\Ldens{a}_{\alpha}, \Ldens{b}_{\alpha})$}}}
\put(-2, 50){\makebox(0,0)[r]{\small  \rotatebox{90}{$\entropy(\Ldens{a}_{\alpha})$}}}
\put(50, 70){\makebox(0,0)[t]{\small dual GEXIT: variables}}
\put(50, 60){\makebox(0,0)[t]{\small $\text{area}=\frac13 h(0.07)$}}
\put(102, 36.6){\makebox(0,0)[l]{\small \rotatebox{90}{$h(0.07) \approx 0.366$}}}
\put(50, 40){\makebox(0,0)[c]{$\Ldens{a}_{\alpha} = \Ldens{c} \conv \sum_{i} \ledge_i \Ldens{b}_{\alpha}^{\conv (i-1)} $}}
}
\put(0,0)
{
\put(0,0){\includegraphics[scale=1.0]{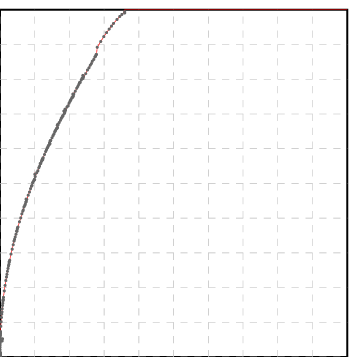}}
\put(50, -2){\makebox(0,0)[t]{\small $\entropy(\Ldens{a}_{\alpha})$}}
\put(-2, 50){\makebox(0,0)[r]{\small \rotatebox{90}{$\gentropy(\Ldens{a}_{\alpha}, \Ldens{b}_{\alpha})$}}}
\put(50, 30){\makebox(0,0)[t]{\small inverse of dual GEXIT:}}
\put(50, 20){\makebox(0,0)[t]{\small variables}}
\put(36.6, 102){\makebox(0,0)[b]{\small $h(0.07) \approx 0.366$}}
}
\put(130,0)
{
\put(0,0){\includegraphics[scale=1.0]{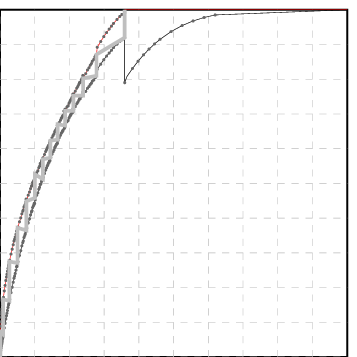}}
\put(50, -2){\makebox(0,0)[t]{\small $\entropy(\Ldens{a}_{\alpha})$}}
\put(-2, 50){\makebox(0,0)[r]{\small \rotatebox{90}{$\gentropy(\Ldens{a}_{\alpha}, \Ldens{b}_{\alpha})$}}}
\put(102, 50){\makebox(0,0)[l]{\small \rotatebox{90}{$\gentropy(\Ldens{a}_{\alpha}, \Ldens{b}_{\alpha+1})$}}}
\put(36.6, 102){\makebox(0,0)[b]{\small $h(0.07) \approx 0.366$}}
}
\end{picture}
\caption{
\label{fig:componentgexit}
Faithful representation of density evolution by two non-overlapping component-wise
GEXIT functions which represent the ``actions'' of the 
check nodes and variable nodes,
respectively. The area between the two curves is proportional to the additive
gap to capacity.
}
\end{figure}

We see that the matching condition still holds for general channels.
There are a few important differences between the general case and the simple
case of transmission over the BEC. For the BEC, the intermediate densities
are always the BEC densities independent of the degree distribution.
This of course enormously simplifies the task. Further, for the BEC, given
the two EXIT curves, the progress of density evolution is simply given
by a staircase function bounded by the two EXIT curves. For the general case,
this staircase function still has vertical pieces but the ``horizontal''
pieces have in general a non-vanishing slope. This is true since the $y$-axis for
the ``check node'' step measures
$\gentropy(\Ldens{a}_{\alpha}, \Ldens{b}_{\alpha+1})$, but
in the subsequent ``inverse variable node'' step
it measures
$\gentropy(\Ldens{a}_{\alpha+1}, \Ldens{b}_{\alpha+1})$.
Therefore, one should think of two sets of labels on the $y$-axis,
one measuring $\gentropy(\Ldens{a}_{\alpha}, \Ldens{b}_{\alpha+1})$,
and the second one measuring $\gentropy(\Ldens{a}_{\alpha+1}, \Ldens{b}_{\alpha+1})$. The ``horizontal'' step then consists of first
switching from the first $y$-axis to the second, so that the labels
correspond to the same density $\Ldens{b}$ and then drawing a horizontal
line until it crosses the ``inverse variable node'' GEXIT curve.
The ``vertical'' step stays as before, i.e., it really corresponds to
drawing a vertical line. All this is certainly best clarified by
a simple example.
\bex[$(3, 6)$ Ensemble and Transmission over $\BSC$]
Consider the $(3, 6)$-regular ensemble and transmission over the $\BSC(0.07)$.
The corresponding illustrations are shown in Fig.~\ref{fig:componentgexit}.
The top-left figure shows the standard GEXIT curve for the check node side.
The top-right figure shows the dual GEXIT curve corresponding to the
variable node side. In order to use these two curves in the same figure,
it is convenient to consider the inverse function for the variable
node side. This is shown in the bottom-left figure. In the bottom-right
figure both curves are shown together with the ``staircase'' like function
which represents density evolution. As we see, the two curves to not overlap
and have both the correct areas.
\eex

As remarked earlier, one potential use of the matching condition
is to find capacity approaching degree distribution pairs. Let us
quickly outline a further such potential application. Assuming that
we have found a sequence of capacity-achieving degree distributions,
how does the number of required iterations scale as we approach capacity.
It has been conjectured that the the number of required iterations
scales like $1/\delta$, where $\delta$ is the gap to capacity.
This conjecture is based on the geometric picture which the
matching condition implies. To make things simple, imagine
the two GEXIT curves as two parallel lines, lets say both
at a 45 degree angle, a certain distance
apart, and think of density evolution as a staircase function.
From the previous results, the area between the lines is proportional
to $\delta$. Therefore, if we half $\delta$ the distance between
the lines has to be halved and one would expect that we need
twice as many steps. Obviously, the above discussion was 
based on a number of simplifying assumptions. It remains to 
be seen if this conjecture can be proven rigorously.  
%
%******************************************************************
%
\section{Conclusion}
\label{sec:conclusion}

Since the introduction of $\exit$ functions for the analysis
iterative coding systems \cite{teB99a,teB99b,tBr00,teB00,teB01}, 
researchers have tried to substantiate
theoretically the empirical area rules that these seemed to satisfy.
In this paper we showed how to {\em prove} these rules in a 
very general setting. The price to pay was to replace $\exit$
functions by $\gexit$ functions. Fortunately, 
$\gexit$ functions are as simple to compute as ordinary $\exit$ functions
and share in general many of their properties.

We also presented several applications of this new tool. 
Most notably: $(i)$ It allows one to prove an upper bound on
the $\MAP$ threshold which is conjectured to coincide with the
actual threshold for several classes of ensembles (e.g. regular ones). 
$(ii)$ Via extended $\BP$ $\gexit$ curves, it provides some constraints
on the relation between $\BP$ and $\MAP$ decoding. These constraints
lead naturally to the Maxwell construction which provides the precise
connection between the two. In particular we found that the $\BP$ 
soft bit estimates are asymptotically exact for a noise range 
{\em above threshold}.
$(iii)$ It implies a matching constraint on component codes of
capacity-achieving systems.

These results open many research directions. It may be worth to list a 
few of them. 

Prove existence, uniqueness and 
regularity properties of  asymptotic $\MAP$ and extended $\BP$ 
$\gexit$ curves. In particular, we expect that the last one is a
smooth single valued function of the entropy of the fixed point density.
The iterative procedure which we presented in Section \ref{sec:HowToEBP} 
only {\em proves} that
for each message entropy there is at least one fixed point of density
evolution. But, empirically, when running this algorithm,
we found that indeed there seems to be a unique such fixed point
and that all these fixed points seem to form a smooth manifold. 
Further,
we proved several  partial results in 
this direction (for instance existence for $\EBP$ curves, uniqueness
for $\MAP$ curves, etc). However, the general question remains open.

Prove that the Maxwell construction indeed provides the correct connection
between $\MAP$ and $\BP$ $\gexit$ curves. As particular case
(which may well be simpler than the general statement),
prove the upper bound (\ref{theo:UBMAP}) is 
indeed tight for some selected ensembles, e.g. for regular ones.

Use the interpolation construction of Section~\ref{sec:Matching}
to prove a lower bound on the number of message passing iterations as a 
function of the gap to capacity.
%
%******************************************************************
%
\section*{Acknowledgment}
The authors would like to thank  Nicolas Macris   
and  Olivier L\'ev\^eque for useful discussions.

A.M. has been partially supported by the EU integrated project EVERGROW.
%
%******************************************************************
%
\appendices

\section{GEXIT Kernel over Gaussian Channels} 
\label{app:gaussiankernel} \label{app:kernellimits} \label{sec:mmse}

This appendix contains a few useful results concerning the 
$\gexit$ kernel for Gaussian channels.

\blemma[$\gexit$ Kernel, $L$-Domain -- $\{\text{BAWGNC}(\ent)\}$]\label{lemma:equivkernel}
Consider the family
$\{\Ldens{c}_{\BAWGNCsmall(\ent)}\}$ of BAWGN channels, where $\ent$ denotes 
the channel entropy. The channel model 
is therefore $Y=X+N$, where $X$ takes values $x\in{\cal{X}}=\{-1,+1\}$ and
$N$ is Gaussian with zero mean and variance $\sigma^2$.
Then the following represent {\em equivalent} kernels:
\begin{align*}
\gexitkl {\Ldens{c}_{\text{\tiny BAWGNC}(\ent)}} z 
&= \displaystyle 
\renewcommand{\cp}{\sigma}  
\frac{ \text{e}^{-z}\int_{-\infty}^{+\infty}\scriptstyle
\frac{  \text{e}^{-\frac{(w \cp^2-2)^2}{8 \cp^2}} %\text{e}^{-\frac{(w-\cp)^2}{4 \cp}}
    }{(\cosh(\frac{w-z}{2}))^2}\text{d}w    } {\int_{-\infty}^{+\infty}\scriptstyle
\frac{  \text{e}^{-\frac{(w \cp^2-2)^2}{8 \cp^2}} 
  }{(\cosh(\frac{w}{2}))^2} \text{d}w}, \tag*{(i)}\\
\gexitkl{' {\Ldens{c}_{\text{\tiny BAWGNC}(\ent)}}} z 
&=  \renewcommand{\cp}{\sigma} 
 \frac{1-\expectation[\expectation[X|Y,\Phi=z]^2]}{1-\expectation[\expectation[X|Y]^2]}, 
\tag*{(ii)}\\
\gexitkl{'' {\Ldens{c}_{\text{\tiny BAWGNC}(\ent)}}} z 
& = \renewcommand{\cp}{\sigma}  
\frac{1-\expectation[X|Y,\Phi=z]}{1-\expectation[X|Y]}. \tag*{(iii)}
\renewcommand{\cp}{\epsilon}
\end{align*}
Hereby, $\Phi$ denotes a further observation of $X$ which is conditionally
independent of $Y$, which is the result of passing $X$ through
a symmetric channel, and which is assumed to
be in log-likelihood form (if we use coding,
$\Phi$ represents the extrinsic estimate 
of $X$ in the $L$-domain).
\elemma

Discussion: This lemma provides several equivalent representations 
of the kernel for the BAWGN channel. The expression (ii) shows the 
relationship between conditional entropy and mean-square error 
(MSE) estimator.  To see this, observe first that the denominator is 
a ($z$ independent) scaling factor depending on our parameterization of
the channel through its entropy $\ent$. Second, observe that 
the numerator $1-\expectation[\expectation[X|Y,\Phi=z]^2]=
\expectation[\expectation[X^2|Y,\Phi=z]-\expectation[X|Y,\Phi=z]^2]$ 
is the mean-square error estimator (which in this framework 
includes the decoding estimate $z$).  
This elegant relationship 
which connects a fundamental information theoretic quantity 
(the conditional entropy, or, equivalently, the mutual information) 
to a measure widely-used in signal processing was first observed
by Guo, Shamai and Verd{\'u} in 
\cite{GSV04,GSV05}. In the above lemma, the channel 
inputs are binary. In Lemma \ref{lemma:gexitawgnEx} we give an alternative way 
of deriving $\gexitkl{{\Ldens{c}_{\text{\tiny BAWGNC}(\ent)}}} z$ 
in the more general context of non-binary channel inputs.

The form (iii) provides a further simplification. This expression, in 
which the numerator shows the magnetization was first stated  
in \cite{Mac05} using the Nishimori identity (in the context of coding,  
this identity was first discussed in \cite{Mon04}).  
 
Before proving Lemma \ref{lemma:equivkernel}, 
let us recall the following well-known fact
which will be used several times in the following:
Consider a BMS channel $p_{Y|X}(y|x)$ and $f(y)$, a measurable function. 
If $f(y)$ is even, then 
$\expectation_Y [f(Y)]=\expectation_{Y|X=1} [f(Y)].$ 
\bproof 
Under the all-one assumption, the channel 
density is 
$\Ldens{c}(w)\defas {\Ldens{c}_{\text{\tiny BAWGNC}(\ent)}}(w) =\frac{\sigma}{\sqrt{8\pi}}\text{e}^{-\frac{(w\sigma^2-2)^2}{8\sigma^2}}$. \\
(i) The kernel as stated in Lemma \ref{lem:gexitlinear} is expressed in
terms of the derivative of $\Ldens{c}(w)$ 
with respect to the channel parameter. 
To get a more pleasing analytic expression we use the fact that
for the Gaussian case we can express this derivative via the 
%Choose the parameterization $\cp\defas2/\sigma^2$ and use
identity
$\frac{\partial\Ldens{c}(w)}{\partial \cp}=-\frac{\partial\Ldens{c}(w)}{\partial w}+
\frac{\partial^2\Ldens{c}(w)}{\partial w^2}$.
Now, use the parameterization 
$\cp\defas2/\sigma^2$.
%For practical computations, 
% the derivative is first taken 
%with respect to a parameter $\cp$. This expression can further be normalized 
% (by the derivative of 
%$\ent=\int_{-\infty}^{+\infty}\Ldens{c}(w) \logtwo(1+\text{e}^{-w})\text{d}w$ 
%with respect to $\cp$) 
%in order to get a derivative of $\Ldens{c}(w)$ with respect to $\ent$. 
%
%A convenient parameterization for this case is $\cp\defas2/\sigma^2$.  
%After some steps of calculus, and using the identity
%$\frac{\partial\Ldens{c}(w)}{\partial \cp}=-\frac{\partial\Ldens{c}(w)}{\partial w}+\frac{\partial^2\Ldens{c}(w)}{\partial w^2}$, we get
Then using twice integration by parts (as in \cite{Mac05}), we get 
\begin{align*}
 \gexitkl {\Ldens{c}_{\text{\tiny BAWGNC}(\ent)}} z \frac{\text{d}\epsilon}{\text{d}\ent}
&= \int_{-\infty}^{+\infty}\frac{\partial\Ldens{c}(w)}{\partial \cp} \log(1+\text{e}^{-w-z})\text{d}w\\
&=\int_{-\infty}^{+\infty}\frac{\partial\Ldens{c}(w)}{\partial w}\frac{\text{e}^{-w-z}}{1+\text{e}^{-w-z}}\text{d}w\\
&~~~~ -\int_{-\infty}^{+\infty}\Ldens{c}(w)\frac{\text{e}^{-w-z}}{1+\text{e}^{-w-z}}\text{d}w\\
& = \int_{-\infty}^{+\infty}\Ldens{c}(w)\frac{-1}{ (1+\text{e}^{w+z})^2}\text{d}w\\
& =  \frac{- \text{e}^{-z}}{4}\int_{-\infty}^{+\infty}\frac{\Ldens{c}(-w)}{(\cosh(\frac{w+z}{2}))^2}\text{d}w.
\end{align*}
The computation of $\frac{\text{d}\epsilon}{\text{d}\ent}$ is exactly the 
same if we set $z=0$.
Therefore, 
\begin{align*}
\gexitkl {\Ldens{c}_{\text{\tiny BAWGNC}(\ent)}} z 
& \defas \frac{ \text{e}^{-z}\int_{-\infty}^{+\infty}\frac{  \text{e}^{-\frac{(w-\cp)^2}{4 \cp}}    }{(\cosh(\frac{w-z}{2}))^2}\text{d}w    } {\int_{-\infty}^{+\infty}\frac{  \text{e}^{-\frac{(w-\cp)^2}{4 \cp}}    }{(\cosh(\frac{w}{2}))^2} \text{d}w  }.
\end{align*}

(ii) First, we claim that the previous expression can be written as
\begin{align*}
\gexitkl {\Ldens{c}_{\text{\tiny BAWGNC}(\ent)}} z 
& = \text{e}^{-z}\frac{1-\expectation[\expectation[X|Y,\Phi=-z]^2]}{1-\expectation[\expectation[X|Y]^2]}.
\end{align*}
To see this, observe that 
\begin{align*}
w+z & \overset{(a)}{=} \log \frac{p_{\frac{2Y}{\sigma^2}|X}(w|+1)}{p_{\frac{2Y}{\sigma^2}|X}(w|-1)} + \log \frac{p_{\Phi|X}(z|+1)}{p_{\Phi|X}(z|-1)}\\
 & \overset{(b)}{=} \log \frac{p_{\frac{2Y}{\sigma^2},\Phi|X}(w,z|+1)}{p_{\frac{2Y}{\sigma^2},\Phi|X}(w,z|-1)} \\
 & \overset{(c)}{=} \log \frac{ p_{X|\frac{2Y}{\sigma^2},\Phi}(+1|w,z) }{ p_{X|\frac{2Y}{\sigma^2},\Phi}(-1|w,z) }, 
\end{align*}
where $(a)$ comes from the definition of $w$ and $z$ in Lemma \ref{lemma:equivkernel}, 
$(b)$ from the independence of $Y$ and $\Phi$ when $X$ is given, and where $(c)$ is the
Bayes rule using $p_X(+1)=p_X(-1)=\frac12$. Therefore, 
\begin{align*}
\tanh(\frac{w+z}{2})  & = \frac{1-\text{e}^{-w-z}}{1+\text{e}^{-w-z}} \\
 & = \frac{p_{X|\frac{2Y}{\sigma^2},\Phi}(+1|w,z) - p_{X|\frac{2Y}{\sigma^2},\Phi}(-1|w,z)  }{p_{X|\frac{2Y}{\sigma^2},\Phi}(+1|w,z) + p_{X|\frac{2Y}{\sigma^2},\Phi}(-1|w,z) }\\
& = \expectation[X|w,z].
\end{align*}
This quantity (which is often called ``soft bit'' as in  \cite{HOP96}) is
 a bit estimate in the $D$-domain and the relationship 
$\expectation[X|w,z]=\tanh(\frac{w+z}{2})$ is in fact well-known. 
%Finally, we have used $(\tanh(\frac{w+z}{2}))^2=1-\frac{1}{(\cosh(\frac{w+z}{2}))^2}$.  
Therefore, since $1-(\tanh(\frac{w+z}{2}))^2=\frac{1}{(\cosh(\frac{w+z}{2}))^2}$,
\begin{align*}
\gexitkl {\Ldens{c}_{\text{\tiny BAWGNC}(\ent)}} z 
& = \text{e}^{-z}\frac{1-\int_{-\infty}^{\infty}\Ldens{c}(w)(\tanh(\frac{w+z}{2}))^2\text{d}w}{1-\int_{-\infty}^{\infty}\Ldens{c}(w)(\tanh(\frac{w}{2}))^2\text{d}w}\\
&  \text{e}^{-z}\frac{1-\expectation_{Y|X=1}[(\tanh(\frac{Y+z}{2}))^2]}{1-\expectation_{Y|X=1}[(\tanh(\frac{Y}{2}))^2]}, 
\end{align*}
and the claim follows since, as discussed above, we can
drop in the last expression the conditioning on $X=1$.

Second, as discussed in Example \ref{ex:nonuniquekernel},
the kernel is in general not unique in the $L$-domain 
and we can use this degree of freedom to get alternative kernels.   
Denote $f(z)\defas\frac{1-\expectation[\expectation[X_1|Y_1,-z]^2]}{1-\expectation[\expectation[X_1|Y_1]^2]}$ and observe that 
$\gexitkl {\Ldens{c}_{\text{\tiny BAWGNC}(\ent)}} z=\exp(-z) f(z)$ with this notation. Then, for any symmetric density 
$\Ldens{a}(z)$, the function $\gexitkl{' {\Ldens{c}_{\text{\tiny BAWGNC}(\ent)}}} z \defas f(-z)$ is also 
a valid kernel for the $L$-domain since $\int_{-\infty}^{+\infty}\Ldens{a}(z) \text{e}^{-z}f(z)\text{d}z= \int_{-\infty}^{+\infty}\Ldens{a}(z) f(-z)\text{d}z$. Therefore, an alternative kernel is 
\begin{align*}
\gexitkl{' {\Ldens{c}_{\text{\tiny BAWGNC}(\ent)}}} z 
&=  \renewcommand{\cp}{\sigma} 
 \frac{1-\expectation[\expectation[X|Y,z]^2]}{1-\expectation[\expectation[X|Y]^2]}
 =\frac{ \int_{-\infty}^{+\infty}\scriptstyle
\frac{ % \text{e}^{-\frac{(w-\cp)^2}{4 \cp}}  
\text{e}^{-\frac{(w \cp^2-2)^2}{8 \cp^2}} 
 }{(\cosh(\frac{w+z}{2}))^2}\text{d}w    } {\int_{-\infty}^{+\infty}\scriptstyle
\frac{  
%\text{e}^{-\frac{(w-\cp)^2}{4 \cp}}   
\text{e}^{-\frac{(w \cp^2-2)^2}{8 \cp^2}}
 }{(\cosh(\frac{w}{2}))^2} \text{d}w  }. 
\end{align*}

(iii) For any symmetric random variable $L$, a straightforward exercise shows that 
 $\expectation[\tanh(L/2)]=\expectation[(\tanh(L/2))^2]$. See, e.g, \cite{Mac05,Mon04}. 
 Applied to the symmetric random variable $L\defas\log \frac{p(Y|+1)}{p(Y|-1)}=\frac{2}{\sigma^2}Y$ 
 under the all-one assumption, this gives us
$\expectation[\expectation[X|Y]^2]=\expectation[\tanh(\frac{L}{2})^2]=\expectation[\tanh(\frac{L}{2})]=\expectation[X|Y]$. 
Therefore the denominator can be easily written as 
 $\frac{1}{1-\expectation[\expectation[X|Y]^2]}=\frac{1}{1-\expectation[X|Y]}$. 
We can not use directly this argument for the term $\expectation[\expectation[X|Y,z]^2]=\expectation[\tanh(\frac{Y}{\sigma^2}+\frac{z}{2})^2]$ at the numerator (the random variable $\frac{2}{\sigma^2}Y+z$ being not symmetric).  
 However, we can  
look 
for an equivalent kernel.
 This is easily done by observing that the values $z$ are provided by the symmetric random 
variable $\Phi_{}$.  The sum of two symmetric random variables is again symmetric, 
therefore $\frac{2}{\sigma^2}Y+\Phi_{}$ is symmetric. See, e.g., \cite{RiU05}. 
We can then use the fact that $\expectation[\tanh(L/2)]=\expectation[(\tanh(L/2))^2]$ 
with  $L\defas \frac{2}{\sigma^2}Y+\Phi_{}$
to write $\expectation_{Y,\Phi_{}}[\expectation_{X}[X|Y,\Phi_{}]^2]=\expectation_{Y,\Phi_{}}[\expectation_{X}[X|Y,\Phi_{}]]$. Therefore, 
\begin{align*}
\gexitkl{'' {\Ldens{c}_{\text{\tiny BAWGNC}(\ent)}}} z 
& = \renewcommand{\cp}{\sigma}  
\frac{1-\expectation[X|Y,z]}{1-\expectation[X|Y]} 
 =
 \frac{ \int_{-\infty}^{+\infty}
\scriptstyle
\frac{%2 
% \text{e}^{-\frac{(w-\cp)^2}{4 \cp}}
\text{e}^{-\frac{(w \cp^2-2)^2}{8 \cp^2}}
 }{1+\text{e}^{\scriptscriptstyle w+z}}    \text{d}w      }{\int_{-\infty}^{+\infty}\scriptstyle
\frac{%2  
\text{e}^{-\frac{(w \cp^2-2)^2}{8 \cp^2}}
}{1+\text{e}^{w}}    \text{d}w} %\tag*{(vi)}
\renewcommand{\cp}{\epsilon}  
\end{align*}
is an equivalent kernel (but pointwise different from 
$\gexitkl{{\Ldens{c}_{\text{\tiny BAWGNC}(\ent)}}} z$ 
and $\gexitkl{' {\Ldens{c}_{\text{\tiny BAWGNC}(\ent)}}} z$). The last equality comes from the fact that $1-\expectation[X|Y,z]=1-\expectation_{Y}[\tanh(\frac{Y+z}2)]=\expectation_{Y}[\frac{2}{1+\text{e}^{Y+z}}]$. %Therefore, 
\eproof

GEXIT and EXIT curves are in general very similar. Next lemma illuminates this fact: it shows 
that, in the limit of small SNR, the kernel for the BAWGNC behaves similarly to the kernel 
for the BSC discussed in Example \ref{ex:lbsckernel}.
\blemma[Limiting Behavior of GEXIT Kernel] Consider the family
$\{\Ldens{c}_{\BAWGNCsmall(\ent)}\}$ of BAWGN channels, where $\ent$ denotes 
the channel entropy: The additive noise $N$ in the model $Y=X+N$ is Gaussian 
with zero-mean and variance $\sigma^2$.  
 Then 
\begin{align*}
 \lim_{\sigma\to\infty}\gexitkabsd {\Ldens{c}_{\text{\tiny BAWGNC}(\ent)}} s & = 1-s^2,\tag*{(i)}\\
 \lim_{\sigma\to0}\gexitkabsd {\Ldens{c}_{\text{\tiny BAWGNC}(\ent)}} s & = 1.\tag*{(ii)}
\end{align*}
In the $|D|$-domain, the kernels are ordered between those two extremal functions.
\elemma
\bproof
First recall the transform formula (\ref{equ:gexitkernelconversion})
and $2\tanh^{-1}(s)=\log\frac{1+s}{1-s}$.
(i) With  expression (iii) of Lemma \ref{lemma:equivkernel} 
we have 
$\gexitkl{\Ldens{c}}{2\tanh^{-1}(s)}=\frac{1-\int_{-\infty}^{+\infty}\Ldens{c}(l)\tanh(l/2+\tanh^{-1}(s)) \text{d} l }{1-\int_{-\infty}^{+\infty}\Ldens{c}(l)\tanh(l/2) \text{d} l }$. Let us restrict ourself to the study of the term $I_\sigma(s)\defas\int_{-\infty}^{+\infty}\Ldens{c}(l)\tanh(l/2+\tanh^{-1}(s)) \text{d} l$. When $\sigma^2\to\infty$, then the distribution of the channel inputs (more exactly of the LLR's in the L-domain) $\Ldens{c}(l) =\frac{\sigma}{2\sqrt{2\pi}}\exp(-\frac{\sigma^2(l-2/\sigma^2)^2}{8})$  becomes a Dirac centered in 0 (since its variance $4/\sigma^2\to0$). For any function continuous in 0, e.g., for the function $k_s:l\mapsto \tanh(l/2+\tanh^{-1}(s)) $, one can indeed replace, without committing much error when $\sigma^2\to\infty$, the integral
 $\int_{-\infty}^{+\infty}\Ldens{c}(l) k_s(l) \text{d}l$ by  
$
\int_{-\infty}^{+\infty}\Ldens{c}(l) k_s(0) \text{d}l. 
$  
See, e.g., \cite{Zem65} for further details.  Therefore 
$$
I_\sigma(z)\underset{\sigma\to\infty}{\longrightarrow}\tanh(0/2+\tanh^{-1}(s))=s.
$$
Using (\ref{equ:gexitkernelconversion}), 
we finally get $$\gexitkabsd{\Ldens{c}}{s}=\frac{1-s}{2}\frac{1+s}{1}+\frac{1+s}{2}\frac{1-s}{1}=1-s^2.$$
(ii) The case $\sigma \to0$ corresponds to the full knowledge of 
the channel input. The kernel in the $|D|$-domain converges point-wise 
to 1. As used for density evolution, see \cite{RiU05}, in this case   $\Ldens{c}(l)$ becomes a Dirac 
at $\infty$ a a similar argument as for (i) can be applied. 

For $\cp\in(0,1)$, the kernels in the $|D|-$domain are ordered because of 
Lemma \ref{lemma:orderingviaphysicaldegradationexit}.
\eproof

As discussed before Lemma \ref{lemma:equivkernel}, 
the pleasing relationship presented 
in \cite{Guo04,GSV05} or \cite{Mac05} emerges for the BAWGNC. 
So far we have restricted ourself 
to the case of binary inputs. But the non-binary case 
as discussed in \cite{Guo04,GSV05,GSV05i} is not much harder. This 
is presented in Lemma \ref{lemma:gexitawgnEx} using our framework.

\blemma[AWGN($\ent$)] \label{lemma:gexitawgnEx}
Consider a length $\n$ code, call it $\graph$.
Assume transmission takes place over a family $\{AWGNC(\ent_i)\}_{i\in[\n]}$ where 
there is a global parameter $\cp$ such that $\ent_i(\cp)=\ent(\cp)$ is the entropy associated to the $i^\text{th}$ 
channel for all $i\in[\n]$. Let this  
parameter be 
$\cp=-2\snr\defas-\frac{2}{\sigma^2}$. Then 
$$
\gexitfi{}(\graph,\cp)= \expectation \left[ \expectation[X_i^2|Y]-\expectation[X_i|Y]^2   \right].
$$
In words, the derivative of the conditional entropy with respect to the particular paramater $\cp$ 
is equal to the Mean-Square Error (MSE) estimator.
\elemma

\begin{proof} We will 
prove the result in general settings when the input alphabet ${\cal{X}}$ 
can be any subset of $\R$. Temporarily, let us denote $\tilde{Y}=X+\tilde{N}$ 
our running Gaussian channel model. $\tilde{N}$ is the additive white Gaussian 
noise with zero-mean and variance $\sigma^2$. Now  
let us normalize this model by $\sigma^2$  to
get the equivalent model $Y=\sqrt{\snr}X+N$ where $\snr=\frac{1}{\sigma^2}$ 
and $N$ is an   additive white Gaussian noise with zero-mean and unit-variance.
In order to be a sufficient statistics, 
the extrinsic MAP estimate $\phi^{}_i=\phi^{}_i(y_{\sim i})$ can no longer 
 be a log-likelihood ratio but, in general, a function of $x_i$, i.e., $\phi^{}_i:x\mapsto\phi^{}_i(y_{\sim i},x)$. 
From (\ref{equ:gexitcompact}), it follows that
\begin{align*}
\gexitfi {}(\graph,\cp) & =  
\int_{\phi^{}_i,y_i, x_i}
p(x_i)
p(\phi^{}_i \mid x_i)
\frac{\text{d}\phantom{\ent}}{\text{d} \cp} p(y_i|x_i)\cdot  \\
& 
~~~~~\cdot \log\left(\int_{x'_i}\frac{p(x'_i|\phi^{}_i)
p(y_i|x'_i)}{p(x_i|\phi^{}_i)p(y_i|x_i)}\text{d}x_i' 
\right) \text{d}x_i \text{d} y_i \text{d} \phi_i^{}.
\end{align*}
To simplify the computations, a few remarks are of order. 
First recall that we have chosen $\cp$ to be $\cp=-2\snr=\frac{-2}{\sigma^2}$. 
Second, observe that the Gaussian density permits us to write 
$\frac{\text{d} p(y_i|x_i)}{\text{d} \cp}   = \frac{x_i}{\sqrt{\snr}}\frac{\text{d}}{\text{d}y_i}p(y_i|x_i).$ 
Therefore, integrating by parts with respect to $y_i$, we get
\begin{align*}
&~~~\gexitfi {}(\graph,\cp) \\
& =  \int_{\phi^{}_i,y_i, x_i}
p(x_i)p(\phi^{}_i \mid x_i) \frac{x_i}{\sqrt{\snr}}
 p(y_i|x_i)\cdot  \\
&  
~~~~~\cdot \frac{\text{d}}{\text{d}y_i}\left\{\log\left(\int_{x'_i}\frac{p(x'_i|\phi^{}_i)
p(y_i|x'_i)  }{p(x_i|\phi^{}_i)p(y_i|x_i)}\text{d}x_i'
\right) \right\} \text{d}x_i \text{d} y_i \text{d} \phi_i^{}\\
& = 
-\int_{\phi^{}_i,y_i, x_i}
p(x_i)p(\phi^{}_i \mid x_i) \frac{x_i}{\sqrt{\snr}}
 p(y_i|x_i)\cdot  \\
&  
~~~~~\cdot  \frac{\int_{x_i'}\ \sqrt{\snr}(x_i'-x_i)p(x_i'|\phi^{}_i)p(y_i|x_i') 
 \text{d}x_i'
  }{\int_{x'_i} p(x'_i|\phi^{}_i)
p(y_i|x'_i) \text{d}x_i' }
  \text{d}x_i \text{d} y_i \text{d} \phi_i^{},
\end{align*}
after having used $\frac{\text{d}p(y_i|x_i')}{\text{d}y_i}=\frac{\text{d}p_{Z_i}(y_i-\sqrt{\snr}x_i')}{\text{d}y_i}=-(y_i-\sqrt{\snr}x_i')p(y_i|x_i')$.
Let us now re-order as $p(x_i'|\phi_i)p(y_i|x_i')=p(x_i'|\phi_i,y_i)p(y_i|\phi_i)$ and use (with a slight abuse of notations) $\frac{y_i+\phi_i}{\sqrt{\snr}}=\expectation_{X_i}\left[X_i|\phi^{}_i,y_i\right]$ to get
\begin{align*}
& ~~~\gexitfi {}(\graph,\cp) \\
 & = -\int_{\phi^{}_i,y_i, x_i}
p(x_i)p(\phi^{}_i \mid x_i) x_i
 p(y_i|x_i)\cdot  \\
&  ~~~~~
\cdot  \frac{p(y_i|\phi_i) (\frac{(y_i+\phi_i) }{\sqrt{\snr}}-x_i)  }{p(y_i|\phi_i)}
  \text{d}x_i \text{d} y_i \text{d} \phi_i^{}\\
& =  \int_{\phi^{}_i,y_i}  p(\phi_i,y_i)\\
&  ~~~~~ 
\cdot \int_{x_i} p(x_i|y_i,\phi_i) \left(x_i^2-\frac{(y_i+\phi_i) x_i}{\sqrt{\snr}}\right)\text{d}x_i \text{d} y_i \text{d} \phi_i^{}\\
& = \int_{\phi^{}_i,y_i}  p(\phi_i,y_i) \\
&  ~~~~~ 
\cdot \left(\expectation_{X_i}\left[X_i^2|\phi^{}_i,y_i\right] - \expectation_{X_i}\left[X_i|\phi^{}_i,y_i\right]^2\right) \text{d} y_i \text{d} \phi_i^{}.  
\end{align*}
This concludes our proof since $\Phi_i$ is a sufficient statistic for $Y_{\sim i}$.
\end{proof}

%In hindsight, it is interesting to note that in probability theory
%a relationship which is equivalent to
%the connection between the derivative of the mutual information
%and the MSE is know under the name ``???'' and is due to
%RussianGuy \cite{}.

%************************************************************
%
\section{Physical Degradation: a Calculus Proof}
\label{sec:AlternativePhysicalDegradation}

In this appendix we provide a direct calculus proof of 
Corollary \ref{lemma:orderingviaphysicaldegradationexit}, exploiting
the explicit representation provided by Lemma \ref{lem:gexitlinear}.
As a byproduct we show that the $\gexit$ kernel in the
$|D|$-domain is non-increasing and concave. This fact is also used in the 
proof of Lemma \ref{lem:GeneralBounds}.

For our purpose it is convenient to represent all quantities
in the $|D|$-domain.
Let $\{\absDdens{c_{\BMSsmall(\ent)}}\}_{\ent}$ denote the
family of $|D|$-densities characterizing the channel family.
Let
$\gexitkabsd {\BMSsmall(\ent)} w$ denote the $\gexit$ kernel
in the $|D|$-domain as introduced in (\ref{equ:gexitkernelconversion}).
We can rewrite it in the form
\begin{align*}
\gexitkabsd {\BMSsmall(\ent)} w  & =
\int_{0}^{1}
\frac{\partial \absDdens{c_{\BMSsmall(\ent)}}(z)}{\partial \ent} \alpha(z, w) \mbox{d}z,
\end{align*}
where
\begin{align*}
\alpha(z, w) = \frac{1}{4} \sum_{i, j = \pm 1} (1+i z)(1+j w) \beta(i z, j w),
\end{align*}
with $\beta(z, w)=\log_2\bigl(1+e^{-2 \tanh^{-1}(z)} e^{-2 \tanh^{-1}(w)} \bigr)$.
Finally, let $\absDdens{a}$ and $\absDdens{b}$ denote the two symmetric densities
in the $|D|$-domain.

The claim of the theorem is then equivalent to the statement
that the $\gexit$ functional
$\int_{0}^{1} \gexitkabsd {\BMSsmall(\ent)} w \absDdens{a}(w) \text{d} w$
preserves the partial order
implied by physical degradation. This means that if
$\absDdens{a} \prec \absDdens{b}$ then
\begin{align*}
\int_{0}^{1} \gexitkabsd {\BMSsmall(\ent)} w \absDdens{a}(w) \text{d} w
& \leq
\int_{0}^{1} \gexitkabsd {\BMSsmall(\ent)} w \absDdens{b}(w) \text{d} w.
\end{align*}
By Theorem 3.4 in \cite{RiU05}, a $|D|$-domain kernel preserves the partial order
implied by physical degradation if it is non-increasing and concave on $[0, 1]$,
i.e., if its first two derivatives are non-positive. This means we need to show that
\begin{align*}
\int_{0}^{1}
\frac{\partial \absDdens{c_{\BMSsmall(\ent)}}(z)}{\partial \ent} \frac{\partial^i \alpha(z, w)}{\partial w^i} \mbox{d}z \leq 0,
\end{align*}
for $i=1, 2$.
By the same Theorem 3.4 the above condition is verified if both
$\frac{\partial^i \alpha(z, w)}{\partial w^i}$ for $i=1, 2$, are convex and
non-decreasing. This in turn is true if
$\frac{\partial^{i+j} \alpha(z, w)}{\partial w^i \partial z^j} \geq 0$ for
$i, j =1, 2$. Now some further calculus shows that
\begin{align}
\frac{\partial \alpha(z, w)}{\partial w}  =  &
\frac{1}{2} \sum_{i=\pm 1} iz \log_2(1+i w z)- \nonumber \\
\phantom{=} & \frac{1}{2}\sum_{i=\pm 1} i\log(1+iw), \label{equ:first} \\
\ln(2) \frac{\partial^2 \alpha(z, w)}{\partial w^2}  = & \frac{z^2}{1-w^2 z^2}-\frac{1}{1-w^2}. \label{equ:second}
\end{align}
Note that equation (\ref{equ:second}) implies that $\frac{\partial^2 \alpha(z, w)}{\partial w^2}$
has a positive expansion in $z$ (except for the constant term). Therefore the derivatives
$\frac{\partial^{2+i} \alpha(z, w)}{\partial w^2 \partial z^i}$, $i = 1, 2$,
are both positive and
by symmetry of the function $\alpha(z, w)$ in its arguments $z$ and $w$ so is  $\frac{\partial^{3} \alpha(z, w)}{\partial w \partial z^2}$.
Finally,
\begin{align*}
\log(2) \frac{\partial^2 \alpha(z, w)}{\partial w \partial z} & =
\frac{1}{2} \ln\frac{1+wz}{1-wz} + \frac{wz}{1-w^2 z^2} \\
& = 2 w z \sum_{i \geq 0} \frac{(i+1)(w^2 z^2)^i}{2 i+1},
\end{align*}
which has a positive Taylor series expansion as well.
This confirms our claim that the $\gexit$ kernel preserves the partial
order implied by physical degradation.
%
%***********************************************************************
%
\section{Proof of Eq.~(\ref{eq:Contraction})}
\label{sec:ProofContraction}

In this appendix we prove the claim (\ref{eq:Contraction}).
First notice that $\batta(\Tc_{\ent}(\Ldens{a}))= B_{\ent}\, 
\lambda(\batta(\rho(\Ldens{a})))$. Since
$0\le \lambda(x)\le 1$ and $\lambda'(x)\le \lambda'(1)$, we have
\begin{align}
|\batta(T_{\ih_1}(\Ldens{a}_{1}))-&\batta(T_{\ih_2}(\Ldens{a}_{2}))|
\le \label{eq:Contraction0}\\ &\lambda'(1)B_{\ih_1}\, 
|\batta(\rho(\Ldens{a}_{1}))-\batta(\rho(\Ldens{a}_{2}))|
+|B_{\ih_1}-B_{\ih_2}|\, .\nonumber
\end{align}
In order to estimate $|\batta(\rho(\Ldens{a}_{1}))-
\batta(\rho(\Ldens{a}_{2}))|$,
define, for $t\in[0,1]$, $\Ldens{a}_{t} = (1-t)\Ldens{a}_1+t\Ldens{a}_2$,
and write
\begin{align}
|\batta(\rho(\Ldens{a}_{1}))-\batta(\rho(\Ldens{a}_{2}))|\le
\int_0^1 \left|\frac{\de \batta(\rho(\Ldens{a}_t))}{\de t}\,\right|\de t\, .
\label{eq:Final1}
\end{align}
The derivative of the Battacharyya parameter
is easily computed (to lighten the
notation we omit hereafter the argument of $\batta(\,\cdot\,)$
in the derivative). 
The result is most conveniently 
expressed in terms of densities of the variable
$u\defas\sqrt{1-\tanh^2(x/2)}$, where $x$ is the log-likelihood ratio
(this quantity is equivalent to the  $|D|$-variable and its
expectation is Battacharyya parameter).
If we  denote the corresponding densities by the same symbols,
we get
\begin{align}
\frac{\de \batta}{\de t}
=\rho'(1)
\int_{0}^1\,&
\sqrt{u_1^2+u_2^2- u_1^2u_2^2}\;\cdot\label{eq:DerivativeExpression}\\
&\phantom{aaa}\cdot (\Ldens{a}_2(u_1)-\Ldens{a}_1(u_1))\, \Ldens{b}(u_2)\,
\de u_1\,\de u_2\, ,\nonumber
\end{align}
where we introduced the density
\begin{align*}
\Ldens{b} \defas \frac{1}{\rho'(1)}\sum_{\rdegree}\rho_{\rdegree}
(\rdegree-1)\,\Ldens{a}_t^{*(\rdegree-2)}\, .
\end{align*}
Using integration by parts with respect to $u_1$ in 
Eq.~(\ref{eq:DerivativeExpression}) and denoting by ${\mathsf A}_1$,
${\mathsf A}_2$ the distributions corresponding to densities  $\Ldens{a}_1$,
$\Ldens{a}_2$, we get
\begin{align*}
\frac{\de \batta}{\de t}
=\rho'(1)
\int_{0}^1\,&\frac{u_1(1-u_2^2)}{\sqrt{u_1^2+u_2^2-u_1^2u_2^2}}\cdot\\
&\phantom{aaa}\cdot ({\mathsf A}_2(u_1)-{\mathsf A}_1(u_1))\, \Ldens{b}(u_2)\,
\de u_1\,\de u_2\, ,
\end{align*}
Since $\Ldens{a}_2$ is physically degraded with respect to 
$\Ldens{a}_1$, ${\mathsf A}_2(u)\ge {\mathsf A}_1(u)$ for any 
$u\in[0,1]$. Furthermore $\int {\mathsf A}_i(v)\, \de v = 
\batta(\Ldens{a}_i)$. Therefore 
\begin{align}
\frac{\de \batta}{\de t}
=\rho'(1)
[\batta(\Ldens{a}_2)-\batta(\Ldens{a}_1)]\, \Xi\, ,\label{eq:Final2}
\end{align}
where
\begin{align*}
\Xi = \int_{0}^1\,\frac{u_1(1-u_2^2)}{\sqrt{u_1^2+u_2^2-u_1^2u_2^2}}\;
f(u_1)\, \Ldens{b}(u_2)\,
\de u_1\,\de u_2\, ,
\end{align*}
and $f$ is a function on $[0,1]$ non negative and with unit integral.
In other words, $f$ is a probability density function.
Since $\sqrt{u_1^2+u_2^2-u_1^2u_2^2}\ge u_1$, we obtain the bound 
\begin{align*}
\Xi &\le \int_{0}^1\,(1-u^2)\; \Ldens{b}(u)\,\de u\\
&=\frac{1}{\rho'(1)}\sum_{\rdegree}\rho_{\rdegree}
(\rdegree-1) \left[\int_{0}^1\,(1-u^2)\; \Ldens{a}_t(u)\right]^{\rdegree-2}\, ,
\end{align*}
where we used the definition of $\Ldens{b}$. If we further 
notice that $\int_{0}^1\,u\; \Ldens{a}_t(u) =\batta(\Ldens{a}_t)\ge
\batta(\Ldens{a}_1)$, we get
\begin{align}
\Xi\le \frac{1}{\rho'(1)}\, \rho''(1-\batta(\Ldens{a}_1)^2)\, .
\label{eq:Final3}
\end{align}
The claim follows by putting together Eqs.~(\ref{eq:Final1}), 
(\ref{eq:Final2}), and (\ref{eq:Final3}).
%
%***********************************************************************
%
\section{$\MAP$ Versus $\BP$ Marginals: Some Technical Details}
\label{app:BPcorrect}

In this appendix we present the proofs which were omitted in 
Sec.~\ref{sec:mapversusbp}.

\begin{proof}[Lemma \ref{lem:GEXITSquareError}]
Let us make a few preliminary remarks.
The first one follows immediately from the definition:
\begin{align}
\E\left\{ \mu_{{\sf Y}}(Y)\mid Z=z \right\} = \mu_{{\sf Z}}(z)\, .
\end{align}
In fact, using the Markov property, the left hand side can be written as
$\E\{\E[X\mid Y]\mid Z=z\}=\E\{\E[X\mid Y, Z]\mid Z=z\}$
that is equal to $\E[X\mid Z=z]\equiv \mu_{{\sf Z}}(z)$.

The second remark is that, by elementary calculus, for any $0\le x_0\le x\le 1$
\begin{align*}
k(x)\le k(x_0)-\frac{1}{2}\, K\, (x^2-x_0^2)\, .
\end{align*}

Finally, for any random variable $W$, taking values in $[0,1]$, we have
(here ${\rm Var}(W)$ is the variance of $W$):
\begin{align*}
\E\,k(W)\le k(\E\, W)-\frac{1}{2}\, K\, {\rm Var}(W)\, .
\end{align*}
In fact, by Taylor expansion $k(W)\le k(w_0)+k'(w_0)(W-w_0)
-\frac{1}{2}\, K\, (W-w_0)$, for any $w_0\in[0,1]$. The claim is proved by 
taking expectation of both sides and setting $w_0= \E W$.

These ingredients are put together as follows 
(here we use the shorthands $\mu_{{\sf Y}}$ and $\mu_{{\sf Z}}$
for, respectively, $\mu_{{\sf Y}}(Y)$ and $\mu_{{\sf Z}}(Z)$)
\begin{eqnarray*}
&&\hspace{-0.5cm}\E[k(|\mu_{{\sf Y}}|)]  =  \E\left\{\E[k(|\mu_{{\sf Y}}|)|Z] \right\}\\
&&\le  \E\left\{k(\E[|\mu_{{\sf Y}}||Z]) -\frac{1}{2}\, K\,
{\rm Var}(|\mu_{{\sf Y}}|\, \mid Z)\right\}\\
&&\le  \E\left\{k(|\E[\mu_{{\sf Y}}|Z]|)-\frac{1}{2}\, K\,
\left(\E[|\mu_{{\sf Y}}||Z]^2-\E[\mu_{{\sf Y}}|Z]^2\right)-\right.\\
&&\left.\hspace{5.cm}
-\frac{1}{2}\, K\,{\rm Var}(|\mu_{{\sf Y}}|\, \mid Z)\right\}\\
&& = \E\left\{k(|\E[\mu_{{\sf Y}}|Z]|)-\frac{1}{2}\, K\,
\E[(\mu_{{\sf Y}}-\E[\mu_{{\sf Y}}|Z])^2|Z]\right\}\\
&& =  E\left\{k(|\mu_{{\sf Z}}|)-\frac{1}{2}\, K\,
\E[(\mu_{{\sf Y}}-\mu_{{\sf Z}})^2|Z]\right\}\\
&& =  E[k(|\mu_{{\sf Z}}|)]-\frac{1}{2}\, K\,
\E[(\mu_{{\sf Y}}-\mu_{{\sf Z}})^2]\, ,
\end{eqnarray*}
which completes the proof.
\end{proof}
\begin{proof}[Lemma \ref{lem:ExtrinsicDistortion}]
We claim (and will prove later) that
\begin{eqnarray*}
\left|\tilde{\mu}^{\BPsmall,\ell}_i(Y)-\tilde{\mu}_i(Y)
\right|\le e^{|l(Y_i)|}\left|\mu^{\BPsmall,\ell}_i(Y)-\mu_i(Y)\right|\, ,
\end{eqnarray*}
where $l(y_i)$ is the log-likelihood associated to the channel output
$y_i$.
If we square and take expectation with respect to $Y$
(recalling that $\mu^{\BPsmall,\ell}_i(Y)$, $\mu_i(Y)$ do not depend upon
$Y_i$), we get
\begin{eqnarray*}
\E\left\{\left|\tilde{\mu}^{\BPsmall,\ell}_i(Y)-\tilde{\mu}_i(Y)
\right|^2\right\}\le C\, 
\E\left\{\left|\mu^{\BPsmall,\ell}_i(Y)-\mu_i(Y)\right|^2\right\}\, .
\end{eqnarray*}
The thesis follows by summing over $i$.

We are left with the task of proving the first claim above.
We recall that the conditional expectations can be represented in terms
of extrinsic log-likelihoods as
\begin{eqnarray*}
\mu_i(y) &= &\tanh\left[\frac{1}{2}\phi^{}_i(y_{\sim i})\right]
\, ,\\
\tilde{\mu}_i(y) &=& \tanh\left[\frac{1}{2}(l(y_i)+\phi^{}_i(y_{\sim i}))\right]\, .
\end{eqnarray*}
Analogous formulae hold if we replace $\mu_i(y)$ (respectively
$\tilde{\mu}_i(y)$) with $\mu^{\BPsmall,\ell}_i(y)$ (respectively
$\tilde{\mu}^{\BPsmall,\ell}_i(y)$) and
$\phi^{}_i(y_{\sim i})$ with $\phi^{\BPsmall,\ell}_i(y_{\sim i})$.
The claim follows immediately form the following 
calculus exercise below.
\end{proof}

\begin{fact}
For any $x_1,x_2,z\in{\mathbb R}$
\begin{align*}
|\tanh(x_1+z)-\tanh(x_2+z)&|\le \\
&e^{2|z|} |\tanh(x_1)-\tanh(x_2)| \, .
\end{align*}
\begin{proof}
Consider, without loss of generality, $x_1>x_2$ and $z<0$.
It is simple to realize that, for any $x\in{\mathbb R}$
\begin{eqnarray*}
1+\tanh(x+z)&\le & 1+\tanh(x)\, ,\\
1-\tanh(x+z)&\le & e^{-2z}(1-\tanh(x))\, .
\end{eqnarray*}
The last statement follows by writing  
$1+e^{2(x+z)}\ge e^{2z}(1+e^{2x})$ and taking the inverse of both sides.

The thesis is proved by multiplying these inequalities, and
integrating over $x\in[x_1,x_2]$.
\end{proof}
\end{fact}
%
%***********************************************************************
%

% bibliography
% ------------
\bibliographystyle{IEEEtran} %plain

\newcommand{\SortNoop}[1]{}

%\begin{biography}{Michael Shell}
%\end{biography}

% bibliography
% ------------
%\bibliographystyle{IEEEtran} %plain
%\bibliography{lth,lthpub}

%\begin{biographynophoto}{John Doe}
%\end{biographynophoto}

% insert where needed to balance the two columns on the last page
%\newpage

%\begin{biographynophoto}{Jane Doe}
%\end{biographynophoto}


%\vfill


\end{document}